\RequirePackage{fix-cm} 
\documentclass[ twoside,openright,titlepage,numbers=noenddot,headinclude,
                footinclude=true,cleardoublepage=empty,abstractoff, 
                BCOR=5mm,paper=a4,fontsize=11pt,
                ngerman,american,%
                ]{scrreprt}

\PassOptionsToPackage{utf8}{inputenc}
	\usepackage{inputenc}

\PassOptionsToPackage{
	listings,%
					 pdfspacing,
					 subfig,beramono,
					 parts}{classicthesis}                       %

\newcommand{\myTitle}{Construction and characterisation of~a~coded-mask gamma camera for beam range~monitoring in proton therapy\xspace}
\newcommand{\mySubtitle}{PhD Dissertation\xspace}

\newcommand{\myName}{Magdalena Ko\l{}odziej\xspace}

\newcommand{\myFaculty}{Faculty of Physics, Astronomy and Applied Computer Science\xspace}

\newcommand{\myUni}{Jagiellonian University\xspace}

\newcommand{\myTime}{February 2025\xspace}
\newcommand{\myVersion}{version 1.0\xspace}

\newcounter{dummy} 
\providecommand{\mLyX}{L\kern-.1667em\lower.25em\hbox{Y}\kern-.125emX\@}

\PassOptionsToPackage{polish,british}{babel} 
	\usepackage{babel}                  

\usepackage{csquotes}

\PassOptionsToPackage{fleqn}{amsmath}       
    \usepackage{amsmath}

\PassOptionsToPackage{T1}{fontenc} 
    \usepackage{fontenc}     
\usepackage{textcomp} 
\usepackage{scrhack} 
\usepackage{xspace} 
\usepackage{mparhack} 
\usepackage{fixltx2e} 
\PassOptionsToPackage{printonlyused,smaller}{acronym} 
    \usepackage{acronym} 
    
\usepackage{tabularx} 
\usepackage{caption}
\usepackage{listings} 
\PassOptionsToPackage{pdftex,hyperfootnotes=false,pdfpagelabels}{hyperref}
    \usepackage{hyperref}  
\PassOptionsToPackage{pdftex}{graphicx}
    \usepackage{graphicx}

\hypersetup{%
    colorlinks=true, linktocpage=true, pdfstartpage=3, pdfstartview=FitV,%
    breaklinks=true, pdfpagemode=UseNone, pageanchor=true, pdfpagemode=UseOutlines,%
    plainpages=false, bookmarksnumbered, bookmarksopen=true, bookmarksopenlevel=1,%
    hypertexnames=true, pdfhighlight=/O,
    urlcolor=webbrown, linkcolor=RoyalBlue, citecolor=webgreen, 
    pdftitle={\myTitle},%
    pdfauthor={\textcopyright\ \myName, \myUni, \myFaculty},%
    pdfsubject={},%
    pdfkeywords={},%
    pdfcreator={pdfLaTeX},%
    pdfproducer={LaTeX with hyperref and classicthesis}%
}   

\makeatletter
\@ifpackageloaded{babel}%
    {%
       \addto\extrasamerican{%
                }%
       \addto\extrasngerman{%
                }%
    }{\relax}
\makeatother

\usepackage{classicthesis} 

\usepackage{indentfirst}
\usepackage{microtype}
\DisableLigatures{encoding = *, family = * }
\usepackage{amssymb}
\usepackage{url,hyperref}
\usepackage{multirow,multicol}
\usepackage[singlelinecheck=false]{caption}
\usepackage{booktabs,float,tabularx,ltablex}
\usepackage[flushleft]{threeparttable}
\usepackage{enumitem}
\usepackage{caption}
\usepackage{subcaption}
\usepackage{comment}
\usepackage{pstricks,graphicx}
\usepackage{amsmath}
\usepackage{lineno}
\usepackage{siunitx} 
\usepackage{eqparbox}

\newfloat{wykres}{tbph}{loc}
\floatname{wykres}{Wykres}
\usepackage{cleveref}
\crefname{paragraph}{Paragraph}{Paragraphs}
\crefname{figure}{Figure}{Figures}
\crefname{section}{Section}{Sections}
\crefname{equation}{Equation}{Equations}
\crefname{table}{Table}{Tables}

\usepackage[acronym,nonumberlist]{glossaries-extra}
\setabbreviationstyle[acronym]{long-short}
\glsdisablehyper
\newacronym{fee}{FEE}{front-end electronics}
\newacronym[shortplural={PCBs},longplural=printed circuit boards]{pcb}{PCB}{printed circuit board}
\newacronym[shortplural={SiPMs},longplural=silicon photomultipliers]{sipm}{SiPM}{silicon photomultiplier}
\newacronym{fov}{FOV}{field of view}
\newacronym{pmma}{PMMA}{polymethyl methacrylate}
\newacronym{daq}{DAQ}{data acquisition}
\newacronym{adc}{ADC}{analog-to-digital converter}
\newacronym{qdc}{QDC}{charge-to-digital converter}
\newacronym[shortplural={PGs},longplural=prompt gammas]{pg}{PG}{prompt gamma}
\newacronym{pgh}{PG}{prompt-gamma} 
\newacronym{pgi}{PGI}{prompt-gamma imaging}
\newacronym{pt}{PT}{proton therapy}
\newacronym{mlem}{MLEM}{maximum-likelihood expectation maximisation}
\newacronym{pet}{PET}{positron emission tomography}
\newacronym{mri}{MRI}{magnetic resonance imaging}
\newacronym{sificc}{SiFi-CC}{\textbf{Si}licon Photomultiplier and Scintillating \textbf{Fi}bre based \textbf{C}ompton \textbf{C}amera}
\newacronym{uqi}{UQI}{universal image quality index}
\newacronym{dfpd}{DFPD}{distal falloff position determination}
\newacronym[shortplural={DFPs},longplural=distal fall-off positions]{dfp}{DFP}{distal fall-off position}
\newacronym{mura}{MURA}{modified uniformly redundant array}
\newacronym{cog}{CoG}{centre of gravity}
\newacronym{mps}{MPS}{multi-parallel slit}
\newacronym{kes}{KES}{knife-edge slit}
\newacronym[shortplural={ROIs},longplural=regions of interest]{roi}{ROI}{region of interest}
\newacronym{dpc}{DPC}{digital photon counter}
\newacronym{llr}{LLR}{low-level reconstruction}
\newacronym{snr}{SNR}{signal-to-noise ratio}
\newacronym{ssp}{SSP}{small-scale prototype}
\newacronym{hit}{HIT}{Heidelberger Ionenstrahl-Therapiezentrum}
\newacronym{cc}{CC}{Compton camera}
\newacronym{bp}{BP}{Bragg peak}
\newacronym{let}{LET}{linear energy transfer}
\newacronym{cm}{CM}{coded mask}
\newacronym{cmh}{CM}{coded-mask}
\newacronym{iqr}{IQR}{interquartile range}
\newacronym{rmse}{RMSE}{root mean squared error}
\newacronym{sm}{SM}{system matrix}
\newacronym{1dcm}{1D~CM}{one-dimensional coded mask}
\newacronym{2dcm}{2D~CM}{two-dimensional coded mask}
\newacronym{asic}{ASIC}{application-specific integrated circuit}
\newacronym{pcie}{PCIe}{Peripheral Component Interconnect Express}
\newacronym{gui}{GUI}{graphical user interface}
\makeglossaries

\usepackage[a4paper, text={140mm,240mm}, headsep=10mm, footskip=12mm, bindingoffset=1.5cm,top=27mm]{geometry}
\titleformat{\chapter}[display]%
  {\relax}{\mbox{}\vskip-3\baselineskip\hfill\color{gray}\chapterNumber\thechapter}{0pt}%
  {\raggedright\spacedallcaps}[\normalsize\vspace*{.8\baselineskip}\titlerule]%

\usepackage{subcaption}

\usepackage[subfigure]{tocloft}
\usepackage{xfrac} 

\newcounter{lofdepth}
\newcounter{lotdepth}
\cftpagenumbersoff{subfigure}

\usepackage{booktabs} 
\usepackage{colortbl} 

\usepackage{bm} 

\numberwithin{figure}{chapter} 
\numberwithin{table}{chapter} 

\usepackage{pdfpages}

\begin{document}
\frenchspacing
\raggedbottom
\selectlanguage{american} 
\pagenumbering{roman}
\pagestyle{plain}


\begin{titlepage}
    \begin{addmargin}[-1cm]{-1cm}
    \begin{center}
        \large  

        \hfill

        \vfill

        \begingroup
            \color{Black}\spacedallcaps{\myTitle} \\ \bigskip
        \endgroup

        \spacedlowsmallcaps{\myName}

        \vfill

        \includegraphics[width=8cm]{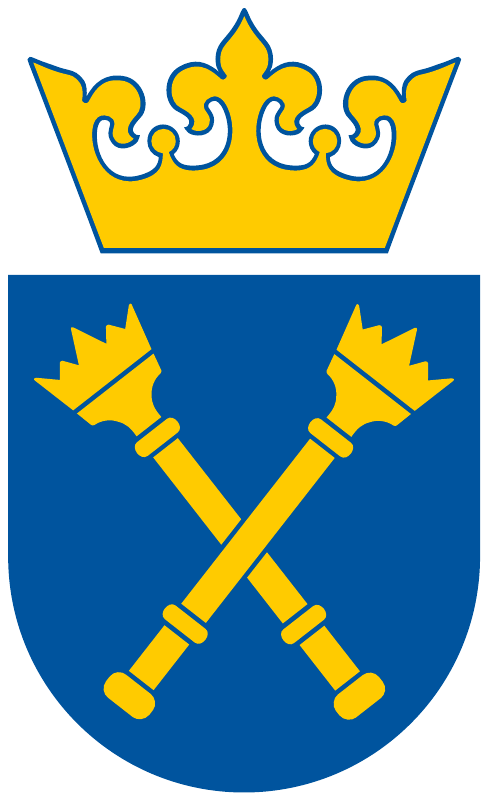} \\ \medskip

        \mySubtitle \\ \medskip 

        Supervisor: dr hab. Aleksandra Wrońska, prof. UJ \\
        \medskip
        \spacedlowsmallcaps{\myFaculty} \\
        \spacedlowsmallcaps{\myUni} \\ \bigskip \bigskip
        Kraków, Poland \\ \medskip
        \myTime

        \vfill                      

    \end{center}  
  \end{addmargin}       
\end{titlepage}

\thispagestyle{empty}

\hfill

\vfill

\begin{titlepage}
    \begin{addmargin}[-1cm]{-1cm}
    \begin{center}
        \large  

        \hfill

        \vfill

        \begingroup
            \color{Black}\spacedallcaps{Budowa i charakteryzacja kamery gamma z maską kodowaną do kontroli zasięgu wiązki w terapii protonowej} \\ \bigskip
        \endgroup

        \spacedlowsmallcaps{\myName}

        \vfill

        \includegraphics[width=8cm]{img/her_pds_c_1.pdf} \\ \medskip

        Praca doktorska \\ \medskip 

        Promotor: dr hab. Aleksandra Wrońska, prof. UJ \\
        \medskip
        \spacedlowsmallcaps{Wydział Fizyki, Astronomii i Informatyki Stosowanej} \\
        \spacedlowsmallcaps{Uniwersytet Jagielloński} \\ \bigskip \bigskip
        Kraków \\ \medskip
        Luty 2025

        \vfill                      

    \end{center}  
  \end{addmargin}       
\end{titlepage}   
\thispagestyle{empty}

\hfill

\vfill

\cleardoublepage
\pdfbookmark[1]{Abstract}{Abstract}
\begingroup
\let\clearpage\relax
\let\cleardoublepage\relax
\let\cleardoublepage\relax

\chapter*{Abstract}
The major advantage of proton therapy over conventional radiotherapy is the dose deposition pattern: unlike X-rays, protons are entirely stopped in patient's tissues with a distinct maximum of energy deposition at the end of their range, the Bragg peak. This enables precise coverage of the tumour volume while sparing nearby healthy tissues. However, accurate control of the proton beam range online during patient irradiation is still considered a challenge. Thus, there are extensive efforts to develop a detector capable of \textit{in vivo} beam range monitoring for proton therapy, among them, the coded-mask gamma camera developed by the SiFi-CC group. The assembly of this detector, as well as its first test under clinical conditions, are the subject of this thesis. The detector consists of a structured collimator made of tungsten, and a stack of thin, elongated scintillation fibres, read out at both ends by silicon photomultipliers. Such a design allows for efficient detection of prompt gammas and reconstruction of a prompt gamma depth profile, which is strongly correlated with the proton beam range. The readout system for the detector was selected in an extensive comparison study of five different systems. After detector assembly, the detector was tested under clinical conditions: a phantom was irradiated with proton beams of energies from \SI{70.51}{MeV} to \SI{108.15}{MeV}, the detector response was processed with low-level reconstruction software, and the maximum-likelihood expectation maximisation algorithm was applied to reconstruct the prompt-gamma depth profiles (images). The distal falloff positions of these profiles were determined and compared with the calculated proton ranges. The image reconstruction parameters were optimised in terms of accuracy of proton range shift determination.  
The results obtained were checked against a Geant4 simulation, showing good consistency. The statistical precision was found to be \SI{1.7}{mm} for $10^8$ protons for a reference position close to the centre of 
the camera field of view. The result is comparable to similar systems investigated by other research groups. However, simulation results suggest that certain hardware amendments to the detector can increase that precision about four times. 

Since this thesis was pursued within the SiFi-CC project, which is a collaborative effort, the form "we" is widely used throughout the text. However, the majority of the work described here is the author's own contribution. In cases where a certain part of the work presented was done by another person, the fact is clearly stated in the footnote. Otherwise, the work was conducted by the author of this thesis.

\newpage 

\ 

\newpage

\begin{otherlanguage}{polish}
\pdfbookmark[1]{Streszczenie}{Streszczenie}
\chapter*{Streszczenie}
Główną zaletą terapii protonowej w porównaniu z konwencjonalną radioterapią jest rozkład dawki: w przeciwieństwie do promieniowania rentgenowskiego, protony są całkowicie zatrzymywane w tkankach pacjenta, gdzie występuje wyraźne maksimum depozycji energii - pik Bragga. Umożliwia to precyzyjne pokrycie objętości guza przy jednoczesnej minimalizacji dawki w otaczających zdrowych tkankach. Wyzwaniem jednak wciąż pozostaje dokładna kontrola zasięgu wiązki protonów w czasie rzeczywistym podczas napromieniowania pacjenta. Dlatego podejmowane są szeroko zakrojone wysiłki w celu opracowania detektora umożliwiającego monitorowanie zasięgu wiązki \textit{in vivo} w terapii protonowej, a jedną z rozważanych opcji jest kamera gamma z maską kodowaną, opracowana przez grupę SiFi-CC. Budowa tego detektora, a także jego pierwszy test w warunkach klinicznych, są przedmiotem niniejszej pracy. Detektor składa się ze strukturalnego kolimatora wykonanego z wolframu i wiązki cienkich, wydłużonych włókien scyntylacyjnych, odczytywanych na obu końcach przez fotopowielacze krzemowe. Taka konstrukcja umożliwia wydajną detekcję natychmiastowego promieniowania gamma i rekonstrukcję profilu głębokościowego tego promieniowania, który jest silnie skorelowany z zakresem wiązki protonów. System odczytu detektora został wybrany w obszernym badaniu porównawczym pięciu różnych systemów. Po zmontowaniu, detektor został przetestowany w warunkach klinicznych: napromieniano fantom wiązkami protonów o energiach od $70,\!51\:\mathrm{MeV}$ do $108,\!15\:\mathrm{MeV}$, odpowiedź detektora została przetworzona za pomocą oprogramowania do rekonstrukcji niskiego poziomu, a do rekonstrukcji profili głębokości natychmiastowego promieniowania gamma (obrazów) zastosowano algorytm MLEM. Na profilach zlokalizowano pozycje spadku dystalnego i porównano je z obliczonymi zasięgami protonów. Parametry rekonstrukcji obrazu zoptymalizowano pod kątem dokładności wyznaczenia przesunięcia zasięgu protonów. Uzyskane wyniki porównano z symulacją wykonaną w środowisku Geant4, wykazując, że oba podejścia dają w dużej mierze spójne wyniki. Uzyskano precyzję statystyczną na poziomie $1,\!7\:\mathrm{mm}$ dla $10^8$ protonów, w pozycji odniesienia blisko środka pola widzenia kamery. Wynik jest porównywalny z podobnymi systemami rozwijanymi przez inne grupy badawcze. Wyniki symulacji sugerują jednak, że pewne modyfikacje sprzętowe detektora mogą zwiększyć tę precyzję około czterokrotnie.

Ponieważ niniejsza praca została zrealizowana w ramach projektu SiFi-CC, który jest pracą zbiorową, forma „my” jest szeroko stosowana w całym tekście. Niemniej jednak większość opisanej tutaj pracy jest własnym wkładem autorki. W przypadkach, gdy pewna część przedstawionej pracy została wykonana przez inną osobę, fakt ten jest wyraźnie zaznaczony w przypisie. W przeciwnym razie praca została wykonana przez autorkę niniejszej dysertacji.
\end{otherlanguage}

\endgroup			
\newpage 

\ 

\newpage
\cleardoublepage
\pdfbookmark[1]{Acknowledgments}{acknowledgments}

\bigskip

\begingroup
\let\clearpage\relax
\let\cleardoublepage\relax
\let\cleardoublepage\relax
\chapter*{Acknowledgments}
I would like to express my sincere gratitude to my supervisor, Prof. Aleksandra Wrońska, for her invaluable support, guidance, and encouraging me to take up challenges. Her expertise, insightful feedback, and inspiring passion for science taught me essential lessons, that helped me grow as a young researcher.
\newline

I am thankful to Katarzyna Rusiecka, for her invaluable practical advice and patience, countless times when she helped me solve problems I was stuck on, as well as all the good days spent together desk-by-desk in B-2-37.
\newline

I would like to thank Ming Liang Wong, for all the helpful advice and fruitful discussions. I genuinely enjoyed the cooperation on the DAQ comparison project, it was really inspiring and taught me a lot.
\newline

My appreciation extends to all the current and former members of the SiFi-CC group with whom I had the pleasure to work and whose input helped to write this thesis, especially Ronja Hetzel, Jonas Kasper, Monika Kercz, Rafał Lalik, Prof. Andrzej Magiera, Linn Mielke, Gabriel Ostrzołek, Prof. Magdalena Rafecas, and Vitalii Urbanevych. 
\newline

I truly appreciate Wojciech Migdał, for invaluable technical support in the lab, even with the last-minute problems.
\newline

I am profoundly grateful to my parents Aleksandra and Jacek, for their unconditional support, encouragement, and inspiration. 
\newline

Finally, I would like to thank my husband Tomasz, for his constant understanding, patience and support throughout this journey.

\endgroup

\pagestyle{scrheadings}
\cleardoublepage
\refstepcounter{dummy}
\pdfbookmark[1]{\contentsname}{tableofcontents}
\setcounter{tocdepth}{2} 
\setcounter{secnumdepth}{3} 
\manualmark
\markboth{\spacedlowsmallcaps{\contentsname}}{\spacedlowsmallcaps{\contentsname}}
\tableofcontents 
\automark[section]{chapter}
\renewcommand{\chaptermark}[1]{\markboth{\spacedlowsmallcaps{#1}}{\spacedlowsmallcaps{#1}}}
\renewcommand{\sectionmark}[1]{\markright{\thesection\enspace\spacedlowsmallcaps{#1}}}
\clearpage

\begingroup 
    \let\clearpage\relax
    \let\cleardoublepage\relax
    \let\cleardoublepage\relax


        
    

       
    \refstepcounter{dummy}
    \pdfbookmark[1]{Acronyms}{acronyms}
    \markboth{\spacedlowsmallcaps{Acronyms}}{\spacedlowsmallcaps{Acronyms}}
        
\endgroup

\includepdf{blankpage.pdf}
\includepdf{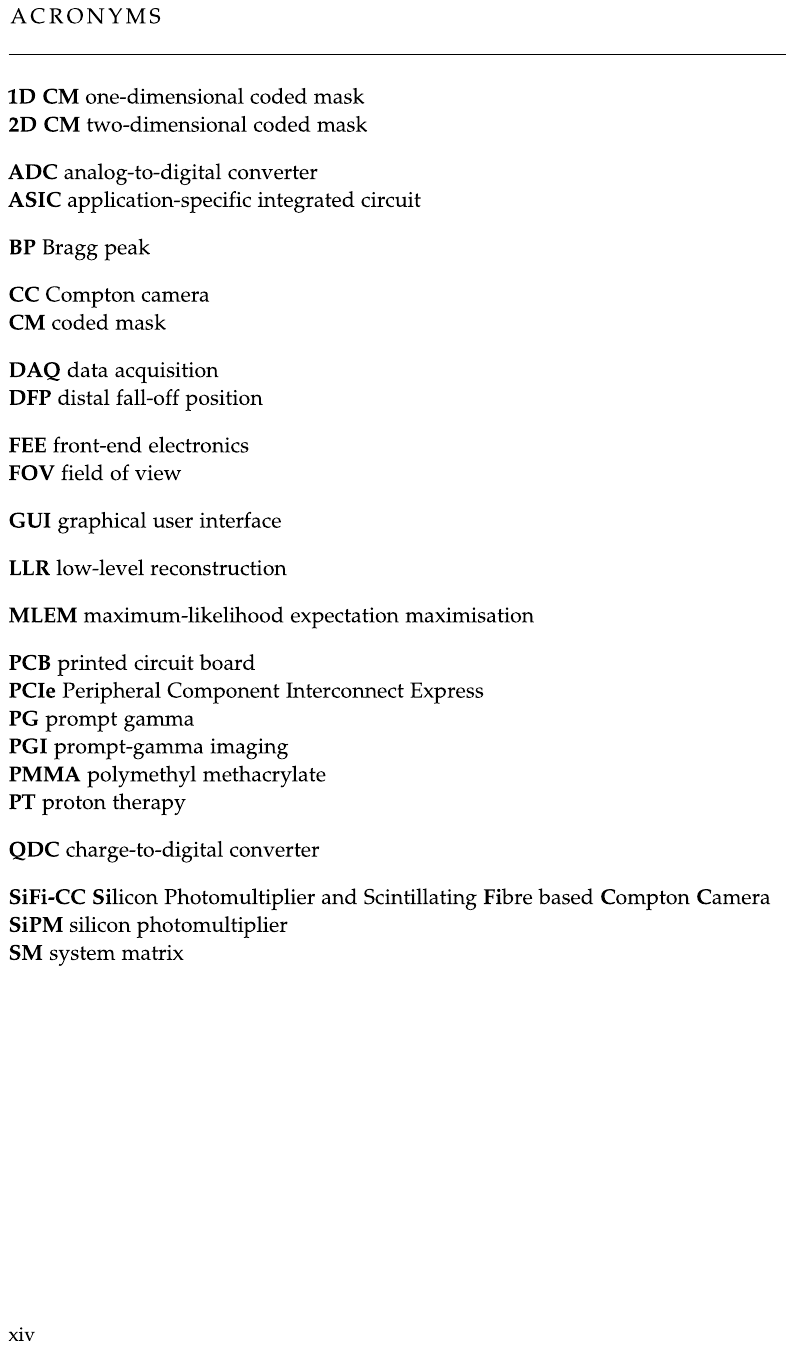}
\cleardoublepage\pagenumbering{arabic}
\cleardoublepage

\chapter{Introduction}
\setcounter{page}{1}

\section{Motivation}
Cancer is the second most common cause of death in Europe~\cite{Eurostat2021}. A study on the US population showed that about 40.5\% of men and 38.9\% of women will receive a cancer diagnosis at some point of their lives~\cite{Siegel2021}. Therefore, extensive efforts are directed towards finding a reliable method for treating various types of cancer. The most widely used treatments are chemotherapy, surgery, and radiotherapy. In the last one, various types of radiation are utilised: photons, electrons, or ions, mainly protons or carbon ions. \Gls{pt} has an advantage over more common X-ray therapy,  which is a superior dose-deposition scheme. An X-ray beam passes through the patient's body, depositing most energy just under the surface, then the amount of deposited energy decreases with depth in tissue. In turn, protons stop at a certain depth in tissue and deposit most of their energy at the end of their range. This dose maximum is called the~\gls{bp}. However, as a consequence of this scheme, very accurate beam range monitoring methods are needed to provide safe and precise treatment. A \gls{cm} gamma camera, which is a subject of this thesis, is one of the approaches to develop such a monitoring method, based on imaging with \gls{pgh} radiation. \Gls{pgh} quanta are one of the by-products of tissue irradiation with a proton beam. They are emitted from the patient's body as a result of inelastic nuclear reactions and the following deexcitation of tissue nuclei. As the distribution of these gammas in tissue is strongly correlated with the proton range, by registering the \gls{pgh} distribution one can conclude about the proton range. Having a fast, reliable method of proton range determination would allow to improve \gls{pt} and widen the scope of possible tumour locations that can be effectively treated with it.

\section{Proton therapy}
\subsection{Overview}
\begin{figure}
\centering
\begin{subfigure}[t]{0.60\textwidth}
\centering
\includegraphics[width=\textwidth]{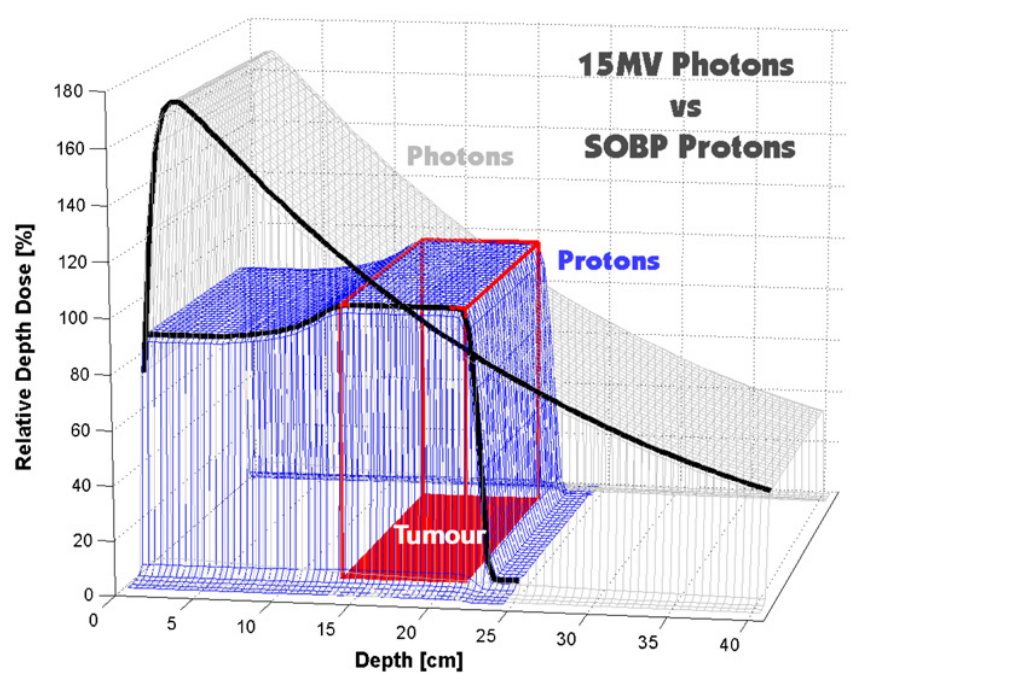} 
\caption{ } \label{fig:braggPeak}
\end{subfigure}
\begin{subfigure}[t]{0.39\textwidth}
\centering
\includegraphics[width=\textwidth]{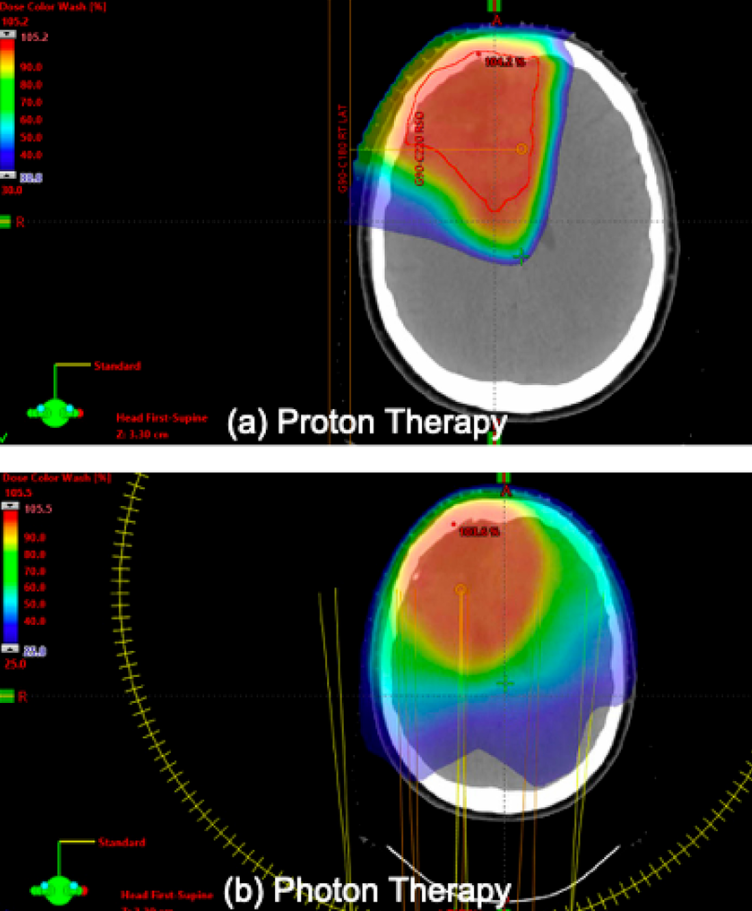} 
\caption{ } \label{fig:protonsVsPhotons}
\end{subfigure}
\caption{(a) Relative deposited dose in \% of maximum dose vs. depth in tissue. Grey - photons (X-rays), blue - protons (spread-out \gls{bp}), the tumour region is marked in red. (b) Example treatment plans for a brain tumour for proton (top) and X-ray (bottom) therapy. The colour scale denotes the relative dose deposited. The red contour in the top picture marks the tumour region. Both figures are adapted from \cite{Lapen2023}.}
\end{figure}

The concept of treating tumours with protons, as well as with heavier ions, relies on their several favourable features. Unlike X-rays, they do not pass all the way through the irradiated body but instead are stopped at a certain depth in tissue, depositing most of their energy in that region. This deposition scheme is well described by the Bethe-Bloch formula (see~\cref{sec:stoppingPower}) and the said maximum is called the \gls{bp}~\cite{MaLomax2012}. With this feature, treatment could be planned so that crucial organs located behind the tumour are fully spared, i.e., receive zero delivered dose. Moreover, any tissue on the way of protons receives less dose than the tissue at the end of their range, which is another advantage of \gls{pt} compared to conventional X-ray therapy. A comparison of dose-depth profiles for protons and X-rays can be found in~\cref{fig:braggPeak}. The higher the beam energy, the deeper the tissue penetration. This is particularly beneficial in deep-seated tumour treatment. Another favourable feature of protons is their relative biological effectiveness (RBE), which is about 10\% higher than for X-rays~\cite{Jones2016}. All these effects contribute to a much better dose conformality than in X-ray therapy, as can be seen in Figure~\ref{fig:protonsVsPhotons}, where dose delivery schemes (so-called treatment plans) for proton and X-ray therapy are compared. 

\subsection{Interaction of protons with tissue}
The physical basis of \gls{pt} is the interaction of protons with human tissue. There are four main types of proton interactions with tissue~\cite{Paganetti2012, Newhauser2015}: 
\begin{enumerate}
    \item Coulomb interaction with electrons bound in atoms, resulting in ionisation (inelastic scattering), 
    \item Coulomb interaction with atomic nuclei (elastic scattering),
    \item nuclear reactions,
    \item brehmsstrahlung.
 \end{enumerate}
Coulomb interaction with bound electrons does not change the trajectory of the proton, but a fraction of its energy is transferred to each of the electrons, and hence the proton velocity decreases. It is the main factor causing the stopping of protons in the tissue, thus determining the proton beam range.
 
 Elastic scattering causes a deflection in the proton trajectory as a result of repulsive Coulomb force. In the context of \gls{pt}, this effect causes lateral broadening of the beam, also blurring its lateral borders.  
 
 When a nuclear reaction occurs, the proton enters the nucleus and, as a result, secondary particles are emitted from the nucleus: protons, neutrons, and heavier ions, as well as gamma rays. This effect causes the primary proton to be effectively removed from the beam. The \gls{pgh} quanta being emitted in this process are crucial for real-time \gls{pgh}-based \gls{pt} monitoring methods. 
 
 Interactions due to brehmsstrahlung are associated with primary proton energy being partially lost due to deceleration caused by deflection by the nucleus and emitted in the form of a brehmsstrahlung photon. However, their significance for \gls{pt} is negligible as they are dominated by the remaining three mechanisms.
 
\subsection{Radiobiology}
The crucial parameters characterising the radiation, related to its dose distribution in the tissue and biological consequences are: linear stopping power, linear energy transfer (LET) and relative biological effectiveness (RBE). 

\subsubsection{Linear stopping power}\label{sec:stoppingPower}
The energy loss associated with the interactions described in the previous section is quantified with the Bethe-Bloch formula \cite{Bethe1930, rpp2022}: \begin{equation}\label{eq:betheBloch}
    \biggl \langle - \frac{dE}{dx} \biggr \rangle = Kz^2 \rho \frac{Z}{A}\frac{1}{\beta^2} \biggl [ \frac{1}{2}\mathrm{ln}\frac{2m_ec^2\beta^2\gamma^2W_{max}}{I^2}-\beta^2-\frac{\delta(\beta\gamma)}{2} \biggr ],
\end{equation}  
where: 
    \newline $E$ - particle energy,
    \newline $x$ - distance travelled by the particle in the medium,
    \newline $K=4\pi N_A r^2_e m_e c^2$ - a constant,
    \newline $N_A$ - Avogadro's number, 
    \newline $r_e$ - classical electron radius,
    \newline $z$ - charge number of the particle,
    \newline $\rho$ - density,
    \newline $Z$ - atomic number of the medium,
    \newline $A$ - atomic mass of the medium,
    \newline $c$ - speed of light,
    \newline $v$ - the particle velocity,
    \newline $\beta=v/c$,
    \newline $\gamma = 1/\sqrt{1-\beta^2}$,
    \newline $m_e$ - electron mass,
    \newline $W_{max}$ - maximum possible energy transfer to an electron during a single collision,
    \newline $I$ - mean excitation energy,
    \newline $\delta(\beta\gamma)$ - density effect correction to ionisation energy loss.
    \newline

The equation defines the linear stopping power, the units are MeV/cm. The mass stopping power is the linear stopping power divided by the density, the units are then \SI{}{\MeV~\g^{-1} \cm^2}.
In the leading term of the Bethe-Bloch formula (\cref{eq:betheBloch}), the dependence of the energy transferred to the medium by the particle per unit path length is inversely proportional to the squared particle velocity. Thus, the lower the particle energy, the higher the energy deposition in that region. This dependence explains the shape of the dose-depth curve. Also, for heavier particles (e.g. $^{12}$C), the distal falloff of the \gls{bp} is steeper than for protons, which is a result of smaller range straggling. This feature varies inversely to the square root of the particle mass. The same dependence occurs for the lateral profile of the beam. The lateral broadening of the dose distribution can be approximated by a Gaussian, but for higher accuracy, one needs to consider the Moliere's theory, which takes into account additional scattering at larger angles, making the lateral penumbra wider than for a simple Gaussian. 

\subsubsection{Linear energy transfer}
Linear energy transfer~\cite{MaLomax2012} is defined as the energy locally transferred to the medium per unit track length (that is, depth) in the medium. 
The LET increases with $Z$, e.g. $^{12}$C ions have higher LET than protons. Higher LET means the dose is more localised, which is favourable for therapy. Dose-averaged LET is a quantity widely used when assessing radiation quality.

\subsubsection{Relative biological effectiveness}
X-ray biological effectiveness is considered a reference, the effectiveness of other types of radiation is assessed relative to it. RBE is defined as a ratio of the reference radiation (X-ray) dose absorbed to the given radiation dose at which the biological effect would be identical (isoeffective), assuming that all other conditions are the same for the examined and reference radiation~\cite{Paganetti2014}. Currently, the proton RBE in clinics is assumed constant along the particle path and equal to 1.1. However, there are arising claims that RBE in \gls{pt} should be variable \cite{Jones2016, Paganetti2015, Paganetti2019, Lhr2018}. For carbon ions, RBE varies along the particle path and ranges from about 2 in the entrance region to about 4 near the end of the particle range. RBE depends on several variables, most significant ones being dose fractionation changes, LET, $\alpha/\beta$ ratio~\cite{MaLomax2012}.

The latter are the parameters of the linear-quadratic (LQ) model \cite{McMahon2018}, which relates the dose delivered to cell survival: $s=e^{-\alpha d -\beta d^2}$, where $s$ is the probability of survival and $d$ is the dose delivered. $\alpha$ and $\beta$ parametrise the radiosensitivity of a given cell type. The term $\alpha$ corresponds to cell death after being hit by a single particle, while the term $\beta$ represents killing after multiple hits. Tissues with different $\alpha/\beta$ ratio react differently to radiation. There are cell repair effects that occur some time after irradiation and are more pronounced in the tissue with a lower $\alpha/\beta$ ratio. This characteristic can be used for more effective treatment, by dividing the total dose into smaller portions in time (dose fractionation), under the condition that the tumour tissue has a higher $\alpha/\beta$ ratio than the healthy tissue~\cite{MaLomax2012}.

\subsection{History}\label{sec:history}
The idea of using protons for cancer treatment was first mentioned in Robert Wilson's paper in 1946 \cite{Wilson1946}, where he discussed the potential of the use of accelerated protons for therapeutic irradiation, described the effects of protons' interactions with tissue and theoretical limits of available precision resulting from these interactions, attributed to range straggling and the angular spread. The first treatment of a human patient with accelerated protons was performed at a cyclotron in Lawrence Berkeley Laboratory \cite{Lawrence1958}. That was an irradiation of the pituitary gland. Soon after, Larsson and others irradiated animals with a beam of modified shape by focusing magnets and sweeping coils  \cite{Larsson1958}, which formed a basis for the spread-out Bragg peak (SOBP) technique and therefore the passively scattered beam. In 1962, Kjellberg et al. treated four patients with brain tumours by irradiation with a 160~MeV proton beam \cite{Kjellberg1962}. In the following decades, patients were treated with \gls{pt} at a small scale in several research facilities i.e. in the US, Sweden and Russia. In 1990, the first clinical facility was commissioned at the Loma Linda University Medical Center in California. Since then, \gls{pt} centres were predominantly located near hospitals rather than in research facilities. A decade later, commercial companies emerged which offered off-the-shelf \gls{pt} systems, which enabled further development of \gls{pt}. The number of \gls{pt} facilities worldwide has been growing steadily over the last few decades, reaching 121 in June 2024, with close to \num{350000} patients treated \cite{PTCOGStats}. \gls{pt} is nowadays an established method of cancer treatment, along with surgery, chemotherapy, conventional radiotherapy and the most recent immunotherapy. 

\subsection{Current status and challenges}
\begin{figure}
\centering
\includegraphics[width = \textwidth]{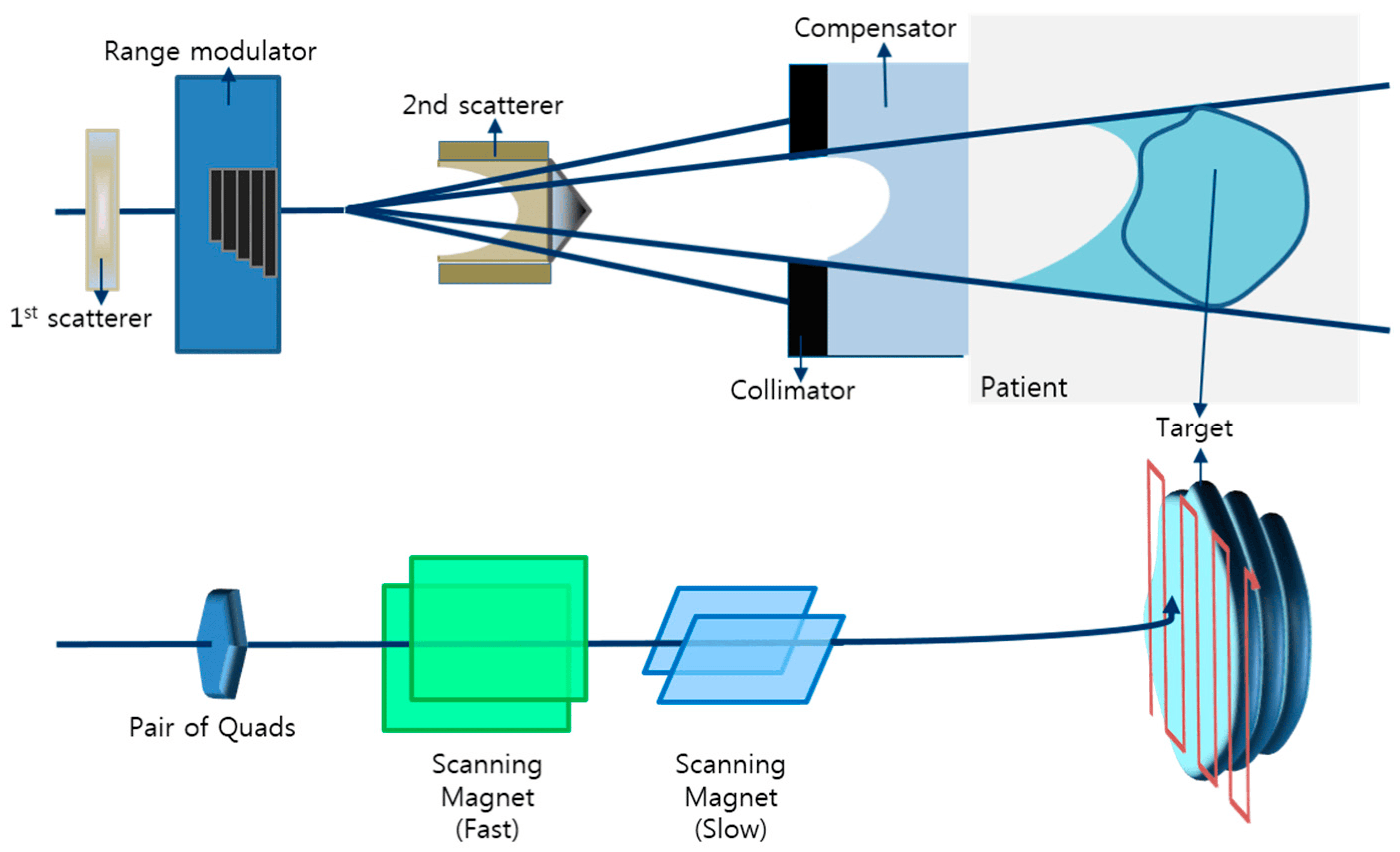}
\caption{Passive scattering (top) and active scanning (bottom) beam delivery systems. Adapted from~\cite{Son2018}.}
\label{fig:passiveScatteringVsActiveScanning}
\end{figure}

Over the years of \gls{pt} progress, various techniques and technologies have been developed. Regarding the beam delivery system, there are two main options: passive scattering~\cite{Koehler1975, Koehler1977} and active scanning~\cite{Kanai1980, Pedroni1995}. The general scheme of both is presented in Figure~\ref{fig:passiveScatteringVsActiveScanning}. Passive scattering is an older, but still widely used method. To be able to deliver the dose precisely to the tumour, the beam undergoes scattering to increase its field, passes through a range modulator, then it is shaped in passive collimators matching the tumour shape. Subsequently, range compensators are applied to correct for tissue density irregularities on the way of the beam. After applying such a shaping procedure, the lateral beam profile exactly matches the tumour shape. Uniform irradiation of the tumour along its proximal-distal axis is performed by superimposing beams of several energies (with a range modulator wheel or a Ridge filter), forming the SOBP. In turn, active scanning, also known as intensity-modulated proton therapy (IMPT), relies on a system of magnets that can narrow the cross section of the beam and bend the beam to direct it to the desired voxel. A beam shaped this way is called a pencil beam, thus another name of this modality is pencil beam scanning (PBS). The tumour volume is segmented into voxels and the pencil beam is directed and delivered to each voxel separately. There are several disadvantages of the passive scattering: the proximal dose conformality is quite poor, there is a need to machine patient-specific hardware for each treatment, secondary neutrons are produced as a result of beam scattering on the passive elements. The above issues are not the case for PBS. There, the proximal edge of the beam can be shaped as needed and there is much less neutron background thanks to the lack of shielding or passive elements. Another approach to beam delivery, which is in the early clinical stage, is FLASH radiotherapy \cite{Lin2021}. In this case, the entire dose is delivered to the tumour with a single, high-rate beam "shot" (minimum~\SI{40}{Gy/s}). It has been demonstrated that such a delivery scheme reduces radiation damage to healthy tissues, as compared to regular \gls{pt}. This phenomenon is called the FLASH effect. The first in-human test of the FLASH \gls{pt} was performed in 2022~\cite{Daugherty2022, Mascia2023}.

Proton beams in the treatment facilities are typically accelerated using either a cyclotron or a synchrotron. They can be delivered to the treatment room via a fixed-position horizontal nozzle or a rotational gantry. Both the accelerators and the beam delivery systems are complex and large-scale. Therefore, a \gls{pt} facility is much more expensive than an X-ray one. The high cost is one of the main factors that limit the common usage of \gls{pt}. Another strong limitation is the accuracy of beam range monitoring. Due to the particular dose deposition scheme in \gls{pt}, even a minor change in the patient anatomy relative to the original treatment plan can result in a significant extra dose deposited in healthy tissue. Thus, various measures are applied to minimise the risk of dose inaccuracy, i.e. patient immobilisation, laser markers on the outer surface of the body or adjusting the irradiation time to breathing cycle if a tumour is located in the lung region. However, currently applied \gls{pt} monitoring methods provide feedback only after an irradiation session, which allows one to adjust treatment plan for the next treatment session, but not during the current one. In view of the above conditions, an online method of beam range monitoring, providing feedback in real time, would be the most beneficial. However, currently there is no such a method applied routinely in clinics, though the first clinical trials are ongoing~\cite{Berthold2021}. This has been indicated to be one of the key factors that limit the applications of \gls{pt}~\cite{NUPECC2014}. The approaches to developing such a method are summarised in the following section.

\subsection{Proton therapy monitoring}
Methods for \textit{in vivo} range verification in \gls{pt} have been reviewed in \cite{Parodi2018}. There, the authors distinguish two general-purpose methods utilising secondary radiation: positron emission tomography (PET) and \gls{pgh} detection. Several types of secondary radiation emitted from the patient's body during radiation treatment correlate with the range of proton beam in tissue, so they can be utilised for monitoring. These can be protons~\cite{Fischetti2020, FlixBautista2024}, neutrons~\cite{Marafini2017}, $\beta^+$ emitters and \gls{pgh} rays (both discussed below), or some combination of them~\cite{Schellhammer2023, Llosa2023}. PET uses the $\beta^+$ emitters, while \gls{pgh} detection utilises \gls{pgh} rays, as the name suggests.

\subsubsection{Positron emission tomography}
One of the by-products of tissue irradiation with protons are $\beta^+$ emitters. They are produced in nuclear fragmentation reactions, which occur along the whole path of an ion in the tissue, up to a few mm before the \gls{bp}. The correlation of their activation pattern with dose spatial distribution depends on the type of primary ion, but it is stronger for heavier ions, e.g. for carbon ions, than for protons. The half-life time of the $\beta^+$ emitters produced in tissue ranges from e.g. 2~minutes for $^{15}$O and 20~minutes for $^{11}$C. Due to those relatively long times, there is an effect of biological washout, which puts a serious limitation on the technique resolution. The effect is amplified by the fact that the pairs of gamma quanta are usually registered only after the treatment session, after moving a patient to another room with a PET scanner. However, there is a recent solution which diminishes the problem of biological washout, proposed by Bisogni et al.~\cite{Bisogni2016}, which involves first in-beam PET for online beam range monitoring, making use of a dedicated dual-head scanner setup. The PET signal is correlated with, but not identical with the dose distribution, so in order to draw conclusions about the dose distribution, one needs to compare the obtained PET image with a Monte Carlo (MC) or analytically calculated expected image. 

\subsubsection{\Gls{pgh} detection}\label{sec:promptGamma}
\begin{figure}
\centering
\includegraphics[width = \textwidth]{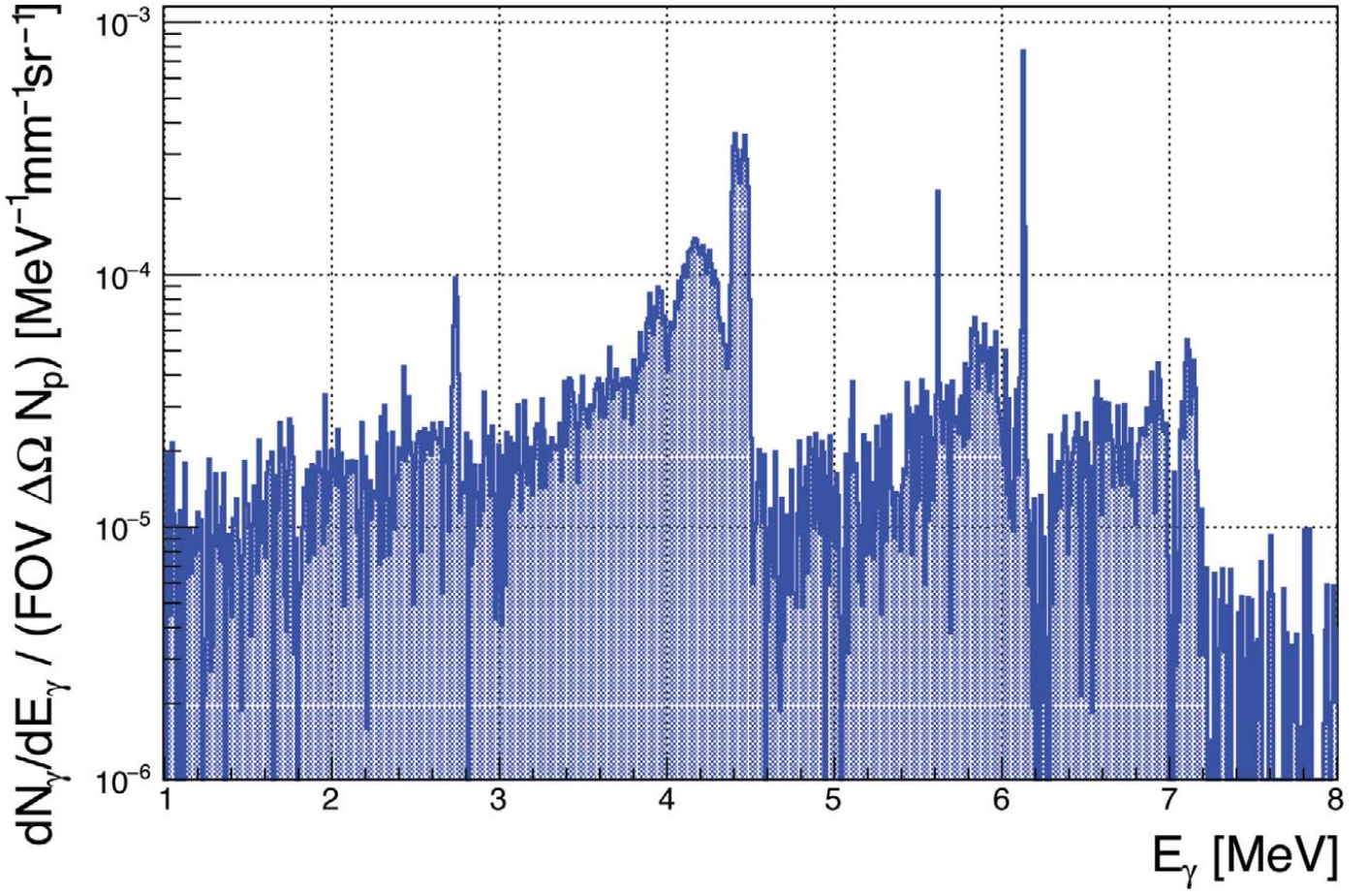}
\caption{\Gls{pgh} spectrum from a 1~mm thick layer of PMMA phantom irradiated with protons of 70~MeV energy, registered with a high-purity germanium detector, normalised to the number of primary protons, corrected for detector acceptance and efficiency. Horizontal axis is the target thickness expressed in the units of proton range in the material. Figure adapted from~\cite{Wronska2021}.} \label{fig:PromptGammaSpectrum}
\end{figure}

\begin{figure}
\centering
\includegraphics[width =\textwidth]{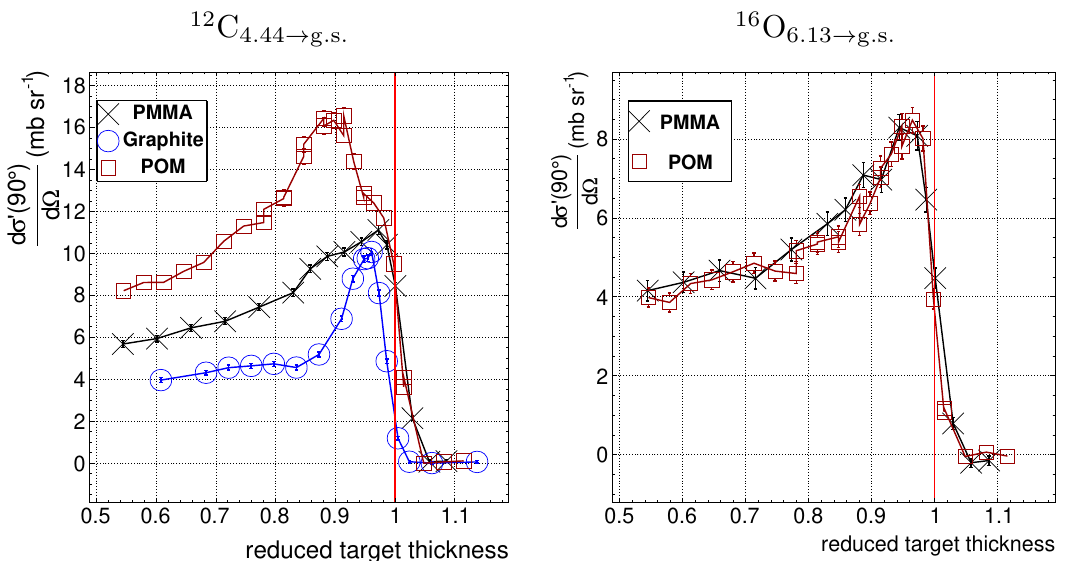}
\caption{Measured depth-distributions of \gls{pgh} rays resulting from deexcitation of $^{12}\mathrm{C}_{4.44\rightarrow \mathrm{g.s.}}$ and $^{16}\mathrm{O}_{6.13\rightarrow \mathrm{g.s.}}$, in PMMA, graphite and POM. The reduced target thickness is the depth in the material, expressed in units of proton range in the material. Figure adapted from~\cite{Kelleter2017}.}
\label{fig:PromptGammaVsProtonRange}
\end{figure}

\Gls{pgh} rays are one of the by-products of the reactions between the impinging protons and nuclei of human tissue. A proton excites a nucleus, which then deexcites, emitting \gls{pgh} rays. They are a valid choice for \gls{pt} monitoring due to the following characteristics:
\begin{enumerate}
    \item they are emitted within a very short time after the proton-nucleus interaction (femto - picoseconds),
    \item thanks to their high energy, their trajectory barely changes when exiting the patient's body. Hence, they carry undisturbed information about the place of interaction,
    \sloppy
    \item their spectrum contains discrete lines originating from nuclei of $^{16}$O at \SI{6.13}{MeV} and $^{12}$C at \SI{4.44}{MeV} (see Figure~\ref{fig:PromptGammaSpectrum}), which facilitates their separation from the continuous neutron background.
\end{enumerate}
In 2006, Min et al. \cite{Min2006} demonstrated that the spectrum of \gls{pgh} rays emitted from the patient's body, particularly the distal falloff, strongly correlates with the range of protons in the patient's tissues. That finding constituted the basis for \gls{pgh} range verification in \gls{pt}, and the field has been growing rapidly ever since. The results obtained in a later experimental study that further explored the details of this correlation \cite{Kelleter2017} are presented in \cref{fig:PromptGammaVsProtonRange}. The methods of \gls{pgh} range verification can be divided into two major groups: imaging and non-imaging techniques. The imaging techniques are, among others, a knife-edge shaped slit camera \cite{Richter2016, Xie2017}, a multi-slit camera \cite{Smeets2016}, and a Compton camera \cite{Koide2018, Draeger2018, Babiano2020, Munoz2021} (see Section~\ref{sec:ComptonCamera}), while the non-imaging techniques include \gls{pgh} timing~\cite{Golnik2014}, \gls{pgh} spectroscopy~\cite{Verburg2014, HuesoGonzalez2018}, and \gls{pgh} peak integrals~\cite{Krimmer2017}. Notably, an online proton range verification system based on \gls{pgh} was clinically applied for the first time by Richter et al.~\cite{Richter2016} in the form of a knife edge-shaped slit camera during a passive-scattered beam treatment. This detector has since been improved and applied to the first-in-human validation of proton range~\cite{Berthold2021}. The first clinical application under pencil beam scanning conditions, during the whole treatment session, was presented in~\cite{Xie2017}. The slit camera has an intrinsic limitation of statistics, which is challenging in clinical applications. Thus, multi-slit setups were also tested~\cite{Pinto2014, Ku2023}, as well as collimators with multiple knife-edge shaped slits~\cite{Ready2016}. Another type of collimator is a coded mask, described in detail in Section~\ref{sec:CodedMask}, which has superior statistics capabilities and has been tested via simulations and experimentally for \gls{pt} monitoring applications in \cite{Sun2020a, Hetzel2023}. The findings described in this thesis are extending and building on the concept described in the latter publication. 

\section{Gamma scintillation detectors}
A common solution used to detect gamma quanta, but also other particles, is a scintillator, which emits a flash of light when ionising radiation deposits energy in it. Registration of this signal requires a photomultiplier, which converts this flash of light into an electric signal. An early version of a gamma camera for medical imaging was proposed by H.~O.~Anger in 1958~\cite{Anger1958}. The core part of such a detector consists of a scintillating crystal and several photomultiplier tubes (PMTs). The concept was later extended in~\cite{Anger1963}. The field has since been developing intensively. Nowadays, there is a variety of scintillating materials available, optimised for various use cases. The photomultiplier tubes have since been replaced, at least in most of the use cases, by silicon photomultipliers (SiPMs). The following briefly characterises the typical building blocks of a typical scintillation detector, i.e., scintillating crystals and silicon photomultipliers.

\subsection{Scintillation crystals}
A scintillator is a special kind of material that emits a flash of visible or near-visible light when exposed to ionising radiation~\cite{Leo1994}, as a result of deexcitation of the scintillator molecules. The process is called scintillation or radioluminescence. A scintillator suitable for radiation detection applications is characterised with:
\begin{enumerate}
    \item linear response to energy,
    \item short response time and lack of long-time activation (this feature is essential for high-rate applications),
    \item transparency to the wavelength at which the scintillation light is emitted,
    \item high light yield.
\end{enumerate}
Scintillators can be divided into two main categories: organic and inorganic. A crucial difference between them is that while organic scintillators are usually based on polymers whose elemental composition is dominated by light elements (H, C, O), inorganic scintillators are usually made of materials with high effective atomic number $Z_\mathrm{eff}$ (so-called high-Z materials). Therefore, typically the inorganic scintillators have shorter radiation length, greater stopping power, and high light output, thus being the optimal choice for X-ray and gamma-ray detection. Inorganic scintillators are in the form of crystals, some examples of the commonly used ones are NaI(Tl), $\mathrm{LaBr}_3$, LSO. 

The mechanism of scintillation in inorganic crystals can be explained based on the electronic band model~\cite{Leo1994}. A crystal in the ground state has approximately all energy levels populated in the valence band and approximately no levels occupied in the conduction band. Between these bands, there is the bandgap, i.e. an energy range without electron states. To enable light emission at desired wavelengths, the crystal must be doped with atoms that provide localised levels for the centres of luminescence in the bandgap. When a gamma ray hits the scintillator, an electron can be excited from one of the lower energy levels to the valence band or, more often, to a much higher energy level~\cite{Tavernier2009}. The electron will then lose its additional energy by exciting another electrons and so on. Thus, a single gamma hitting the scintillator produces multiple electron-hole pairs or separate electrons and holes. Then, the electrons and holes reach the luminescence centres and scintillation light is produced. One needs to note that this is a simplified model, the more precise one involves crystal impurities and the Stokes shift~\cite{Tavernier2009}. 

One of the features characterising the signal produced in the scintillation crystal is its decay time, which depends on the lifetime of a certain excited state. Usually, a short and long decay components are present in a plot of relative light emission versus time. They correspond to fluorescence and phosphorescence effects, respectively. Usually, the short component is dominant. 

A scintillation crystal should, most desirably, have a linear response to energy~\cite{Tavernier2009}. It is true in the energy region of above \SI{1}{MeV}, however, at lower energies certain deviations from linearity occur. The type of particle interacting with the scintillation crystal also affects its output, e.g. alpha particles are usually detected with a few times lower sensitivity than electrons~\cite{Tavernier2009}. The shape of the scintillation pulse produced by various particles can also vary, which is an underlying principle of a method of particle identification called pulse shape discrimination~\cite{Roush1964}.

Gamma rays interact with matter in three dominant ways: photoelectric effect, pair production, and Compton scattering. While in the first two, the gamma is completely absorbed and its energy transferred to a particle (particles), in the last one, it leaves only part of its energy to an electron and can undergo further interactions or escape the material, not depositing its full energy there. Thus, in an efficient gamma-ray detector, the first two processes should dominate. This is the case for high-Z materials.

\subsection{Silicon photomultipliers}
A silicon photomultiplier is a solid-state silicon detector. It is essentially an array of hundreds to thousands of single-photon avalanche photodiodes (SPADs) or avalanche photodiodes (APDs) connected in series to quenching resistors. APDs and SPADs have similar principles of operation, but this section focuses on APDs. An APD together with a quenching resistor form a single pixel (microcell) of a SiPM. The size of such a single pixel (microcell) is typically 10~$\mu$m to 100~$\mu$m~\cite{SiPMHamamatsuWeb}. An APD is basically a semi-conductor p-n junction. Each APD operates in Geiger mode, which means that the bias voltage is above the breakdown voltage of the APD. 

The operation principle of the APD is the following: if a photon is absorbed by the photodiode, the ejected charge carrier (an electron or a hole) triggers an avalanche and a continuous Geiger discharge starts. In such a process, 10$^5$ to 10$^6$ charge carriers are produced, forming an electrical pulse. The discharge is quenched by the resistor, so that the APD can return to its initial state and register another photon. The summed electrical pulses from all activated APDs during a single "event" form an output SiPM signal,  proportional to the number of photons that reached the SiPM. Due to its internal structure, a SiPM lower detection threshold can be as low as a single photon, while the upper detection threshold is limited by the number of microcells. However, the saturation effect is already visible when the number of photons reaching a SiPM becomes comparable with the number of microcells. Then the effect becomes more pronounced with increasing number of activated microcells, up to a point where an increase in the number of incident photons has no impact on the pulse height anymore, due to full saturation. 

SiPMs, unlike PMTs, are insensitive to magnetic fields. They are also smaller in size, and a lower bias voltage is required for their operation. Like all other light-sensitive devices, they must be operated in light-tight conditions; otherwise, they can be damaged due to large current increase. SiPMs can detect photons from the region close to the visible range, typically 300-900~nm \cite{Georgel2022}. The active surface of a typical SiPM ranges from one to several mm$^2$. Several characteristic parameters of the SiPMs are:
\begin{enumerate}
    \item Gain, which is the number of charge carriers produced in a single microcell. The gain depends linearly on the bias voltage (in a limited range of accepted overvoltage). It is also temperature-dependent, which may lead to the need to use a cooling system.
    \item Breakdown voltage, the highest voltage that can be applied before the current increases exponentially in the photodiodes. 
    \item Photon detection efficiency (PDE), which is a measure of the ability of the SiPM to detect photons. PDE varies depending on the wavelength of the incident light. It is the ratio of photons detected and photons impinging on the detector.
    \item Dark count rate, the number of counts per unit time when there is no light reaching the SiPM from outside, caused by thermal activations inside the SiPM.
    \item Afterpulse probability. A single incident photon can sometimes produce more than one electric pulse. The additional pulses are called afterpulses and contribute to SiPM noise.
    \item Crosstalk probability. Crosstalk occurs when a secondary photon is produced after a hit in one microcell and is detected by one of the neighbouring cells. This effect also contributes to the SiPM noise. 
\end{enumerate} 

The main areas in which SiPMs are used are low-light applications, e.g. PET, also with fast timing (time of flight - TOF PET), radiation detection in high-energy physics, single-photon measurements in spectroscopy, light detection and ranging (LIDAR), quantum experiments \cite{Acerbi2019}. 

\subsection{Requirements for a PG  detector for PT monitoring}\label{sec:introduction_requirementsForPTMonitoring}
From a clinical point of view, the \gls{pgh} detector needs to meet the following criteria, in order to be applicable for \gls{pt} monitoring~\cite{Pausch2020}:
\begin{itemize}
    \item The \gls{pgh} detector must be compatible with the beam delivery system. If there is a gantry, the detector should be gantry mountable.
    \item The monitoring process should not prolong the total treatment time, as it would limit the number of patients that can be treated and thus is not economically justified.
    \item The clinical workflow demands the irradiation time to be as short as possible, in order to reduce patient's strain and the risk of positioning errors due to patient's movement. Thus, the \gls{pgh} detector must be operable under high, clinical rates and cope with non-uniform beam time structure.
\end{itemize}
The first criterion puts a significant limit of the detector size and weight. The second one demands a fast data processing scheme allowing for getting results in real time. The last criterion sets constraints on the detector design, which are further discussed in this section.

Typically, the number of incident protons is about $10^8$ for a single, distal spot in PBS. Irradiation of one spot lasts about \SI{10}{ms}~\cite{Krimmer2018}. The resulting yield of \gls{pgh} rays is $1-3 \times 10^7$, emitted in full solid angle, resulting in the rate of $1-3 \times 10^9$~cps. Assuming angular detector acceptance of the order of $10^{-3}$, the typical detector load is of the order of $1-3 \times 10^6$. Thus, the possible detector load and system throughput need to be maximised in a \gls{pgh} detector design. Efficient event selection criteria should also be applied, to eliminate as much background as possible. Moreover, the beam's time structure is irregular: huge leaps of the load occur. The detector has to be stable under such conditions. Finally, the \gls{pgh} detector should be segmented, having multiple pixels with independent readout, as this not only enables position sensitivity but also increases the possible data throughput, as the count rate is distributed among readout channels. 

These criteria in view of our proposed detector design are addressed in~\cref{sec:ComptonCamera,sec:CodedMask,sec:DAQ}. 
\chapter{The SiFi-CC project}
The work presented in this thesis has been done within the \gls{sificc} project~\cite{SiFiCC}. 
The aim of the project is to build a novel \gls{pgh} detector for \textit{in vivo} proton therapy monitoring. The detector can be set up in one of the two modalities: as a \gls{cc} or as a coded-mask (CM) camera. The first one allows in principle for a reconstruction of a three-dimensional \gls{pgh} distribution, while the other one is limited to one or two dimensions, depending on the shape of the mask and detector properties.

The \gls{cm} option requires less material, less data acquisition channels and involves a simpler image reconstruction algorithm. Hence, the detector in the \gls{cm} modality was assembled and tested in the first place, and it is the focus of this thesis. In this chapter, the detector concept and design are described, along with its previous versions and optimisation of the components. 
\section{Detector modalities}
\subsection{Compton camera}\label{sec:ComptonCamera}
\begin{figure}
\centering
\begin{subfigure}[t]{0.3\textwidth}
\centering
\includegraphics[width=\textwidth]{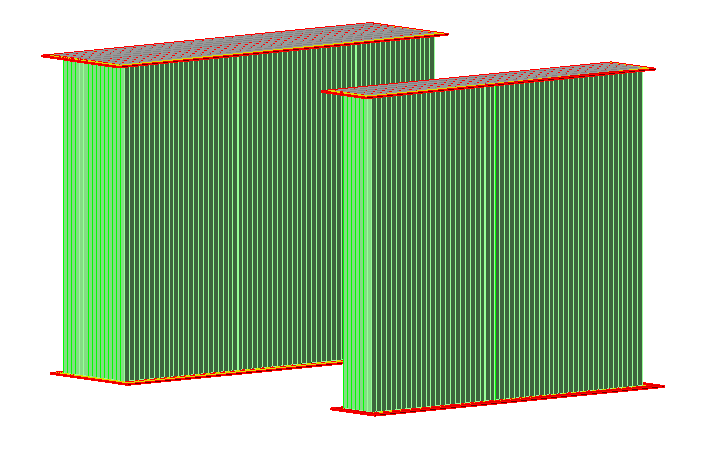} 
\caption{CC} \label{fig:CCSimPicture}
\end{subfigure}
\begin{subfigure}[t]{0.2\textwidth}
\centering
\includegraphics[width=\textwidth]{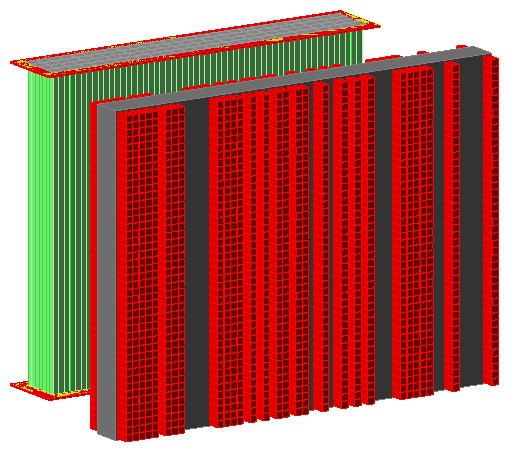} 
\caption{1D~CM} \label{fig:CM1DSimPicture}
\end{subfigure}
\begin{subfigure}[t]{0.2\textwidth}
\centering
\includegraphics[width=\textwidth]{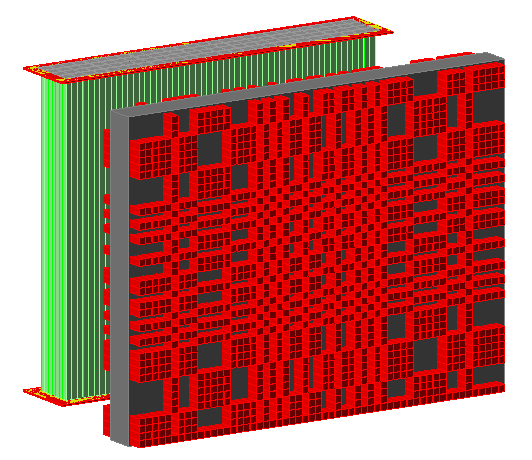} 
\caption{2D~CM} \label{fig:CM2DSimPicture}
\end{subfigure}
\caption{Illustration of the SiFi detector modalities.}
\label{fig:CCandCMSim}
\end{figure}

A Compton camera is a type of radiation detector that exploits Compton scattering. Among all approaches to \gls{pgi}, it is the only detector that allows obtaining a 3D spatial distribution of the observed radiation source with a single device~\cite{Pinto2024}. Originally, \glspl{cc} were applied in astrophysics~\cite{Schnfelder1973}, later they were also optimised for use in medical applications~\cite{Todd1974}, characterisation of nuclear waste~\cite{Phillips1997} and homeland security~\cite{Sweeney2014compton}. Typically, a \gls{cc} is composed of two detector modules: a scatterer, where a gamma is Compton-scattered and the recoil electron is absorbed, and an absorber, where the scattered gamma is absorbed. The two modules register the positions of the interaction points and energy of the electron and the scattered photon in coincidence. Having this information and knowing the kinematics of Compton scattering (Equation~\ref{comptonTheta}), one can reconstruct a cone of possible directions of the primary gamma. 
\begin{equation}
   \theta = \arccos\left(1+m_ec^2\left(\frac{1}{E_{\gamma'}+E_e}-\frac{1}{E_{\gamma'}}\right)\right)\label{comptonTheta}.
\end{equation}
There, $E_{\gamma'}$ is the energy of the scattered gamma, $E_e$ is the energy of the recoil electron, $E_{\gamma'}+E_e$ is the energy of the primary gamma, and $\theta$ is the angle between the direction of the primary and the scattered gamma, which defines the cone. Having multiple such cones, a spatial distribution of the source can be retrieved by using an appropriate image reconstruction algorithm. 

There are various approaches to the design of \glspl{cc} for medical imaging. Multiple scatterer modules were proposed in~\cite{Roellinghoff2011, Richard2009}. In the latter study, the absorber was built of scintillation crystals, while the scatterers were in form of silicon detectors. Kurosawa et al. proposed a camera in which Compton scattering is induced in a gas chamber~\cite{Kurosawa2012}. There, the authors present the first PG distribution after proton irradiation that was measured with a \gls{cc}. Draeger et al. performed an experiment in clinical conditions, demonstrating feasibility of three-dimensional imaging with a \gls{cc}~\cite{Draeger2018}; the detector modules were based on CdZnTe crystals. Another solution tested under proton beam conditions are several generations of the MACACO detector~\cite{Llosa2013, Barrientos2024}. There, the detector modules are made of LaBr$_3$ and LYSO scintillation crystals. In~\cite{Koide2018}, the authors report on a \gls{cc} with 3D position sensitivity, in which the detector modules are composed of multiple cubical Ce:GAGG crystals; the detector was tested under proton beam conditions.

Our approach to \gls{cc} design (see~\cref{fig:CCSimPicture}) involves constructing both detection modules out of thin scintillating fibres, read out at both ends by \glspl{sipm} (a detailed description of the design can be found in~\cref{sec:DetectorComponents_Scintillators}). The high granularity of the detection modules ensures a low pile-up probability. Moreover, two out of three coordinates of the particle interaction point can be retrieved directly if one identifies the fibre in which the interaction took place, provided that the readout channels are independent. The third coordinate of the interaction point is reconstructed from the charge ratio of the signals registered at both ends of the fibre. The fibres are made of a dense, high-$Z$ material (LYSO:Ce,Ca), to ensure good detection efficiency of PGs. 

In the further parts of this thesis, we use the following nomenclature: we refer to the position of the interaction point in the detector as the "hit position", and by "hit fibre" we mean the fibre in which the interaction took place. 

\subsection{Coded mask} \label{sec:CodedMask}
A \gls{cm} camera is an extension of a pinhole camera concept~\cite{Young1989}. The pinhole camera involves a black, light-tight box with a small hole in one of its walls to obtain an image. This simple concept has certain advantages over lens-based optics: no linear distortion, wide angular field of view, and arbitrary field depth~\cite{Young1989}. A serious limitation of this detection method is that the vast part of the light is lost, so a significant time of measurement is needed to collect enough data. 

To increase the statistics and thus shorten the measurement time, a passive collimator in the form of multiple holes or slits (named coded mask or coded aperture) with known pattern can be used instead of a single pinhole. The collimator modifies the image registered on the detector plane in a known and characteristic way. Then, an appropriate algorithm is applied to reconstruct an actual image from the raw detector response. The \gls{cm} imaging technique was proposed in~\cite{MertzYoung1961} and was originally applied in astrophysics to image far-field, point sources~\cite{Dicke1968}. Examples of image reconstruction algorithms for \gls{cm} imaging are \gls{mlem}~\cite{Shepp1982, Lange1984}, OSEM~\cite{Hudson1994, Mu2016}, and FISTA~\cite{Beck2009, Sun2020a}. The \gls{mlem} was used in our case (see~\cref{sec:imageReco}). A class of masks that is widely perceived as optimal for \gls{cm} imaging is a Uniformly Redundant Array (URA), an overview of which can be found in~\cite{Busboom1998}. These masks are built out of square pixels, and the pattern is constructed based on prime numbers. One of the subgroups of this class is a modified uniformly redundant array (MURA) \gls{cm} pattern, proposed in~\cite{Gottesman1989}. This pattern ensures that the noise terms in the reconstructed image are independent of the structure of the image source~\cite{Gottesman1989}. We have chosen the MURA pattern (more details of the design in~\cref{sec:masks}) for our measurements, see Figures~\ref{fig:CM1DSimPicture} and~\ref{fig:CM2DSimPicture}.

A \gls{cm} gamma camera for near-field applications, in particular proton therapy monitoring, has been studied via simulations by Sun et al. \cite{Sun2020a} and experimentally by the \gls{sificc} group~\cite{Hetzel2023}. 
In the \gls{sificc} project, the \gls{cm} is one of the two modalities in which the detector can be set up, the other one being the \gls{cc}. The \gls{cm} imaging requires only one detection module, hence it was the first modality tested in proton beam conditions. Description of those tests constitutes a major part of this thesis. Once the detector is reconfigured to the \gls{cc} modality, the detection module used in the \gls{cm} one will serve as a scatterer.

\section{Earlier studies and versions of the detector}
\subsection{Optimisation of fibre material and wrapping}\label{sec:optOfFibreMaterial}
On the way to find an optimal detector design and materials, an extensive experimental study was conducted\footnote{The study was conducted by K.~Rusiecka.}~\cite{Rusiecka2021, rusiecka2023sificc}. There, various fibre materials and wrappings were compared in terms of energy resolution, position resolution, timing resolution, attenuation length, and light collection. The first two parameters are discussed here in more detail, as they were the most crucial in the context of the research described in this thesis. 

Let us define the charge collected by the SiPMs on both fibre sides ($L$, $R$) as $Q_L$ and $Q_R$, respectively. The energy deposited in the fibre depends primarily on $\sqrt{Q_L Q_R}$, while the hit position along the fibre depends mainly on $\ln(\sqrt{Q_L/Q_R})$. Based on these observations, position and energy calibration can be performed - for this purpose, two light attenuation models were developed, one of them considering also the light reflected in the fibre~\cite{Rusiecka2021}. The calibration is based on a series of measurements with an electronically collimated, radioactive source, which is placed in different positions along the fibre. For each source position, the charges collected by the SiPMs on both fibre sides are registered. Using the energy and position calibration, one can determine the resolutions, which are defined as follows:
\begin{itemize}
    \item The energy resolution is the relative width of the reconstructed \SI{511}{keV} peak: $\sigma/\mu$, where $\sigma$ is the standard deviation of the Gaussian fit, and  $\mu$ is its mean position.
    \item The position resolution is the full width at half maximum of the distribution of reconstructed positions. 
\end{itemize}
In this study, the SiPMs coupled to the fibre had approximately 9 times larger surface than the fibre end. The fibre ends were positioned centrally with respect to both SiPMs.

The conclusion from the study of fibre properties was that the fibre that provides the best tradeoff between position and energy resolution is a LYSO:Ce fibre wrapped in Al foil, the shiny side facing the fibre surface. The dimensions of the fibres compared in that study were \SI{1}{}$\times$\SI{1}{}$\times$\SI{100}{\mm\cubed}. The obtained parameters were: energy resolution 8.56\% at \SI{511}{keV}, position resolution \SI{32}{\mm}. 

\subsection{Optimisation of the detector setup geometry}
Another study performed within the \gls{sificc} group, which was the basis for the detector construction, was a Geant4 simulation study\footnote{The simulation study was performed by J.~Kasper.}, where the geometry of the experimental setup was optimised~\cite{KasperPhD}. The optimised parameters were:
\begin{enumerate}
    \item the distance between the radiation source and the first module: \SI{150}{mm}
    \item the distance between the two \gls{cc} modules: \SI{120}{mm}
    \item the number of layers of the first \gls{cc} module: 16
    \item the number of layers of the second \gls{cc} module: 36
\end{enumerate}
Using the optimised parameters, the detector performance in such configuration was evaluated. It was found that the detector is able to detect a 5-mm shift of the range of proton beam with a resolution of \SI{2}{\mm} at the statistics corresponding to \num{5e8} protons. The detector imaging sensitivity in this configuration is $(5.58\pm0.01)\times10^{-5}$.   

\subsection{Small-scale prototype of a detector module}\label{sec:smallScalePrototype}
\begin{figure}[!htb]
     \centering
     \begin{subfigure}[b]{0.45\textwidth}
         \centering
     \includegraphics[width=\textwidth]{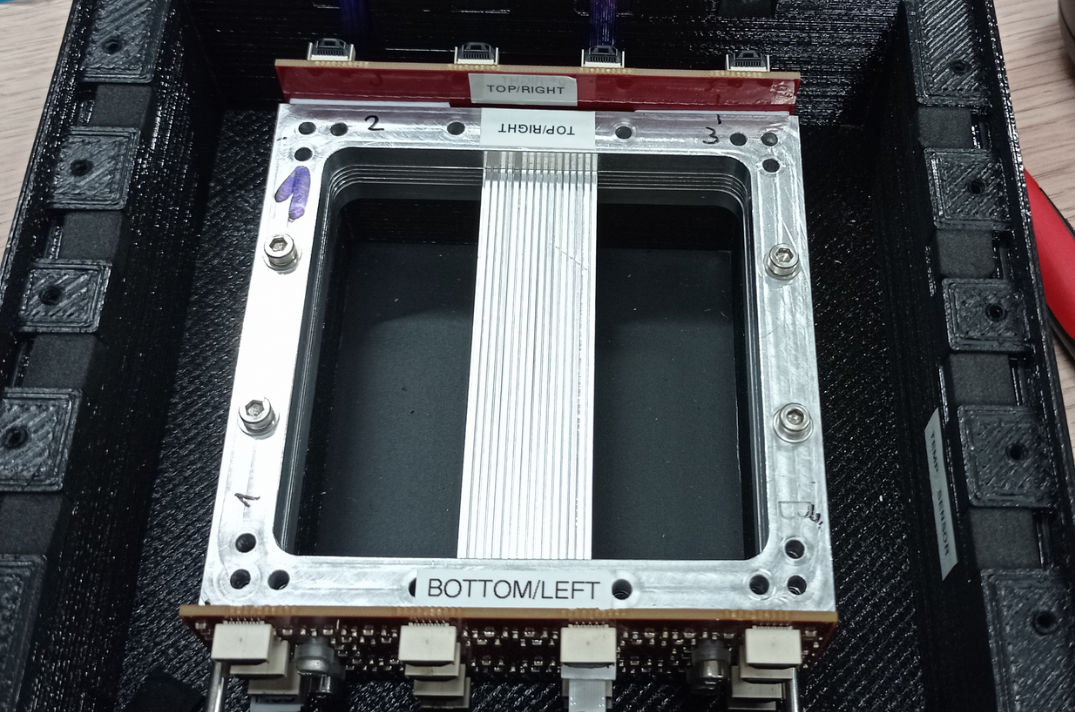}
     \end{subfigure}
     \hfill
     \begin{subfigure}[b]{0.45\textwidth}
         \centering
        \includegraphics[width=\textwidth]{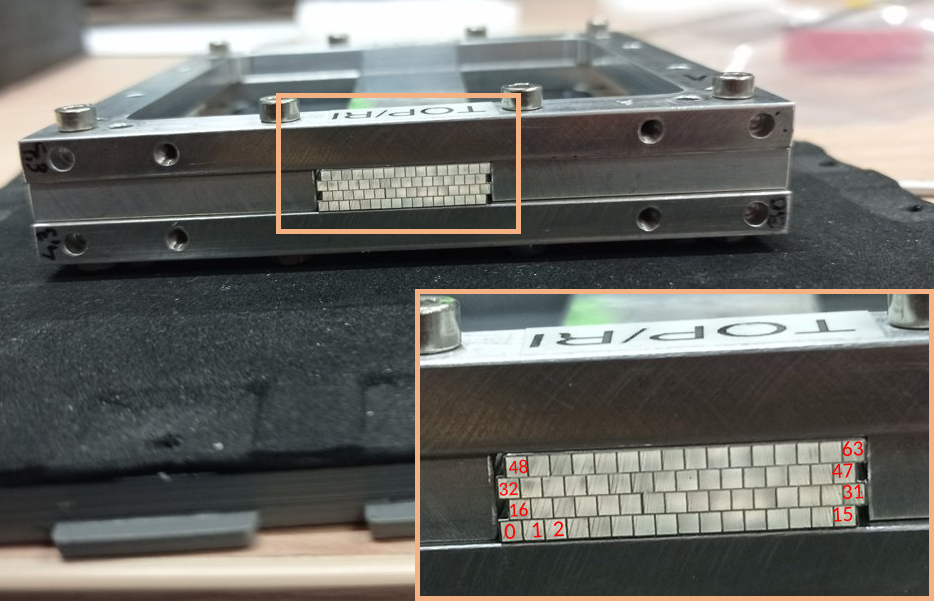}
     \end{subfigure}
        \caption{Small-scale prototype of the \gls{sificc} detector module. Left panel: top view, red \glspl{pcb} housing the SiPMs visible; right panel: side view, fibre stack arrangement visible.}
        \label{fig:smallScalePrototypePhoto}
\end{figure}

The studies described above formed the basis for the construction of a small-scale prototype of a detector module~(\cref{fig:smallScalePrototypePhoto}). It consisted of four layers of 16 fibres each, of the properties determined as optimal in the fibre study (see~\cref{sec:optOfFibreMaterial}). All fibres were coupled at both ends to SiPMs~\cite{KetekSiPMs} with a matching active surface size ($1\times$\SI{1}{\mm\squared}). One fibre end was coupled to one SiPM (so-called 1-to-1 coupling), to ensure direct fibre identification: if the SiPMs at both ends of a given fibre register a signal, it is assumed that this fibre was hit. Optical pads made of Elastosil RT 604~\cite{ElastosilPads} served as a coupling between the fibre end and the SiPM surface. The SiPMs were read out by a Caen DT5742 digitiser~\cite{DT5742} (16 SiPMs at a time, due to the limited number of DAQ channels). 
The prototype was calibrated fibre-by-fibre, in the same way as the single fibres in~\cref{sec:optOfFibreMaterial}. The energy and position resolutions obtained were 10.58\% and \SI{91.49}{mm}, respectively~\cite{rusiecka2023sificc}. Such position resolution was not sufficient for the purpose of determining the hit position along the fibre, and significantly worse than in the single fibres study. The main difference between the small-scale prototype setup and the single fibre setup (\cref{sec:optOfFibreMaterial}) was the larger SiPM size in the latter, which allowed for collection of light scattered at larger angles. This light undergoes multiple internal reflections and thus experiences a larger optical path, being strongly attenuated. Hence, it is mostly responsible for the position sensitivity and restricting the scattering angle leads to worse position resolution.

This issue was addressed in another experiment, using readout based on light sharing, also described in~\cite{rusiecka2023sificc}. The fibres were read out by Philips Power Tile digital sensors~\cite{Frach2009}, which, contrary to the SiPMs of the active surface of $1 \times$\SI{1}{mm \squared}, allowed the light from each fibre to spread and be registered by several pixels of the photosensor. Then, the centre-of-gravity method was employed to determine the two coordinates of the interaction point in the plane of the fibre ends. Position and energy calibration procedures were performed analogously as for the 1-to-1 coupled setup, and the following values were obtained: an energy resolution of \SI{7.73}{\percent}, and a position resolution~\SI{34.28}{mm}. Both the resolutions were significantly better than in the 1-to-1 coupled setup. This supported the conclusion that collecting the light scattered at larger angles leads to better position resolution; moreover, it improves the energy resolution, as more scintillation photons are registered.

Based on the prototype study, conclusions were drawn for the full-size setup. To improve the light collection, which is connected to the energy resolution, we decided to increase the fibre cross section up to \SI{1.94}{}$\times$\SI{1.94}{\mm\squared}. In order to improve position resolution, we used 4-to-1 coupling between the fibres and SiPMs, as this enables collection of light scattered at large angles (SiPM area is larger than fibre area, like in the single fibre studies described in~\cref{sec:optOfFibreMaterial}). The cost of this modification, however, is the necessity to decode fibre hits from SiPM hits, which is not as straightforward as in the 1-to-1 coupling case. Another argument for the transition to 4-to-1 coupling was a limited budget, and such a geometry requires four times less \gls{daq} channels than the 1-to-1 geometry.

\subsection{Small-scale prototype with a CM}
\begin{figure}[!htb]
\centering
\includegraphics[width = \textwidth]{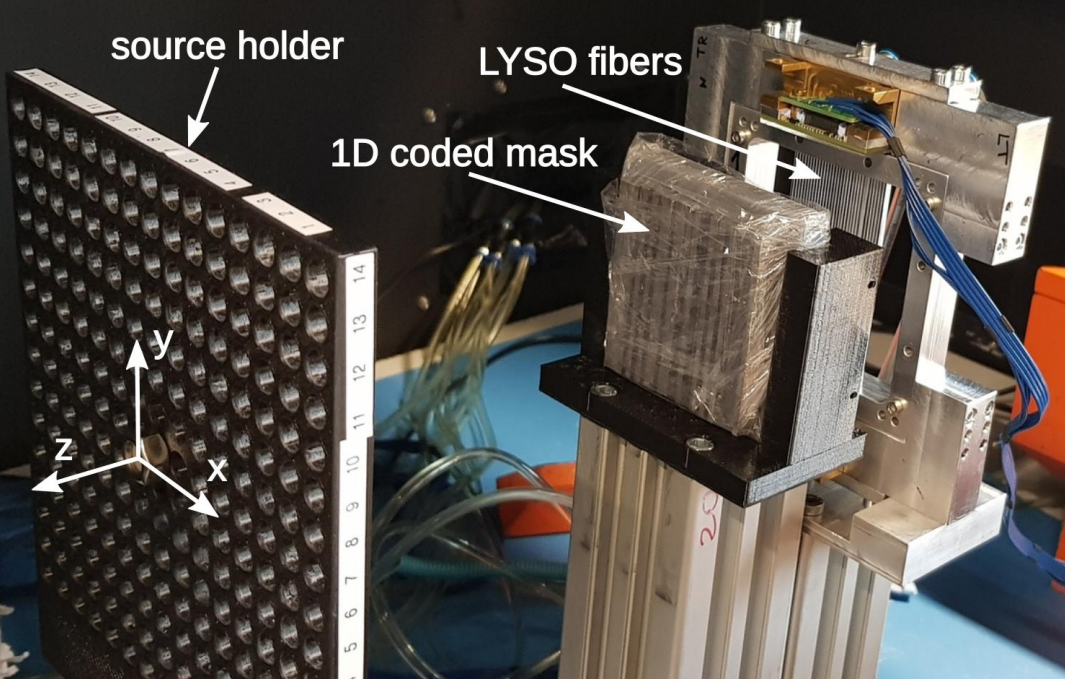}
\caption{Experimental setup for 1D measurements with the small-scale prototype of a \gls{cm} gamma camera. Adapted from~\cite{Hetzel2023}.}
\label{fig:SSP_expSetup}
\end{figure}

The small-scale prototype was also incorporated into a \gls{cm} imaging setup and tested under laboratory conditions to demonstrate the proof of principle for the near-field \gls{cm} application. The findings are reported in~\cite{Hetzel2023}. The fibre matrix was rearranged into two layers, 32 fibres each. Digital SiPMs by Philips Digital Photon Counting were used for readout and the Hyperion system~\cite{Weissler2015} served as the \gls{daq}. With this setup and a 1D~CM (see~\cref{fig:SSP_expSetup}), the images for point-like radioactive sources were reconstructed with a mean standard deviation of \SI{1.14(18)}{\mm}. 

The mean reconstructed positions show a consistent offset of about \SI{1.03(14)}{\mm}, which is caused by a systematic effect due to the limited alignment accuracy of the setup elements. 

A further demonstration of the proof of principle for the \gls{cm} approach to \gls{pgi} was a simulation study of a full-scale setup, also described in~\cite{Hetzel2023}. It yielded a mean \gls{dfp} determination precision of \SI{0.72}{\mm}. This value was obtained for the beam energy range of \SI{85.9}{}-\SI{107.9}{MeV}, with the statistics of $10^8$~protons. The image reconstruction algorithm used was \gls{mlem}. 

The conclusion from both the experimental and the simulation study was that the \gls{cm} imaging is a feasible option for proton therapy monitoring, and the precision of the method is comparable with the precision of knife-edge-shaped and multi-slit cameras - other solutions being investigated for this purpose (see~\ref{sec:promptGamma}). 

\section{Detector components}
In the following, the building blocks of the first full-scale detector module are presented.
\subsection{Scintillators}\label{sec:DetectorComponents_Scintillators}
The active part of the detector module is a block of scintillating fibres, stacked in 7 layers, 55 fibres each, 385 fibres in total. We refer to a fibre either by its absolute ID (0-384), as indicated in the bottom plot of~\cref{fig:fibersOnSiPMs}, or by its address. The address requires three numbers instead of one, but has the benefit of being more informative about the actual position of a given fibre within the fibre stack. The address is comprised of module \textit{m}, layer \textit{l}, and fibre \textit{f}, which are defined as follows:
\begin{enumerate}
    \item module: ID of the detector module; 0 - scatterer, 1 - absorber. In the course of this thesis, as we are using only one module (the scatterer), thus the module ID is always 0.
    \item layer: ID of the fibre layer, increasing with distance from the radioactive source (along the \textit{z} axis); range 0-6.
    \item fibre: fibre number within a layer, range: 0-54. A group of fibres with the same value of the fibre number is called a \textit{column}.
\end{enumerate}
The structure of the fibre block (top view) is marked in red in~\cref{fig:fibersOnSiPMs}.
The dimensions of a single scatterer pixel (i.e. a bare, unwrapped fibre) are $1.94\times 1.94 \times \SI{100}{\mm\cubed}$, the fibres are stacked together so that the long walls touch each other. Each fibre is wrapped in an aluminium foil type 1060 \cite{AlFoil1060}, \SI{30}{\micro\m} thick. The pitch between subsequent pixels is \SI{2}{\mm} in both \textit{x} and \textit{z} directions (\textit{x} is the dimension along a single layer, \textit{z} is the one across the layers (see~\cref{fig:fibersOnSiPMs}),
with the exception of the gaps after each 8 fibre columns: there, the pitch is \SI{100}{\micro\m} larger, which comes from the requirement to match the distance between the borders of SiPMs from two different arrays. The remaining space between the fibres after wrapping with the foil was filled with an ESR foil (\SI{0.085}{mm} thick) and a glue layer (\SI{0.015}{mm} thick) to match the required pitch. The fibres are made of LYSO:Ce,Ca. The entire stack of fibres was manufactured by Taiwan Applied Crystals~\cite{TACwebsite}. The LYSO:Ce,Ca material and the wrapping material were chosen based on a previous study within the \gls{sificc} project; see~\cref{sec:optOfFibreMaterial}. 

\subsection{SiPM arrays}
\begin{figure}[!htb]
\centering
\includegraphics[width = \textwidth]{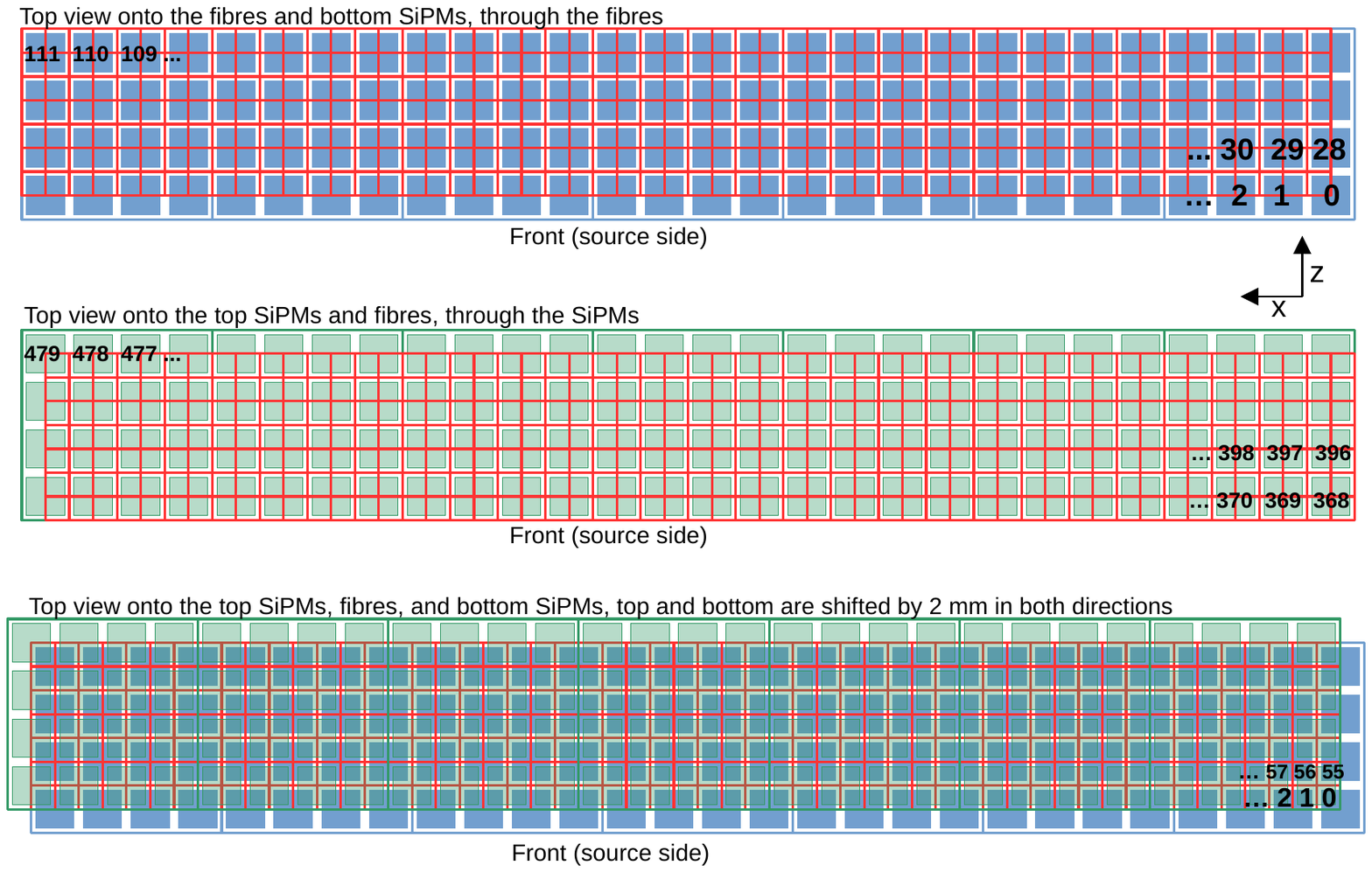}
\caption{Geometry of the fibre stack constituting the sensitive part of the constructed detection module. Top panel: the bottom SiPM array with respect to the fibres (bottom SiPMs in blue, fibre contours in red), with bottom SiPMs numbering scheme; middle panel: the top SiPM array with respect to the fibres (top SiPMs in green, fibre contours in red), with top SiPMs numbering scheme; bottom panel: both the top and bottom SiPM arrays and the fibres, with fibres numbering scheme. The bottom panel illustrates the architecture of 4-to-1 coupling - 4 fibres coupled to one SiPM on each side, with one SiPM board shifted by \sfrac{1}{2} of a SiPM pitch with respect to the other SiPM board, in both \textit{x} and \textit{z} directions.}
\label{fig:fibersOnSiPMs}
\end{figure}
In this study, Broadcom AFBR-S4N44P164M  $4 \times 4$ SiPM arrays~\cite{BroadcomSiPMArrays} were used. They were coupled to fibre ends on both sides of the scintillation fibre stacks. The size of each SiPM pixel was \SI{4}{}$\times$\SI{4}{\mm\squared} (with the photosensitive area of $3.72\times\SI{3.62}{\mm\squared}$), so one SiPM was coupled to four fibre ends. There were two custom-made \glspl{pcb}, one on the top side of the fibre stack and one on the bottom side. Each \gls{pcb} housed 7 SiPM arrays arranged in a row. The geometry details of the fibre stack are presented in~\cref{fig:fibersOnSiPMs}. There, in the bottom panel, one can see that the SiPM arrays are shifted with respect to one another by half a pitch in the diagonal direction. Such an arrangement is advantageous in the process of identification of the fibre that generated the signal: each fibre is coupled to a unique pair of SiPMs. If there is an event in which only those two SiPMs registered a signal, one can immediately deduce which fibre was active.

\subsection{Silicon rasters} 
\begin{figure}[!htb]
     \centering
     \begin{subfigure}[b]{0.48\textwidth}
         \centering
     \includegraphics[width=\textwidth]{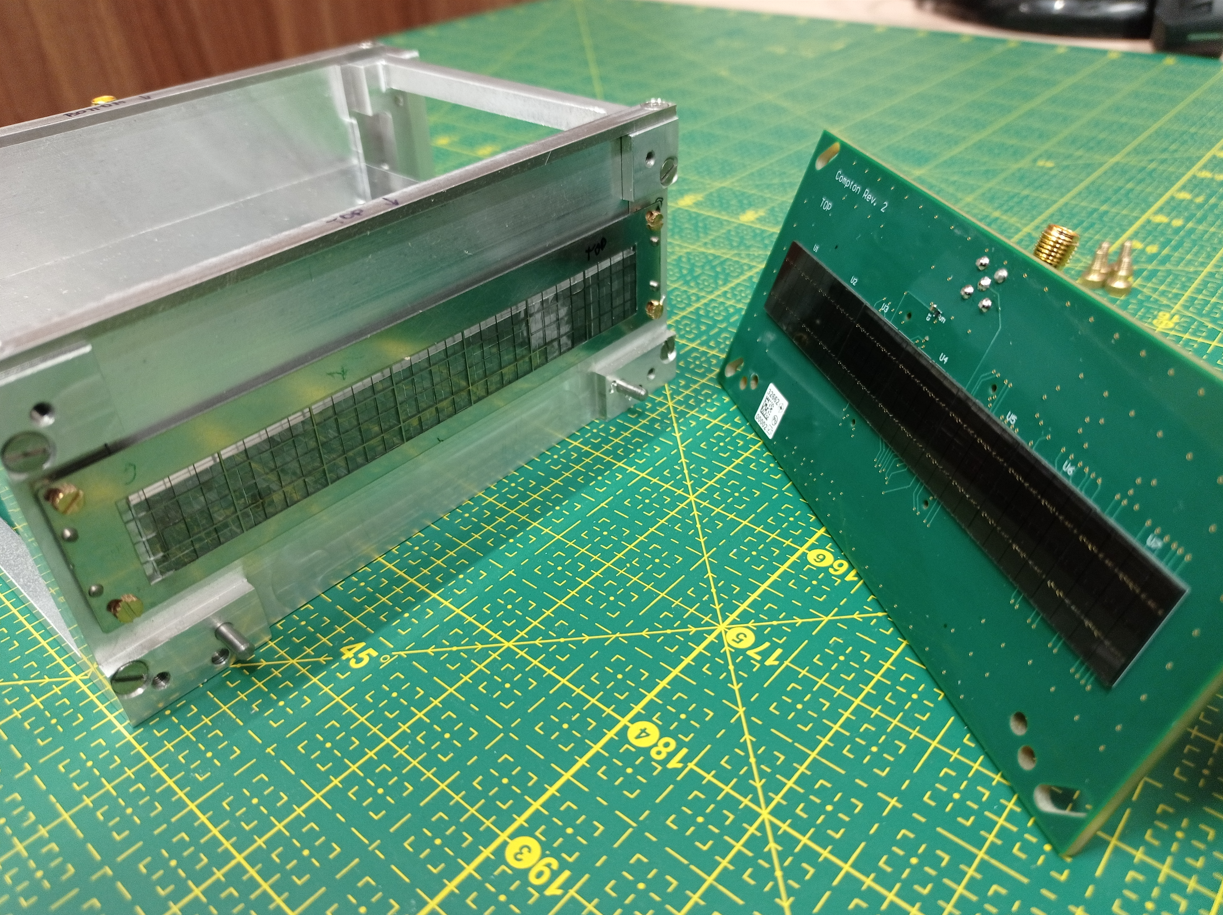}
     \end{subfigure}
     \hfill
     \begin{subfigure}[b]{0.48\textwidth}
         \centering
        \includegraphics[width=\textwidth]{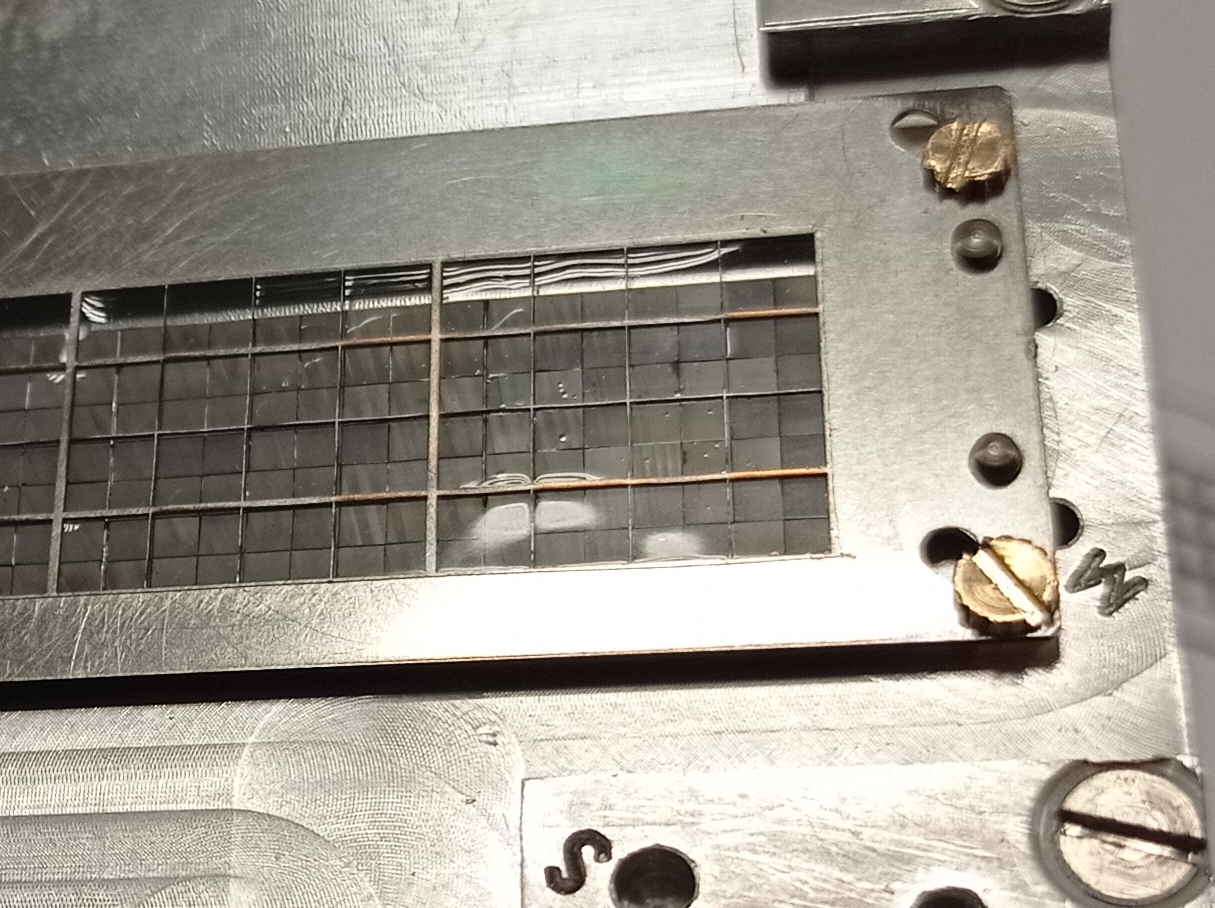}
     \end{subfigure}
        \caption{Left panel: the fibre stack with the attached optical-interface raster visible on the left, and the SiPM-housing PCB visible on the right; right panel: zoom on the raster.}
        \label{fig:rasterPhoto}
\end{figure}

In a previous study~\cite{Rusiecka2021}, several types of coupling between fibre ends and SiPMs were examined and it was concluded that the coupling that provides both good light transmission and stability over time are optical pads made of transparent silicon or equivalent material. Thus, in the present setup, we used \SI{0.5}{\mm} thick aluminium rasters filled with silicon material Elastosil RT 604~\cite{ElastosilPads} (see~\cref{fig:rasterPhoto}). The rasters were placed between the fibre ends and the SiPM arrays. The openings in rasters match exactly the SiPM borders, to reduce inter-SiPM crosstalk. 

\subsection{Masks}\label{sec:masks}
The coded mask is the MURA pattern of the 467th order. The 2D mask (\cref{fig:CM2DSimPicture}) is made up of 57 central pixels from the entire array horizontally and 45 central pixels vertically. The 1D mask (\cref{fig:CM1DSimPicture}) has the same dimensions, but the central row pattern is repeated over the 45 vertical pixel rows. Ensuring opacity for \gls{pgh} is nontrivial, thus the \gls{pgh} collimators are usually built out of thick, high-$Z$ materials. In our case, the opaque part of the mask is made of 20-mm-thick tungsten. 
A 3D printed mould with the mask pattern, made of 3D Jake Aqua Clear resin \cite{ResinAcquaClear}, was designed as a mask skeleton. The mould is~\SI{13}{mm} thick and has 10-mm-deep holes for the pixels that should be filled with mask material. The resin is a low-$Z$ material, so it can be treated as approximately transparent for the high-energy gammas. The raster is filled with removable tungsten rods of size \SI{2.26}{}$\times$\SI{2.26}{}$\times$\SI{20}{\mm\cubed}, building the opaque pixels. Due to this design, the mask can be easily reconfigurable. When the mask is put vertically, it has to be supported on the front side by an additional thin, removable sheet of \gls{pmma} so that the tungsten rods do not slide out.

\section{Front-end electronics and data acquisition system} \label{sec:DAQ}
A \gls{fee} and \gls{daq} system suitable for both modes (\gls{cc} and \gls{cm}) of the \gls{sificc} detector should preferably have the following features:
\begin{itemize}
    \item dead time not larger than a few \SI{}{\us}
    \item independent processing of data from each channel
    \item programmable coincidence scheme (allowing for channel-to-channel coincidences)
    \item scalability.
\end{itemize}

In order to find the optimal system, after a thorough market survey, we have selected several \gls{daq} systems that meet at least part of the criteria listed above. Those were: A5202, DT5742 (as a reference only), KLauS6b, TOFPET2c, TwinPeaks+TRB5sc. Their properties were investigated experimentally in a comparative manner to select the optimal \gls{daq} system for the \gls{sificc} project.\footnote{The \gls{daq} comparison study was done by the author of this thesis in collaboration with M.~L.~Wong and they contributed equally to the presented work.} The methodology, results and conclusions from these tests are described in detail in this section, and the major points have been published in~\cite{WongKolodziej2024}.
\subsection{SiPM readout systems}
\subsubsection{A5202}
FERS-5200~\cite{fers5200} is a scalable readout system for various detector types. The A5202~\cite{A5202_Manual_2023} (see~\cref{fig:DAQsystems_photo}a) is a 13-bit, 64-channel unit of this system, designed for SiPM readout. The core parts of such a unit are two 32-channel Citiroc-1A \glspl{asic}~\cite{DatasheetCitiroc1A}. The \gls{asic} can operate in several modes, depending on the use case:
\begin{itemize}
    \item Counting mode: the system counts self-triggers of every channel separately. The maximum counting rate is \SI{20}{Mcps}. It is the fastest of all modes, however, no energy or time information is provided.
    \item Timing mode: time stamps in each channel are registered, along with time-over-threshold, from which an approximate information on the pulse amplitude (and thus energy) can be extracted. The timing resolution in this mode is about \SI{250}{\pico\second} RMS. The acquisition is performed in packets - a packet is filled until it reaches a programmed size, then it is read out.
    \item Spectroscopy mode: optimised for energy measurement. The energy is retrieved from the pulse amplitude. This mode features a global trigger: once the trigger condition is met, the \gls{adc} conversion starts in all channels. The cost of accurate energy information is a significant dead time: \SI{10}{\micro\second} due to conversion.
    \item Spectroscopy and timing mode: combines the two preceding modes.
\end{itemize}
In this study, mainly the spectroscopy mode was used, except for the timing properties test, where we made use of the spectroscopy and timing mode. 
A major advantage of this system is the moderately flexible coincidence scheme, allowing for precise selection of the events of interest (e.g. channel-to-channel coincidences). However, the channel numbers that can be matched are hard-coded (0 \& 32, 1 \& 33 etc.) which requires appropriate matching of physical channels to \gls{asic} channels, only then this feature can be exploited. Another advantage of the system is a tunable gain, with two ranges available: high gain and low gain. Each of them has 64 steps. The setup comes with the Janus software, one of its convenient features being the live plotting of the collected spectra. Up to 16 A5202 units can be daisy-chained, larger numbers of channels require the use of a data concentrator DT5215~\cite{CAEN_concentrator_board}, to which up to 128 A5202 units can be connected.

\subsubsection{DT5742}
The DT5742~\cite{DT5742} (see~\cref{fig:DAQsystems_photo}b) is a desktop-form digitiser, its basic mode of operation is waveform recording. It offers 16 input channels and one channel for external trigger, 12-bit ADC and \SI{5}{GS\per\s} sampling frequency. The device is based on the DRS4 switched capacitor array chip \cite{DT5742_chip}. The maximum range of the signal amplitude is \SI{1}{Vpp}, but also \SI{2}{Vpp} range is available on demand at the time of ordering. The user can select from 4 available frequencies: \SI{5}{\giga \hertz}, \SI{2.5}{\giga \hertz}, \SI{1}{\giga \hertz}, and \SI{0.75}{\giga \hertz}. The acquisition window is always 1024 samples. Once the trigger condition is met, the \gls{adc} conversion starts for all channels, causing dead time of \SI{110}{\micro \second} or \SI{181}{\micro \second}, depending on whether the trigger channel is also digitised. In our studies, we used external trigger, but other options are also available (software trigger, self-trigger, low-latency trigger). Due to limited scaling options, the device was used only as a reference. It is nevertheless a good solution for applications requiring full waveforms, e.g. for detecting pile-ups or noise or examining signal shape. An earlier performance study of the DT5742 can be found in~\cite{Wang2022}. 

\subsubsection{KLauS6b}
The KLauS6b (see~\cref{fig:DAQsystems_photo}c) is an \gls{asic} originally prepared for the Analogue Hadronic Calorimeter (AHCAL)~\cite{Laudrain:2022fmh} and ScECAL~\cite{CALICE:2013zlb} within the CALICE~\cite{CALICE} \cite{Briggl2014} collaboration. It is manufactured in UMC \SI{180}{\nano\meter} CMOS technology. It is designed to operate with SiPMs with high density of cells. Such SiPMs feature a high dynamic range, but a low intrinsic gain of the order of $10^5$. KLauS6b can be either auto- or externally triggered. There are 36 channels per unit, with selectable \gls{adc} precision of 10 or 12 bits. There are two charge integration branches, high or low gain. Each branch consists of a passive integrating circuit and a low-pass filter. The charge information is conveyed by the amplitude of the output signal from either of the branches~\cite{Yuan2019}. One can select the preferred branch in software, then only output from this branch is digitised. The available interfaces to send data off the chip are I$^2$C (up to 20~Mbit/s) or LVDS (up to 160~Mbit/s). The main dead-time contributions are the front-end processing time and the \gls{adc} conversion time, which are estimated by the producer to be below \SI{800}{ns}~\cite{Yuan2019}. The acquisition software, operating on Linux, is provided by the producer.

\subsubsection{TOFPET2c}\label{sec:TOFPET2c}
The TOFPET2c (see~\cref{fig:DAQsystems_photo}d) is a SiPM readout and digitisation system~\cite{DIFRANCESCO2016194}. It has been optimised for time-of-flight (TOF) measurements with PET. Its main advantages are low power consumption and low noise. The system consists of a FEB/A board housing an~\gls{asic}, an interface board FEB/I, FEB/D (which manages powering, configuration and synchronisation of the ASICs) and a clock board (which, together with the FEB/D, allows for coincidence detection). The \gls{asic} is produced in the CMOS \SI{110}{nm} technology. The basic unit (the \gls{asic}) has 64 channels with independent readout. Each channel features a quad-buffered TDC and a charge integrating 10-bit \gls{adc}. The dynamic range declared by the producer is \SI{1500}{pC}. Up to 16 \gls{asic} units can be connected to a front-end board (FEB/D)~\cite{FEBD} via interface boards (FEB/I). The system can be scaled up further by plugging multiple FEB/Ds to a~Clock\&Trigger unit. The data are transferred via an optical link to a \gls{pcie} data acquisition board and then to a PC. A Linux-operated acquisition software, provided by the PETSys company, is used to manage this process. The system is a complete product that is commercially available and can be used off-the-shelf.

\subsubsection{TwinPeaks+TRB5sc}
\begin{figure}
     \centering
     \begin{subfigure}[b]{0.48\textwidth}
         \centering
         \includegraphics[width=\textwidth]{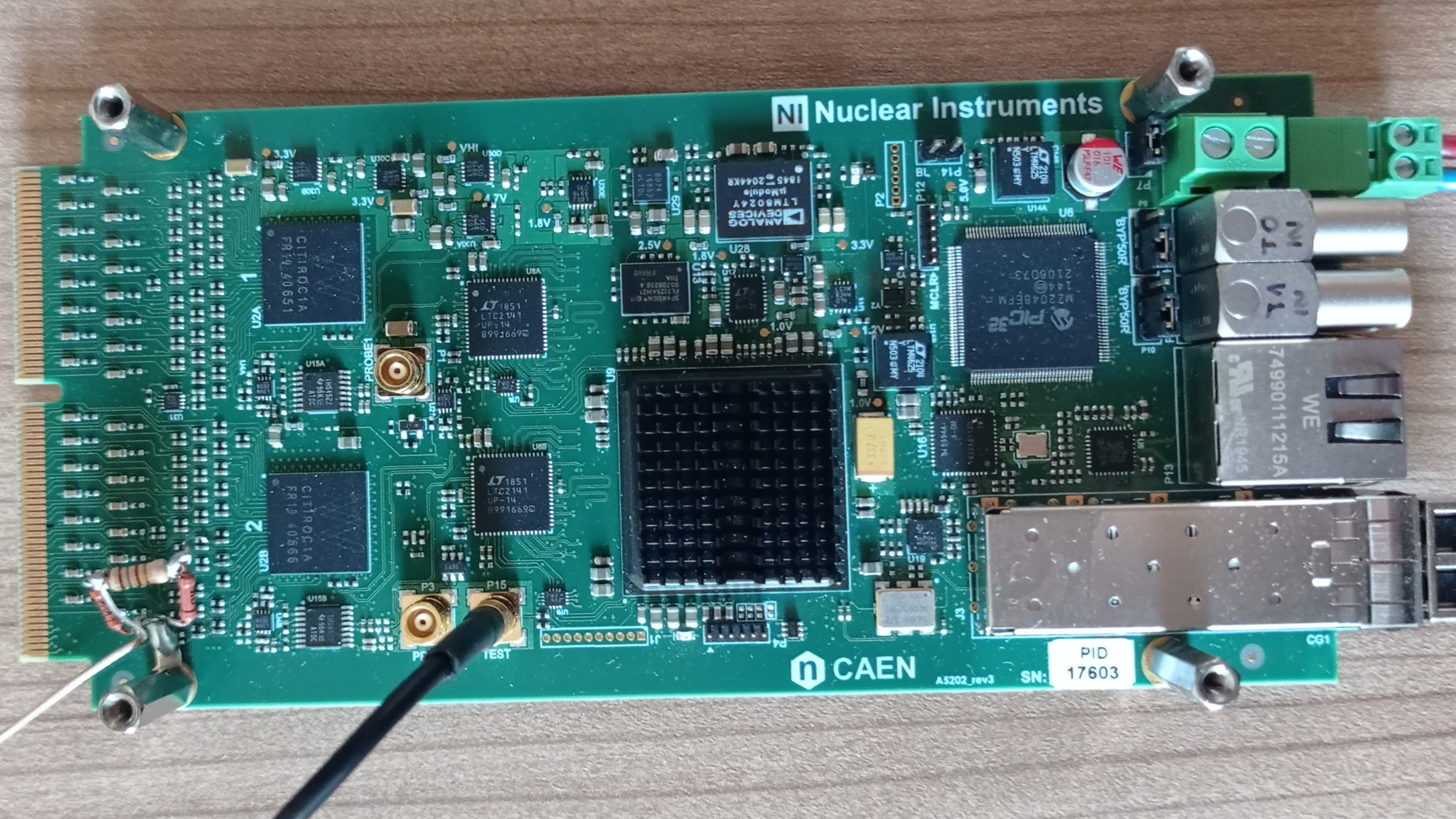}
         \caption{A5202}
         \label{fig:A5202}
     \end{subfigure}
     \hfill
     \begin{subfigure}[b]{0.48\textwidth}
         \centering
         \includegraphics[width=\textwidth]{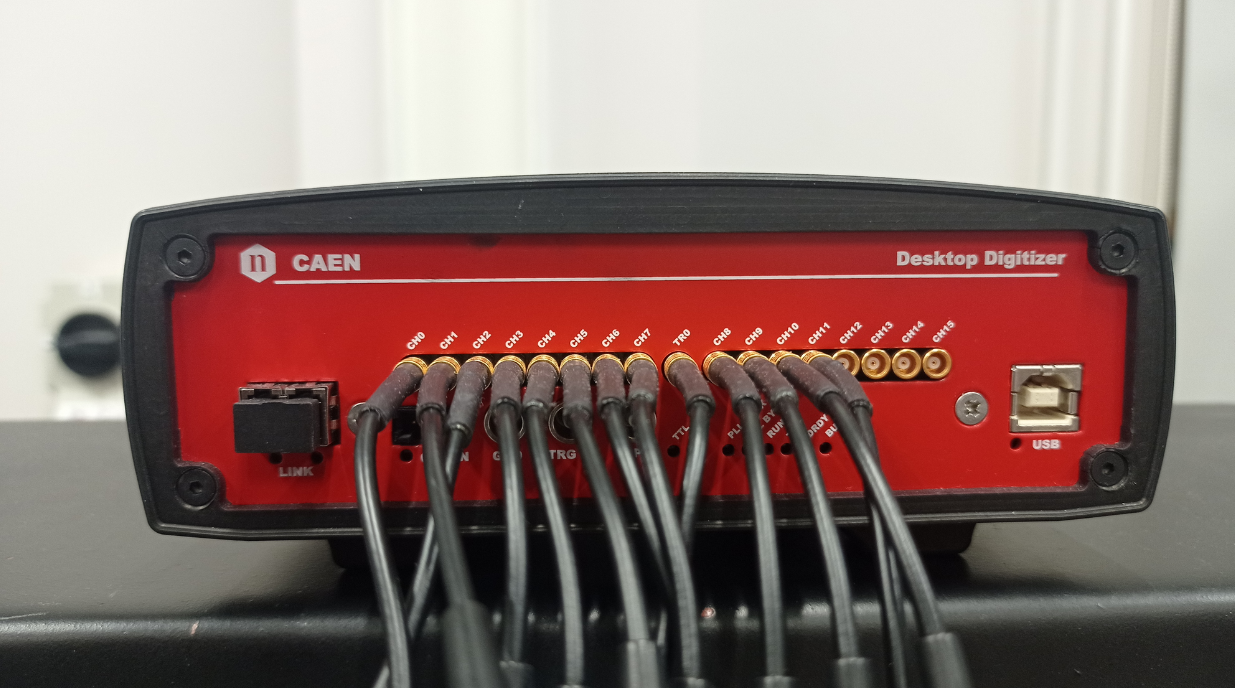}
         \caption{DT5742}
         \label{fig:DT5742}
     \end{subfigure}
     \hfill
     \begin{subfigure}[b]{0.48\textwidth}
         \centering
         \includegraphics[width=\textwidth]{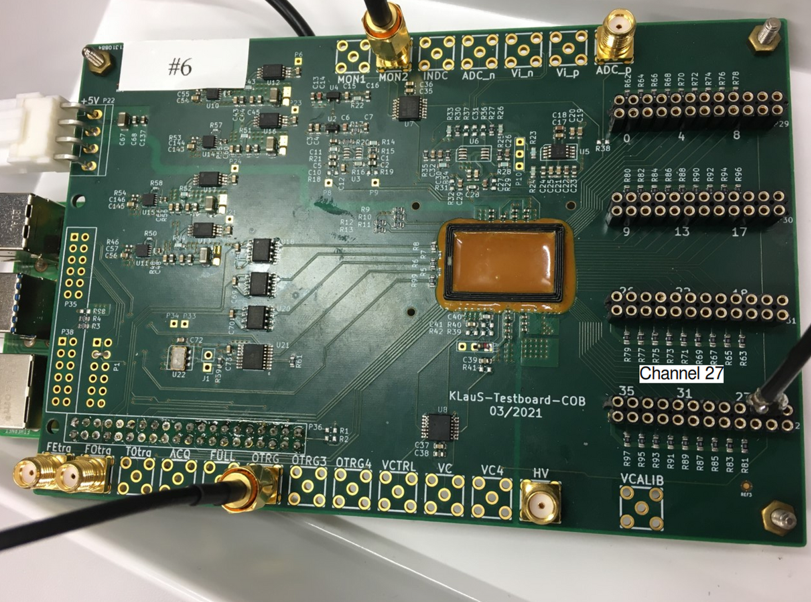}
         \caption{KLauS6b}
         \label{fig:KLauS6b}
     \end{subfigure}
     \hfill
     \begin{subfigure}[b]{0.48\textwidth}
         \centering
         \includegraphics[width=\textwidth]{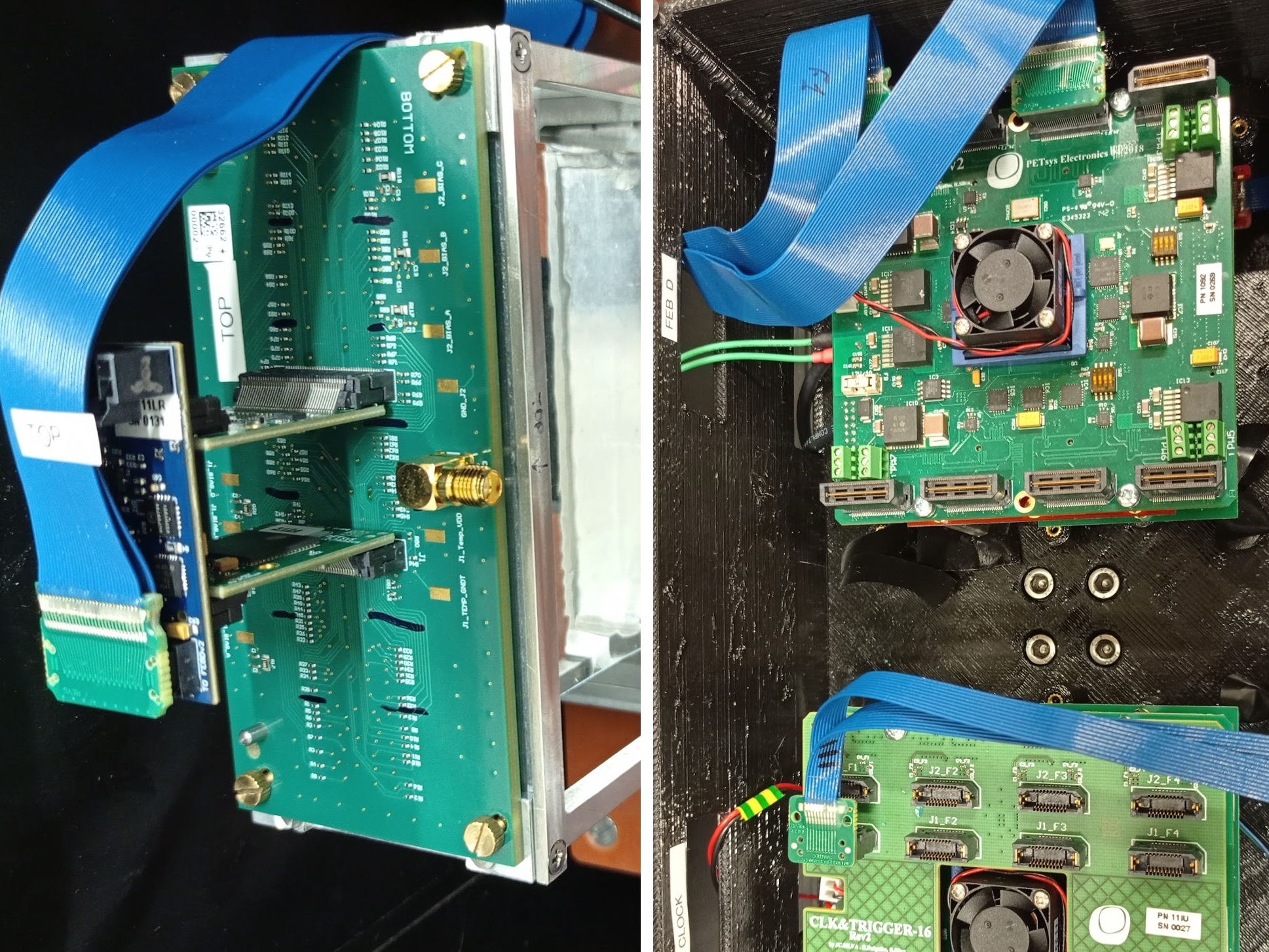}
         \caption{TOFPET2c}
         \label{fig:TOFPET2c}
     \end{subfigure}
     \hfill
     \begin{subfigure}[b]{\textwidth}
         \centering
         \includegraphics[width=\textwidth]{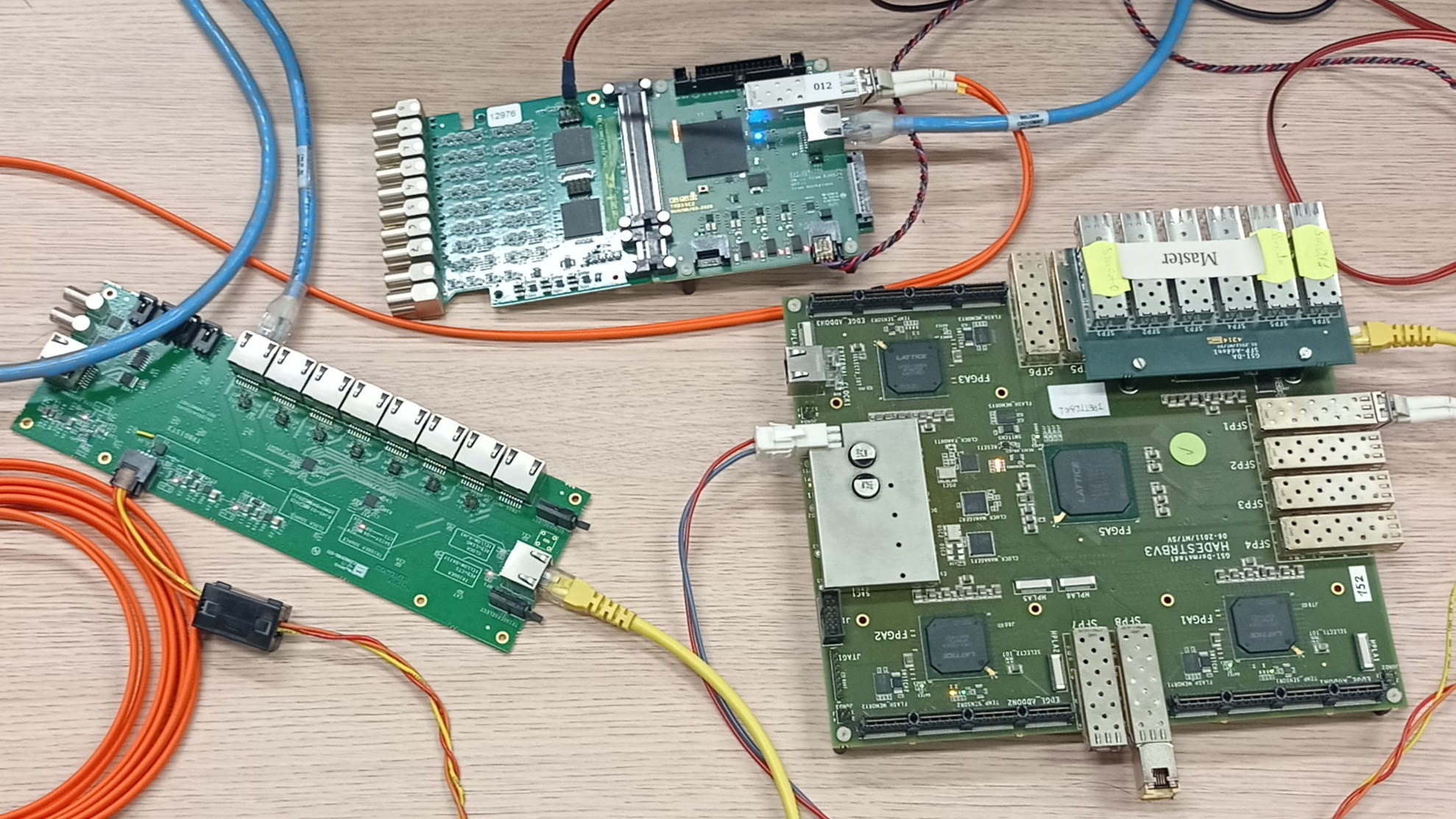}
         \caption{TwinPeaks+TR5Bsc}
         \label{fig:TwinPeaks_TR5Bsc}
     \end{subfigure}
        \caption{The data acquisition systems.}
        \label{fig:DAQsystems_photo}
\end{figure}

This readout system (see~\cref{fig:DAQsystems_photo}e) comprises a few devices: TwinPeaks~\cite{GSI_report_TwinPeaks}, TRB5sc~\cite{TRBfamily}, and TRB3~\cite{Neiser:2013yma}. The first one is an add-on board specially developed for the DESPEC experiment at GSI/FAIR~\cite{RUDIGIER2020163967}. It is a high-resolution charge-to-time amplifier. It is connected to TRB5sc, which features an ECP5 FPGA that can perform high-resolution TDC conversion. Multiple sets of TwinPeaks+TRB5sc boards can be connected via the TrbNet~\cite{Michel:2010ffa} protocol to TRB3, in order to build a scalable system of a tree-like structure. The TRB3 motherboard includes a trigger system that can send a timing signal to all connected boards, collect data recorded by them, and transfer the data to event-building software. There are 16 channels per TwinPeaks board, with ADC precision of 8 bits. The system comes with Linux-operated, custom software, DABC~\cite{Adamczewski-Musch:2015arx}.

\subsection{Experimental setup}
\begin{figure}
\centering
\includegraphics[width = \textwidth]{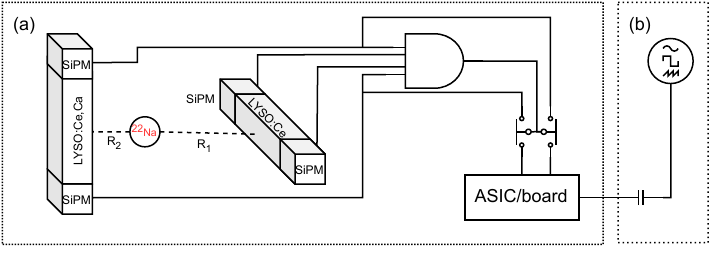}
\caption{Experimental setups for \gls{daq} comparison tests with a single fibre. Adapted from~\cite{WongKolodziej2024}.}
\label{fig:SetupSingleFibre}
\end{figure}

We designed and constructed two setups to test the performance of the readout systems. The first setup involved a bunch of 64 LYSO:Ce,Ca fibres of size \SI{1.28}{}$\times$\SI{1.28}{}$\times$\SI{100}{\mm\cubed} each, produced and stacked by Taiwan Applied Crystals \cite{TACwebsite} in 4 layers, 16 fibres per layer. Each fibre was coupled at both ends to a pair of AFBR-S4K11C0125B SiPMs~\cite{KetekSiPMs} (formerly PM1125-WB produced by KETEK GmbH), which were read out by the tested readout system. A radioactive $^{22}\mathrm{Na}$ source was placed at half the length of the fibre stack and another LYSO:Ce crystal was placed opposite to the fibre stack as a reference~(\cref{fig:SetupSingleFibre}a). The only registered signals were the ones in coincidence with the reference crystal, forming the electronic collimation setup similar to the one in \cite{Anfre2007}. The other setup used for readout systems testing was an SiPM-like pulse generator~(\cref{fig:SetupSingleFibre}b), consisting of a 81160A Pulse Function Arbitrary Generator \cite{PulseGeneratorWeb} and either a \SI{33}{pF} or a \SI{600}{pF} capacitor (depending on the readout system).
 
\subsection{Performance metrics}\label{sec:DAQ_performanceMetrics}
The readout systems were compared in terms of five major features: energy resolution, dead time, dynamic range, efficiency, and coincidence timing resolution.
\begin{figure}[!htb]
\centering
\includegraphics[width = \textwidth]{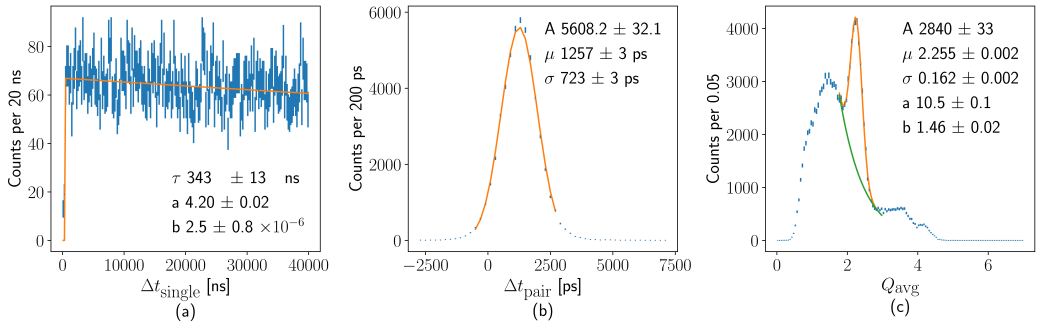}
\caption{Methodology of the \gls{daq} comparative study. Adapted from~\cite{WongKolodziej2024}.}
\label{fig:methodology}
\end{figure}
\begin{itemize}
    \sloppy
    \item \textbf{Energy resolution} The QDC or ADC (depending on the readout system) spectra were recorded with a $^{22}\mathrm{Na}$ source~(\cref{fig:methodology}c). The averaged $Q_{\mathrm{av}}$ ($A_{\mathrm{av}}$) is defined as the geometric mean of QDCs (ADCs) recorded by the SiPMs on both sides. 
    In the $Q_{\mathrm{av}}$ ($A_{\mathrm{av}}$) spectrum, the neighbourhood of the \SI{511}{keV} peak was fit with a combination of a Gaussian describing the peak, and an exponent with a constant offset representing the background. The standard deviation (sigma) of the Gaussian, normalised by its mean position, defines the energy resolution (ER).
    \item \textbf{Dead time} We define the dead time $\tau$ of the system as the minimum time between two subsequent registered hits. We determine this parameter by plotting a histogram of time difference between consecutive hits in one of the channels and determining the width of the gap near zero, by fitting the following function: 
    \begin{equation}
    y=\begin{cases}
        0 & \text{if } \Delta t_\mathrm{single}<\tau, \\
        e^{a-b(\Delta t_\mathrm{single}-\tau) } & \text{otherwise}.
    \end{cases}        
    \end{equation}
The $\tau$, $a$ and $b$ are the fit parameters. An example of such a fit is presented in~\cref{fig:methodology}a. The dead time in this case was found to be \SI{0.343(13)}{\micro s}.
    \item \textbf{Dynamic range} To quantify the dynamic range of a readout system, we probe it by sending artificial pulses of different amplitudes and thus, different injected charge. We use a square pulse from a generator and put it through a capacitor, to mimic the SiPM signal shape. The experimental setup is depicted in~\cref{fig:SetupSingleFibre}b. We plot the injected charge ($Q_{\mathrm{in}}$) vs. the response of the system (in \gls{adc}, QDC or TOT units, depending on the system; details of the procedure can be found in~\cite{WongKolodziej2024}) and look for $Q_{\mathrm{in}}$ value where the dependence deviates from linearity (see~\cref{fig:dynamicrange}). Such a $Q_{\mathrm{in}}$ value defines the upper limit of the dynamic range of the investigated system.
    \begin{figure}
    \centering
    \includegraphics[width = \textwidth]{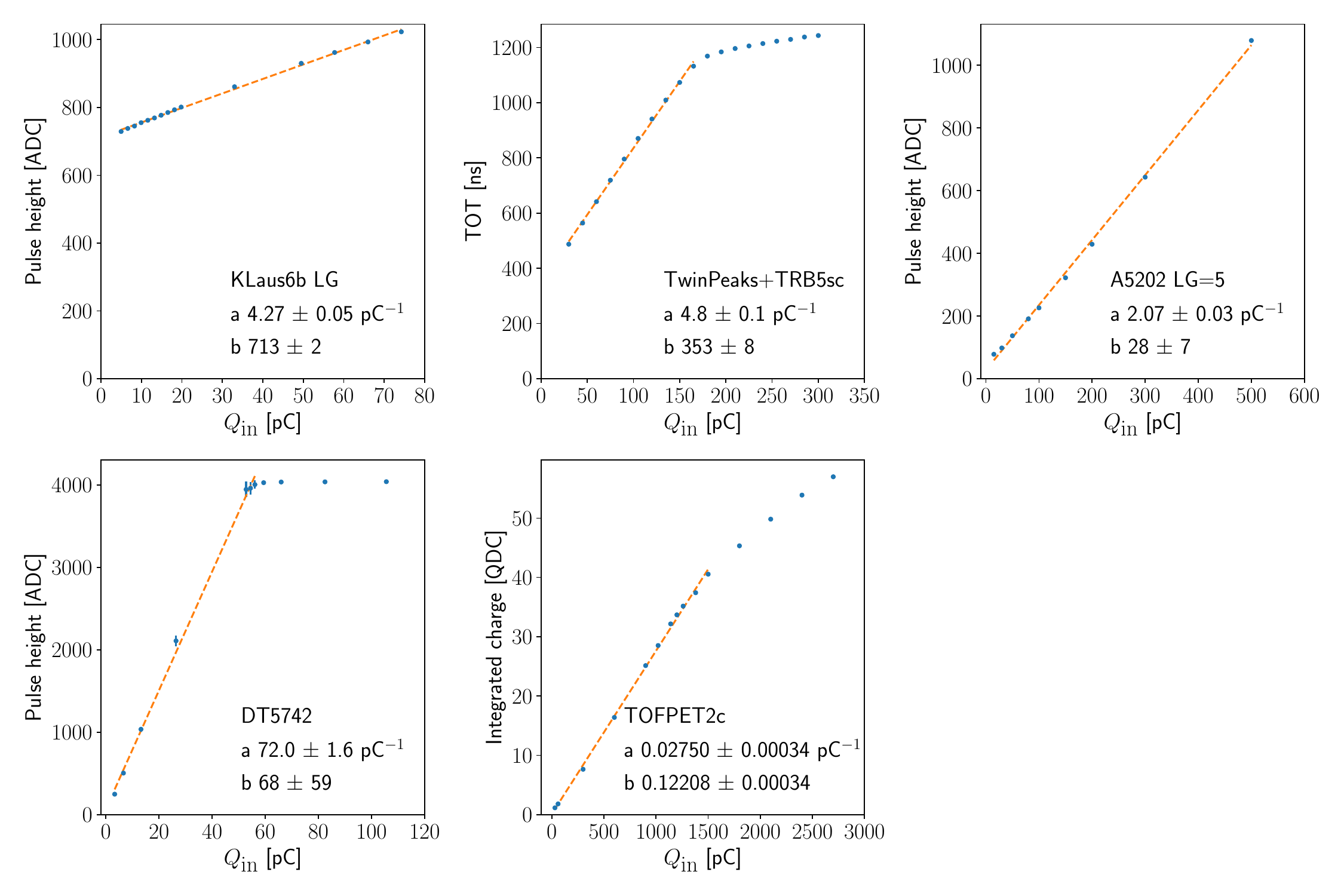}
    \caption{Dynamic range for all tested \gls{daq} systems. Adapted from~\cite{WongKolodziej2024}.}
    \label{fig:dynamicrange}
    \end{figure}
    \item \textbf{Efficiency} For efficiency measurement, we use the same setup as for dynamic range determination, but in this case the signal amplitude is fixed to \SI{300}{mV}, while the frequency of the signals is gradually increased. We define the efficiency as a ratio of the registered and injected (true) rate $r$. It is quantified by fitting a sigmoid: 
    \begin{equation}
        \epsilon = \frac{1}{1+e^{a+b(r-r_0)}},
    \end{equation}
    where $r_0$ is the middle point of the falloff, while $a$ and $b$ are the remaining fit parameters and the efficiency limit is defined as the rate at which the efficiency curve drops to 90\%. 
    \item \textbf{Coincidence timing resolution} We define the coincidence timing resolution of the system (CTR) by plotting a time difference between the SiPM signals at both fibre ends. To diminish the walk effect, for this plot we choose only events from within 3$\sigma$ around the \SI{511}{keV} peak in the $Q_{\mathrm{av}}$ ($A_{\mathrm{av}}$) spectrum. The coincidence timing resolution is the standard deviation of the Gaussian fit to such a plot. For this analysis, the setup in~\cref{fig:SetupSingleFibre}a was used.
\end{itemize}

\subsection{FEE and DAQ: results}
We have evaluated the readout systems in terms of the performance metrics described in~\cref{sec:DAQ_performanceMetrics}. The values obtained are collected in~\cref{tab:measurements_comparison} and in~Figures~\ref{fig:dynamicrange} and~\ref{fig:efficiency}.
\begin{itemize}
\item \textbf{Energy resolution} In terms of energy resolution, TOFPET2c performed the best: the resolution obtained with this system was 7.2(1)\%. It is followed by DT5742 with 20\% poorer resolution (for exact values, see~\cref{tab:measurements_comparison}), then A5202 (30\% poorer), TwinPeaks+TRB5sc (60\% poorer), and finally KLauS6b (75\% poorer).  
\item \textbf{Dead time}
Two systems presented comparably low dead time: TOFPET2c and KLauS6b (\SI{0.343(13)}{\micro\s} and \SI{0.352(77)}{\micro\s}, respectively). The next system in this category was TwinPeaks+TRB5sc with \SI{0.870(9)}{\micro\s}, followed by A5202 (\SI{44(4)}{\micro\s}) and DT5742 (\SI{443(97)}{\micro\s}). 
\item \textbf{Dynamic range}
The system that performed the best in this category was again TOFPET2c with a dynamic range of up to \SI{1899}{pC}. This is much higher than the next best system, A5202, for which the dynamic range was 4 times smaller. The rest of the systems' dynamic ranges were: 9 times smaller than TOFPET2c (TwinPeaks+TRB5sc), 25 times smaller (KLauS6b), and 30 times smaller (DT5742). 
\item \textbf{Efficiency}
The efficiency plots are presented in~\cref{fig:efficiency}. 
\begin{figure}[!htb]
\centering
\includegraphics[width = 0.6\textwidth]{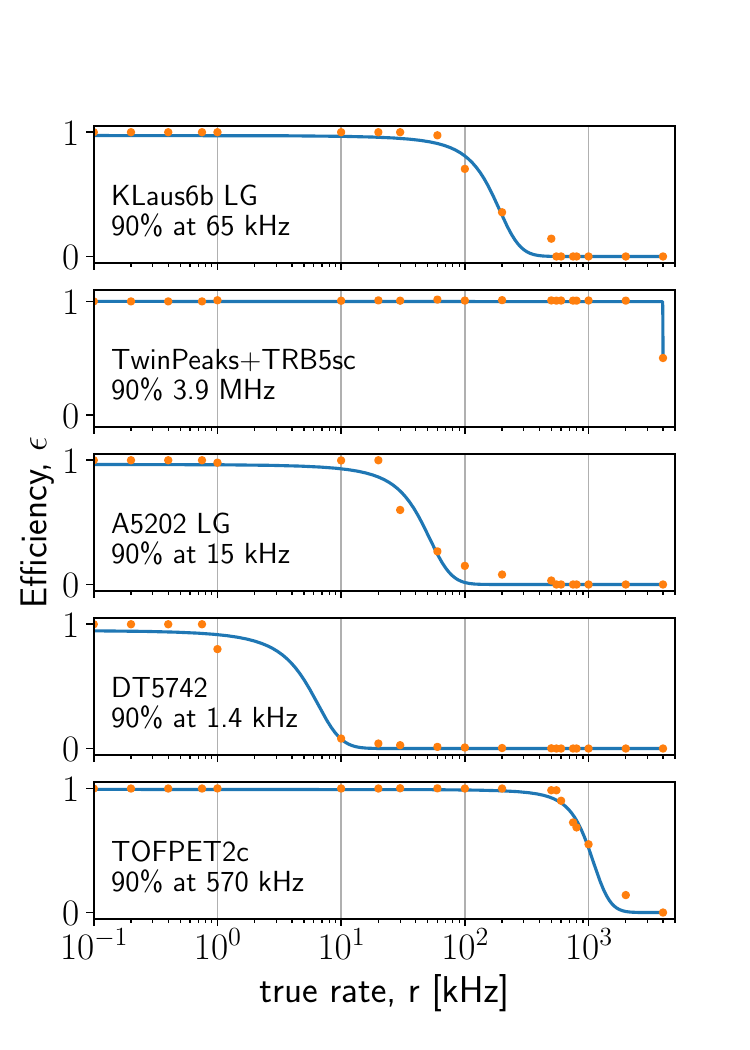}
\caption{Efficiency for all tested DAQ systems. Adapted from~\cite{WongKolodziej2024}.}
\label{fig:efficiency}
\end{figure}
The winner in this category was TwinPeaks+TRB5sc, which can operate with full efficiency at \SI{3.9}{MHz} rate. TOFPET2c is the second-best with 7~times lower limit of full efficiency operation. It is followed by KLauS6b, A5202 and DT5742, with respectively 60~times, 260~times, and 2800~times lower maximal rate than the winner.
\item \textbf{Coincidence timing resolution}
The obtained coincidence timing resolution was the best for TOFPET2c: \SI{0.723(3)}{ns}. The next best result was obtained for DT5742 (\SI{1.152(1)}{ns}), followed by KLauS6b (\SI{1.53(7)}{ns}). The last ones in this category were A5202 and TwinPeaks+TRB5sc, obtaining respectively coincidence timing resolution of \SI{3.0(2)}{ns} and \SI{10.5(3)}{ns}.
\end{itemize}
\begin{table}
\caption{Summary of the obtained values of performance metrics, as defined in~\cref{sec:DAQ_performanceMetrics}, for the examined \gls{daq} systems. ER - energy resolution, DR - dynamic range, Rate - rate at 90\% efficiency, CTR - coincidence timing resolution, TP - TwinPeaks. The $\dag$ marks the systems for which the deviation from linearity was not observed.}
    \centering
    \begin{tabular}{rcccccc}\toprule
        System & ER [\%]  &  Dead time [\SI{}{\micro\second}] & DR [pC]  &  Rate [kHz] & CTR [ns] \\   \midrule \rowcolor[gray]{.95}
       KLauS6b  & \eqmakebox[cint][r]{12.4(1)}              & \eqmakebox[cint][r]{0.352(77)}          & \eqmakebox[cint][r]{74.25$^{\dag}$}           & \eqmakebox[cint][r]{65}                            & \eqmakebox[cint][r]{1.53(7)} \\               
        TP+TRB5sc             & \eqmakebox[cint][r]{11.1(13)}     & \eqmakebox[cint][r]{0.870(9)}              & \eqmakebox[cint][r]{202}            & \eqmakebox[cint][r]{3900}                                 & \eqmakebox[cint][r]{10.5(3)}   \\   \rowcolor[gray]{.95}       
        A5202           & \eqmakebox[cint][r]{9.4(3)}                & \eqmakebox[cint][r]{44(4)}                  & \eqmakebox[cint][r]{500$^{\dag}$}            & \eqmakebox[cint][r]{15}                      & \eqmakebox[cint][r]{3.0(2)}       \\                    
         DT5742            & \eqmakebox[cint][r]{8.55(4)}                & \eqmakebox[cint][r]{443(97)}                & \eqmakebox[cint][r]{62}                          & \eqmakebox[cint][r]{1.4}                   & \eqmakebox[cint][r]{1.152(1)}  \\  \rowcolor[gray]{.95}                
        TOFPET2c           & \eqmakebox[cint][r]{7.2(1)}               & \eqmakebox[cint][r]{0.343(13)}         & \eqmakebox[cint][r]{1899}                          & \eqmakebox[cint][r]{570}                       & \eqmakebox[cint][r]{0.723(3)}  \\   \bottomrule                                                           
    \end{tabular}                                   
    \label{tab:measurements_comparison}
\end{table}

\subsection{FEE and DAQ: summary and discussion}
The comparison of \gls{fee}+\gls{daq} systems was initially just an auxiliary piece of work necessary before the purchase of an \gls{fee}+\gls{daq} system for the \gls{sificc} project. However, it evolved into a more general study, the results of which have been summarised in~\cite{WongKolodziej2024} and are valid for any application requiring the use of high-rate \glspl{sipm} readout. We selected the most promising solutions available on the market or in the development process and thoroughly examined them. 

In the context of finding the \gls{daq} system for the \gls{sificc} project, TOFPET2c turned out to be optimal. It performed the best of all the systems in four out of five categories (dead time, energy resolution, dynamic range, coincidence timing resolution), being the second best choice for the last category (efficiency). However, it should be noted that none of the systems was optimised particularly for the \gls{sificc} project, but for other applications (not necessarily fully consistent with our use case), so the systems' parameters varied vastly among one another, as various features were prioritised by the producers. For example, KLauS6b was optimised for high-density SiPMs, which can achieve high dynamic range at the cost of a relatively low single-pixel gain (10$^5$). In our study, it obtained second-lowest dead time, which makes it a good choice for high-rate applications. A5202, even though it did not obtain competitive results in our comparison, has a very good peak separation in the single photoelectron spectrum. Thus, it can be useful when the focus of the study is in the low-intensity region and the rate is not too high. TwinPeaks+TRB5sc proved to have excellent efficiency, much higher than any other system. Thus, it is the optimal choice for applications where rate capability is crucial, and the coincidence timing resolution is not prioritised. Finally, DT5742 (which was included as a reference system, due to limited scalability options) has the second-best energy- and coincidence timing resolution. It is a good choice for applications with a small number of readout channels, where full waveforms are required, e.g. for SiPM performance studies or particle identification via pulse-shape discrimination.

\section{Setup assembly}
\begin{figure}[!htb]
     \centering
     \begin{subfigure}[b]{0.45\textwidth}
         \centering
         \includegraphics[width=\textwidth]{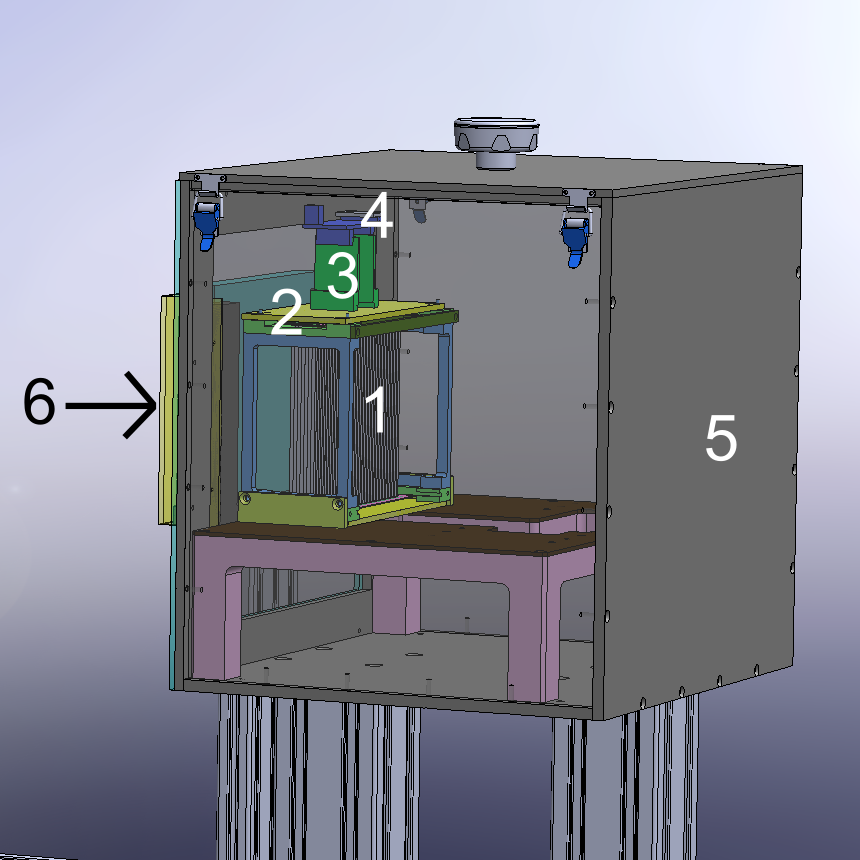}
         \caption{ }
         \label{fig:supportStructure1}
     \end{subfigure}
     \hfill
     \begin{subfigure}[b]{0.45\textwidth}
         \centering
         \includegraphics[width=\textwidth]{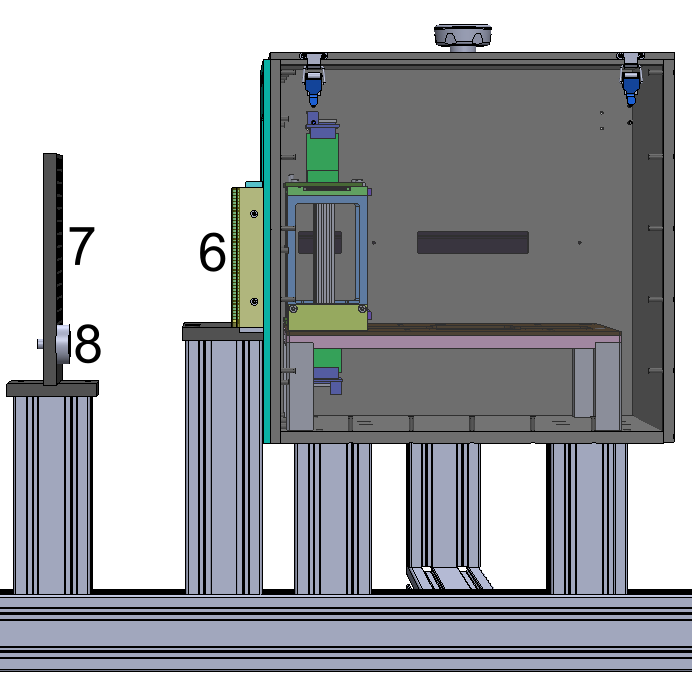}
         \caption{ }
         \label{fig:supportStructure2}
     \end{subfigure}
     \hfill
        \caption{The experimental setup viewed at two different angles. (a) Fibre stack module (1, grey); SiPM \glspl{pcb} (2, yellow); parts of the TOFPET2c system: FEB/A (3, green) and FEB/I (4, purple); light-tight box (5, dark grey); mask (6). (b) Side view of the experimental setup in a radioactive source-measurement configuration. Colour coding the same as in the left panel; additionally, there is the  source raster (7, black) and the radioactive source (8, grey) in an example position below the setup central axis.} 
        \label{fig:supportStructure}
\end{figure}

The scatterer module consists of a stack of scintillation fibres, \glspl{pcb} housing SiPM arrays and silicon rasters which couple the fibre ends to the SiPM arrays. For the tests, the module was mounted in a light-tight container, which can be seen in~\cref{fig:supportStructure}. 
The container is made of black polyoxymethylene (POM) of \SI{9}{\mm} thickness, all the feedthroughs for cables were optically isolated with black tape. The top wall is a removable lid, for easy setup manipulation. The front wall of the container is made of black fabric~\cite{blackoutFabric}, to minimise the scattering effect on the way to the detector. We checked the light-tightness of the container in two steps. First, by shining a strong torch onto the box from the outside and checking visually if any light leaked to the inside. After this test revealed no leaks, we took two short measurements of the SiPM signal, with the room light on and off, and compared if the registered count rate differed in the two situations. No such effect was observed, so we concluded that the container was sufficiently light-tight for our purpose. All parts of the experimental setup were mounted on a support structure, which was made of aluminium profiles with sub-millimetre positioning precision. The assembled experimental setup is presented in~\cref{fig:supportStructure}. The height of the aluminium legs holding the light-tight box, the mask and the radioactive source raster (i.e., a raster allowing for mounting the radioactive source in different positions) were adjusted so that the central points of the setup elements are at the same level (the same position in \textit{y}) and at the same \textit{x} position (horizontal, perpendicular to the setup axis \textit{z}). The distance between the central points of the radioactive source and the mask was \SI{170}{\mm}, the distance between the central points of the mask and the fibre stack was \SI{63}{\mm}. The fully-coded \gls{fov} resulting from the detector geometry was \qtyproduct[product-units=bracket-power]{176x100}{\mm}. However, in the measurements described in this thesis, the analysis was restricted to \gls{fov} of \qtyproduct[product-units=bracket-power]{140x100}{\mm}. The weight of the mask was \SI{3}{kg}. The setup used in tests with the proton beam was analogous to the one for the radioactive source measurements, with the exception of a cuboidal \gls{pmma} phantom being placed instead of the radioactive source and the radioactive source raster. The geometric centre of the phantom was aligned with the central points of the mask and the fibre stack. The phantom was not placed on the aluminium support structure, but on a dedicated \gls{pmma} table instead.
\chapter{Experiments}
The experiments presented in this chapter, as well as the data analysis and results presented in the further chapters of this thesis, have also been partially described in \cite{Kolodziej2025preprint}; the preprint has been submitted to Physics in Medicine and Biology in January 2025. The author of this thesis is also the first and corresponding author of the said preprint.
\section{Laboratory tests}
\subsection{Calibration}\label{sec:Experiments_Calibration}
\subsubsection{Position calibration along the fibre}
As mentioned in the detector description~(\cref{sec:ComptonCamera}), the reconstruction of the hit position along the $x$ and $z$ axes of the fibre stack is performed by identifying the fibre that was hit. Along the $y$ axis (i.e. along the fibre), determination of the hit position is more complex, as one needs to reconstruct it from charges collected by \glspl{sipm} at both fibre ends. This is based on the fact that the light collection on the \gls{sipm} depends on the point of interaction along the fibre, due to the attenuation of the scintillation light within the fibre. The procedure of $y$-position calibration is described in~\cite{rusiecka2023sificc}. The measure of this procedure's accuracy is the position resolution along the fibre. In principle, the position along the fibre could also be determined based on the time information from both SiPMs, but the obtained time resolutions result in a position resolution of about \SI{7}{cm} only, thus we discarded this method.

Before the measurements with the proton beam, we performed a preliminary study of the light attenuation along the fibre (which is connected to the position resolution). The study showed that the light attenuation (and therefore the position resolution along $y$) in the present experimental setup was not sufficient, which would also impede the performance of the \gls{2dcm} setup~(\cref{fig:CM2DSimPicture}). Thus, we decided to focus on examining the performance of the \gls{1dcm} setup~(\cref{fig:CM1DSimPicture}). There, the fibres can be treated as pixels and the mask pattern is the same along $y$, so the position along $y$ is not needed.

The $y$-position calibration of the detector was performed only after the proton beam measurements\footnote{The position calibration was conducted by K.~Rusiecka}. It was done by placing an electronically collimated $^{22}$Na source (the electronic collimation principle can be found in~\cite{Anfre2007}) in different positions along the fibre (at \SI{10}{\mm} intervals) and registering \gls{sipm} signals from both ends of the fibre. Each measurement was \SI{15}{\min} long to collect enough statistics. With the electronic collimation scheme, the length of the irradiated slice of the fibre is determined by the geometry of the reference detector and source-detector distances. In this case, for a single source position a fibre slice of \SI{2.4}{\mm} length was irradiated, which also defined the hit position uncertainty. Due to the 4-to-1 coupling between fibres and \glspl{sipm}, for a calibration measurement one needs to select only single-fibre events, i.e. those in which there was only one fibre active, i.e. with one top and one bottom \gls{sipm} active. In the data, there were also other event classes present, with multiple active fibres. The details of how they were handled can be found in~\cref{sec:fibreHits}. The calibration measurement was done for all fibres simultaneously; data samples representing responses of individual fibres were identified and the calibration procedure for each fibre was performed separately. The detailed calibration procedure was developed within the group during previous fibre tests and is described in~\cite{Rusiecka2021}. A custom model of light propagation along the fibre is fit to the data. The hit position along the fibre and the energy deposit are reconstructed event-by-event, based on that model, and the charge information from \glspl{sipm} at both fibre ends. The distribution of the hit position residuals and energies yields the position and energy resolutions. From this measurement, an energy resolution of 6.5$\pm$\SI{0.5}{\percent} and a position resolution of 74$\pm$\SI{10}{\mm} were obtained. These values differed significantly from what was expected based on extensive tests of both single fibres and small-scale fibre stacks. The energy resolution improved, while the position resolution deteriorated, which suggests that the attenuation of optical photons in the fibres of the full-scale module was much smaller than in the previously tested fibres. After an extensive experimental survey, we identified the reason for such a change of fibre properties: the manufacturer has changed the type of the wrapping aluminium foil (8011~\cite{AlFoil8011} to 1060~\cite{AlFoil1060}) without any notice, when moving to thicker fibres (cross section of 1.28$\times$\SI{1.28}{\mm\squared} to cross section of 1.94$\times$\SI{1.94}{\mm\squared}).

The issue with position resolution was addressed in the construction of the next version of the scatterer (and the absorber for the CC mode): firstly, the aluminium foil type 8011 will be used; secondly, it was determined that covering one side of the fibre with white paint before wrapping it in the aluminium foil greatly increases the performance in terms of position resolution. Preliminary tests on single fibres\footnote{The tests were conducted by K.~Rusiecka and B.~Pióro.} yielded very promising results: position resolution of 12~mm, and energy resolution of 8.3\%. Thus, future versions of the detector will be assembled in this way.

\subsubsection{Gain alignment of SiPMs}\label{sec:energyCalibrationPerSiPM}
\begin{figure}[!htb]
         \centering
         \includegraphics[width=\textwidth]{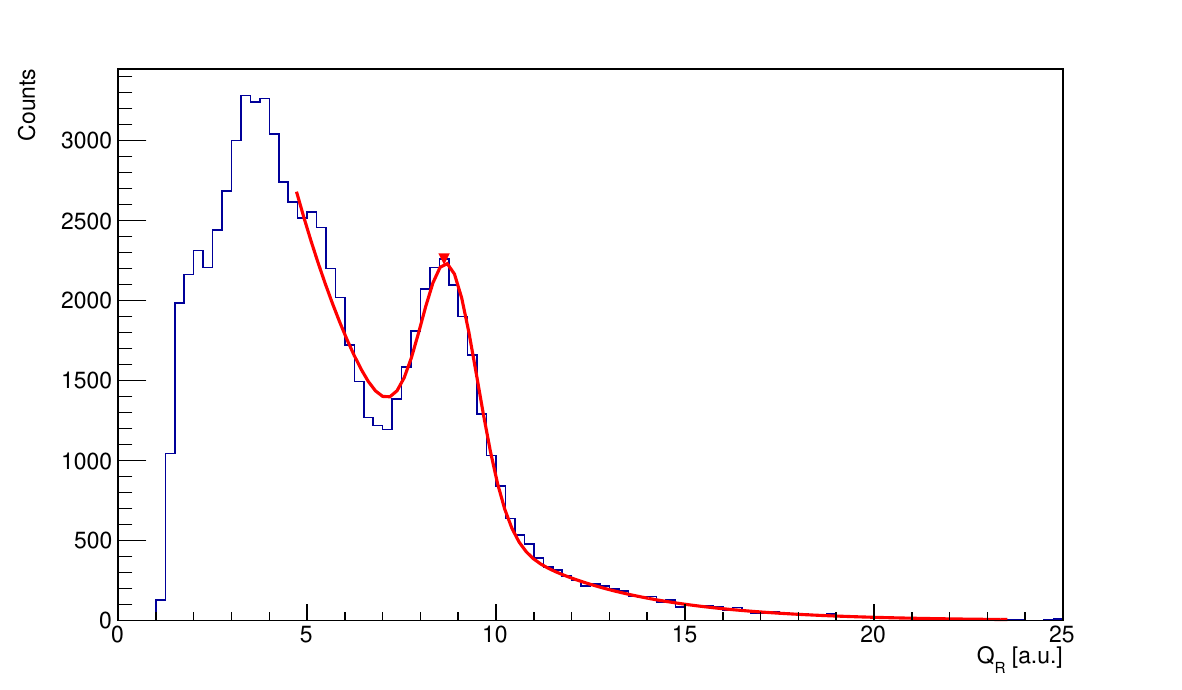}
         \caption{An example QDC spectrum (for \gls{sipm} 5 in layer 0 on the top side of the detector) with the fit function for gain alignment.}
         \label{fig:SiPMGainAlignment}
 \end{figure}

The alignment of \glspl{sipm} gains is a simplified form of energy calibration for the analysis of the \gls{1dcm} data. The \glspl{sipm} register charge in arbitrary QDC units. In order to rescale it to energy units and align the \glspl{sipm}' gains, an auxiliary measurement was done with a radioactive $^{68}$Ge/$^{68}$Ga source. The energy spectrum from this source contains the characteristic annihilation peak of \SI{511}{keV} energy. We assumed there was no offset: zero charge corresponded to zero energy. We plotted charge spectra for all \glspl{sipm} separately (see example spectrum in~\cref{fig:SiPMGainAlignment}) and fit a Gaussian (with standard parameters $\mu$ and $\sigma$) on an exponential background in the \SI{511}{keV} peak region. The mean value of the peak position for all \glspl{sipm} in QDC units was 9.563(18)~a.u. The determined peak positions per \gls{sipm} were then used to rescale the QDC values to energy. The energy calibration is incorporated in the \gls{llr} software.

\subsection{Dead SiPM issue}\label{sec:deadSiPMsIssue}
 The assembly of the PCBs housing the \gls{sipm} arrays was done by an external company, that did not comply with the soldering method described by the producer in the product specification. This resulted in the fact that gradually, more and more SiPMs were losing electrical contact and became inactive. As a consequence, during all measurements involving the Broadcom \glspl{sipm}~\cite{BroadcomSiPMArrays} described in this thesis, there were acceptance gaps present (see the IDs of the dead \glspl{sipm} in~\cref{tab:deadSiPMs} and the IDs of fibres that were connected to the dead \glspl{sipm}\footnote{For simplicity, we refer to them as "dead fibres" or "dead pixels" in the further parts of the thesis.} in~\cref{tab:deadFibers}). We acknowledge that such a fault of the experimental setup limits the detector performance in various aspects. Therefore, we took measures to compensate for this issue: 
\begin{itemize}
    \item as most gaps occurred on the sides, we have studied the effect of eliminating these detector parts in the analysis,
    \item we have adapted the simulations to the experimental conditions by filtering out the responses of the \glspl{sipm} that were dead during the experiment.
\end{itemize}
Moreover, we studied to what extent the issue of dead \glspl{sipm} deteriorates the image reconstruction quality by comparing the results obtained for full simulation (\cref{sec:unfiltered_simulation}) with those obtained for the simulation with dead \glspl{sipm} excluded (\cref{sec:simulationVsExperiment}).

\section{Measurements with a proton beam}
The objective of the proton beam measurements was to evaluate the performance of the detector in the \gls{1dcm} modality. This was achieved by means of the following characterisation steps:\begin{itemize}
    \item The adequacy of the detector dynamic range for \gls{pgh} registration was evaluated and the optimal bias voltage was determined.
    \item The detector rate capability was assessed by performing a beam intensity scan.
    \item A phantom (see~\cref{sec:phantom}) was irradiated with beams of different energies, \gls{pgh} depth profiles were reconstructed and based on them, the accuracy of the image reconstruction as well as beam range retrieval capabilities were assessed.
\end{itemize}

\subsection{Conditions at Heidelberg Ion Beam Therapy Center}
In January 2023, the SiFi-CC group\footnote{The SiFi-CC group members participating on-site: G.~Farah, A.~Fenger, R.~Hetzel, B.~Ko\l{}odziej, M.~Ko\l{}odziej, K.~Rusiecka, M.~L.~Wong, A.~Wro\'nska} performed a series of tests of the scatterer module at the Heidelberg Ion Beam Therapy Center (HIT)~\cite{Haberer2004, HITCenterWeb}. The experiments were carried out in an experimental hall equipped with a horizontal beam line. Part of the experimental hall, with the beam nozzle visible, can be seen in~\cref{fig:SetupPhotoFacingTheNozzle}. The facility offers a wide set of pencil beams of variable energy, lateral beam size (so-called focus), and intensity. Several ion types are available in the facility: protons and carbon ions are used for both therapeutic and research purposes, while helium and oxygen ions are currently only exploited for research. The particles are accelerated by a combination of a linac and a synchrotron of \SI{20}{\meter} diameter, delivered by Siemens~\cite{HITCenterWeb}. For protons, beam energies between \SI{47.80}{MeV} and \SI{219.57}{MeV} are available, in 255 steps. In our tests, we exploited protons in the range of energies between \SI{70.51}{MeV} and \SI{108.15}{MeV}, with the focus range of 14.5-\SI{22.5}{mm}. We also tested a few different beam intensities: $8\times10^7$~protons/s, $6\times10^8$~protons/s, and $3.2\times10^9$~protons/s.

\subsection{Experimental setup}
\subsubsection{Overview}
\begin{figure}[!htb]
    \centering
    \begin{subfigure}[b]{0.45\textwidth}
        \centering
        \includegraphics[width = \textwidth]{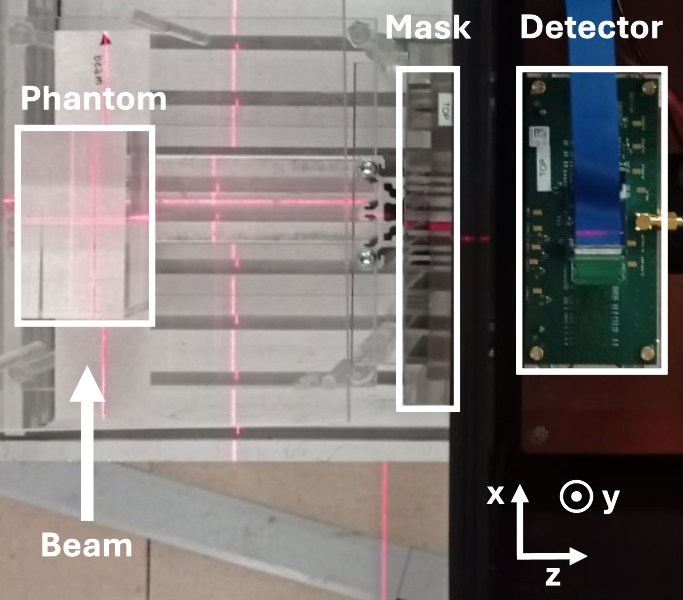}
        \caption{Exp. setup, top view.}
        \label{fig:expSetup}
    \end{subfigure}
    \hfill
    \begin{subfigure}[b]{0.45\textwidth}
        \centering
        \includegraphics[width = \textwidth]{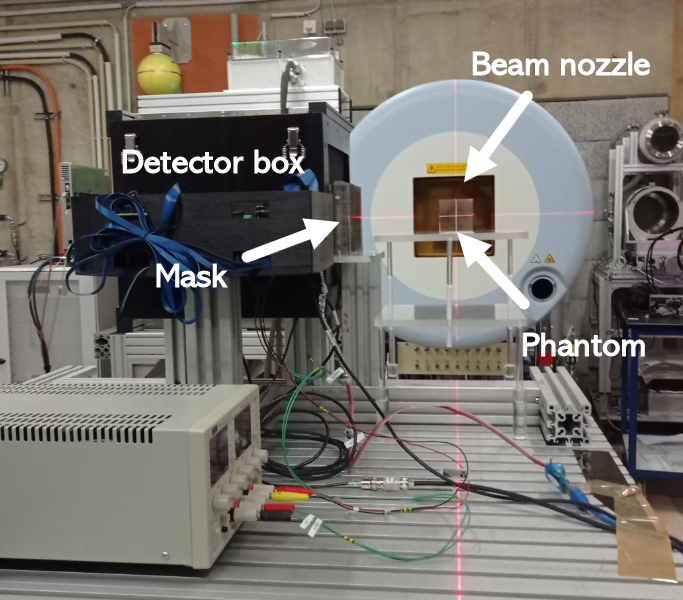}
        \caption{Exp. setup, side view.}
        \label{fig:SetupPhotoFacingTheNozzle}
    \end{subfigure}
    \hfill
    \label{fig:setupPhotos}
    \caption{The experimental setup viewed from the top (a) and from the side (b). Key setup components are marked with white boxes or arrows and labelled.}
\end{figure}

The experimental setup used in the HIT measurements is presented in Figures~\ref{fig:expSetup}~and~\ref{fig:SetupPhotoFacingTheNozzle}. It consisted of a \gls{pmma} phantom, \gls{1dcm}, the detector module (placed in a light-tight box), and the DAQ system. 

The coded mask and the detector module are described in detail in Sections~\ref{sec:masks} and ~\ref{sec:DetectorComponents_Scintillators}, the DAQ system is presented in~\cref{sec:TOFPET2c}, and the phantom in~\cref{sec:phantom}. We also summarise the detector operation in~\cref{sec:detectorOperation}. The distances between the parts of the setup were: \SI{170}{\mm} between the phantom centre and the mask centre, and \SI{63}{mm} between the mask centre and the detector centre. These distances were chosen to be as close to the optimum (found in~\cite{Hetzel2023}) as possible, given the spatial constraints resulting from the presence of the light-tight box and the support structure. The measured mask horizontal size was \SI{128.3}{mm}, while the horizontal size of the detection module was \SI{100.6}{mm}.
\subsubsection{Phantom}\label{sec:phantom}
We used a \gls{pmma} phantom of the dimensions $50\times50\times\SI{90}{mm \cubed}$, of the density~\SI{1.19}{\g/\cm\cubed}. It was positioned centrally in the \gls{fov} of the detector, the longest phantom dimension along the beam direction. The phantom rested on a table also made of \gls{pmma}~(\cref{fig:SetupPhotoFacingTheNozzle}).

\subsubsection{Detector operation}\label{sec:detectorOperation}
The values of parameters of the TOFPET2c \gls{daq} system used can be found in~\cref{tab:DAQparameters} in Appendix. The \gls{daq} was operated manually with a \gls{gui}. The raw data were saved on a local PC and then copied to a data server. All measurement metadata were saved in an SQL database. 

The scheme of data taking with the proton beam was the following: we started the data collection with the TOFPET2c \gls{gui}, requested the beam of desired parameters from the control room, and collected data for the preset time. The measurement time was set according to the expected duration of beam delivery, to embrace the whole beam signal. The beam irradiated the phantom, the detector was placed to the side of the beam axis, in front of the phantom (see~\cref{fig:expSetup}), so that the prompt gammas emitted from the phantom could be registered. 

\subsubsection{Irradiation plan}
\begin{figure}[!htb]
\centering
\includegraphics[width = \textwidth]{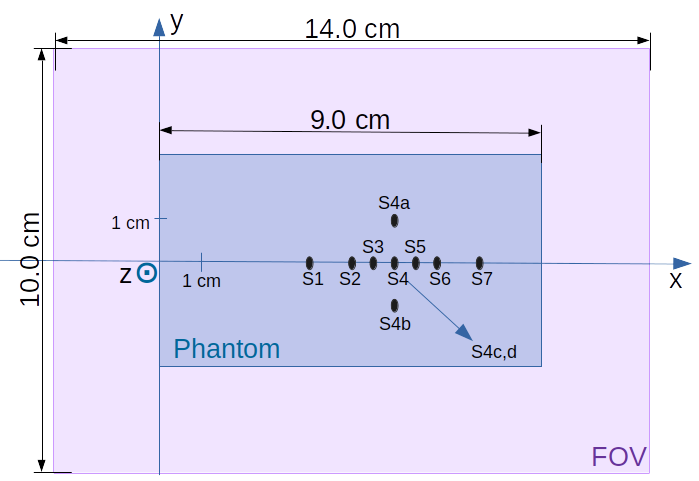}
\caption{Irradiation plan of the measurements with the proton beam in HIT. The proton beam direction is along $z$ axis, the phantom and the detector's \gls{fov} are shown.}
\label{fig:IrradiationPlan}
\end{figure}

The phantom irradiation was done according to the plan (presented in~\cref{fig:IrradiationPlan}), the exact \gls{bp} positions and beam energies for the S1-S7 points (so-called beam spots) can be found in~\cref{tab:beam_parameters}. 
\begin{table}[!ht]
\caption{Summary of measurements (runs) in the \gls{1dcm} modality. The \gls{bp} position was calculated from PSTAR~\cite{PSTAR}. The runs are in a chronological order. \newline}
    \centering
    \begin{tabular}{cccccc}\toprule
        Beam spot & Run ID & E [MeV] & Focus [mm] & BP pos. [mm] & $N_{\mathrm{protons}}$ \\ \midrule \rowcolor[gray]{.95}
        S4 & 567 & 90.86 & 17.2 & (0, 0, 56.093) & $10^{10}$ \\ 
        S4 & 568 & 90.86 & 17.2 & (0, 0, 56.093) & $10^{10}$ \\ \rowcolor[gray]{.95}
        S1 & 569 & 70.51 & 22.5 & (0, 0, 35.63) & $10^{10}$ \\ 
        S2 & 570 & 81.20 & 19.3 & (0, 0, 45.882) & $10^{10}$ \\ \rowcolor[gray]{.95}
        S3 & 571 & 86.14 & 18.2 & (0, 0, 50.992) & $10^{10}$ \\ 
        S5 & 575 & 95.40 & 16.4 & (0, 0, 61.185) & $10^{10}$ \\ \rowcolor[gray]{.95}
        S6 & 576 & 99.78 & 15.7 & (0, 0, 66.269) & $10^{10}$ \\ 
        S7 & 577 & 108.15 & 14.5 & (0, 0, 76.462) & $10^{10}$ \\ \rowcolor[gray]{.95}
        S4a & 578 & 90.86 & 17.2 & (0, 10, 56.093) & $10^{10}$ \\
        S4b & 579 & 90.86 & 17.2 & (0, -10, 56.093) & $10^{10}$ \\ \rowcolor[gray]{.95}
        S4c & 580 & 90.86 & 17.2 & (10, 0, 56.093) & $10^{10}$ \\
        S4d & 581 & 90.86 & 17.2 & (-10, 0, 56.093) & $10^{10}$ \\ \rowcolor[gray]{.95}
        S4' & 582 & 90.86 & 17.2 & (0, 0, 56.093) & $10^{11}$ \\ 
        S2' & 583 & 81.20 & 19.3 & (0, 0, 45.882) & $10^{11}$ \\ \rowcolor[gray]{.95}
        S3' & 584 & 86.14 & 18.2 & (0, 0, 50.992) & $10^{11}$ \\ \bottomrule
    \end{tabular}
    \label{tab:beam_parameters}
\end{table}
The measurements were made with the highest available beam intensity ($3.2 \times 10^9$~protons/s) and the smallest available lateral beam size at the given energy (14.5-\SI{22.5}{mm}). The default proton statistics was $10^{10}$~protons per measurement, and the time of measurement was \SI{30}{s} - longer than the proton irradiation time, as we also wanted to register the background before and after the irradiation. The beam spots were chosen to be quite sparse, to cover a large part of the detector \gls{fov} along the beam axis, and to check if the image reconstruction performance changes with the position within the \gls{fov}. Furthermore, several measurements (S4a-d) were made for the same energy but at a different position along the $x$ or $y$ axes. The measurements for the beam spots S2-S4 were also repeated with 10 times larger statistics ($10^{11}$~protons) than the rest of the measurements (and correspondingly a longer measurement time). They are marked with a prime: S2'-S4'.

\subsection{Online results} 
Several preliminary measurements were performed to ensure the setup's operability under clinical conditions and to find optimal settings for the main measurement series, which was the performance test of the gamma camera in the \gls{1dcm} modality. They are summarised in the following sections.

\subsubsection{Rate capability}\label{sec:experiments_rateCapability}
High rate capability is one of the crucial features of a \gls{pgh} detector for proton therapy monitoring, as described in~\cref{sec:introduction_requirementsForPTMonitoring}. To study the rate capability of our detector, we performed a series of measurements for two beam energies (\SI{90.86}{MeV} and \SI{108.15}{MeV}) and three intensities ($I_1 = 8\times10^7$, $I_2=6\times10^8$, and $I_3=3.2\times10^9$~protons/s), the duration of measurements was \SI{300}{s}, \SI{60}{s} and \SI{40}{s} respectively, to match the beam duration. Then, the data corresponding to beam (so-called spills) were separated from the background (the details of the separation procedure can be found in~\cref{sec:timePreselection}). 

Then, we compared the number of spill counts for different beam intensities. The list of runs for this measurement can be found in~\cref{Tab:rateCapability_listOfRuns} (in the Appendix). The goal of this measurement was to check if there is a substantial decrease in the total counts when moving to higher beam intensities due to dead time. If so, it would be a sign of reaching the limit of the rate capability. 

For the beam energy of \SI{90.86}{MeV}, the number of spill counts for the maximum investigated intensity $I_3$ was 6.3\% lower than for the minimum intensity $I_1$, but the spill counts for the middle intensity $I_2$ was 0.9\% higher than for $I_1$. For both pairs of intensities ($I_1$ and $I_2$; $I_1$ and $I_3$) one can observe that they are not equal within statistical uncertainties (presented in~\cref{Tab:rateCapability_listOfRuns}). Based on this observation, we suspect that there is a significant uncertainty associated with estimating the spill counts in this manner (not taken into account in the statistical uncertainty calculation), as the increase in the number of spill counts is not expected when increasing the intensity from $I_1$ to $I_2$. Basing on the fact that the number of spill counts grows when increasing the intensity from $I_1$ to $I_2$, we assume that at $I_2$, the rate capability limit is not reached.

A similar trend can be observed for the higher beam energy of \SI{108.15}{MeV}: the number of spill counts for $I_3$ was 7.6\% lower than for $I_1$, while the spill counts for the middle intensity $I_2$ was 1.4\% higher than for $I_1$. For both beam energies, the number of counts was lower for $I_3$ than for $I_2$, but it was unknown whether the cause of such behaviour was reaching the rate capability limit. To investigate whether the limit is reached at $I_3$, we compared the relative increase of spill counts between the two investigated energies for $I_2$ and $I_3$. A smaller increase for the higher intensity would mean that the rate capability limit is reached. For $I_2$, the number of spill counts for \SI{108.15}{MeV} is 30\% higher than for \SI{90.86}{MeV}. An analogous comparison for $I_3$ also yields a 30\% increase. Thus, we concluded that the rate capability is sufficient even for the highest beam intensity used clinically on site and we performed the rest of the measurements at that intensity. However, it is not entirely clear what is the cause of the observed fluctuations of the number of spill counts when changing the beam intensity. A discussion of the detector's maximum potential rate capability is presented in~\cref{sec:results_rateCapability}.

\subsubsection{Overvoltage scan} 
\begin{figure}
\centering
\includegraphics[width = 0.7\textwidth]{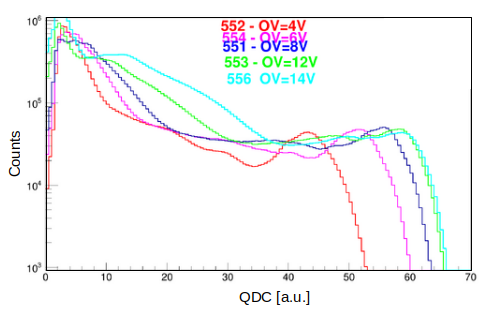}
\caption{Comparison of QDC spectra for different values of overvoltage. No correction of non-linearity was applied.}
\label{fig:overvoltageStudies}
\end{figure}

An overvoltage scan was performed to find the optimal bias voltage to power the \glspl{sipm}. The bias voltage is a sum of two parameters: breakdown voltage, which can vary between 32-\SI{33}{V} according to the producer (for simplicity, we assumed the same fixed value of \SI{33}{V} for all \glspl{sipm}), and overvoltage, which we varied in a range between 4 and \SI{14}{V}. This range was smaller than the maximum safe range provided by the producer (0-\SI{16}{V}), but we did not want to test the extreme values, assuming that they will not be optimal anyway. The differences in \gls{sipm} gain that originate, among others, from the spread of actual breakdown voltage values are corrected for later, as described in~\cref{sec:energyCalibrationPerSiPM}. The runs in this measurement series are listed in~\cref{Tab:overvoltageScan_listOfRuns}. We compared them by inspecting the raw \gls{sipm} charge spectra (see~\cref{fig:overvoltageStudies}). There, one can see that for the overvoltage of 4 and 6~V, the spectra are squeezed and the structure (multiple wide peaks) at lower QDC values ($<10$~a.u.) is not clearly visible. The structure becomes more pronounced at 8~V overvoltage. Going even higher with overvoltage, a saturation effect is observed at higher QDC values ($>60$~a.u.): the spectra for 12 and 14~V overvoltage have the right-most parts of the spectrum very close to each other. Thus, the optimal bias voltage was found to be 8~V and this value was used in all the subsequent measurements. It should be noted that it was rather a rough estimate of the optimal overvoltage, sufficient for this particular measurement campaign, rather than a comprehensive study of the \gls{sipm} properties. A detailed application note on the correlations between various \gls{sipm} parameters, including overvoltage, can be found in~\cite{BroadcomOvervoltageApplicationNote}.

\subsection{Auxiliary measurements}\label{sec:experiments_auxiliaryMeasurements}
Apart from the series of measurements with the coded mask, auxiliary measurements were needed for detector efficiency calculation (see~\cref{sec:efficiencyCorrection}), as well as for simplified energy calibration (see~\cref{sec:Experiments_Calibration}). 

The first auxiliary measurement was a reference measurement with uniform irradiation of the detector plane. For this purpose, a linear radioactive source ($^{68}$Ge/$^{68}$Ga) was placed horizontally in front of the detector, in the middle of its height, at a distance of \SI{226}{mm} from the centre of the detection module. The coded mask was removed from the setup. The radioactive source was of cylindrical shape: \SI{3.2}{mm} of external diameter and 192~mm of length. Its active part formed a smaller cylinder located centrally inside the larger one, of \SI{184}{mm} length and \SI{1.6}{mm} diameter. At the time of measurement, the source activity was \SI{10.7}{MBq}. 

The other auxiliary measurement was the background in the absence of the beam, registered over 30~minutes after the last beam run, to exclude the background component from activation of the experimental setup parts. In such a case the detector response is due to the internal activity of the scintillation fibre material, LYSO:Ce,Ca. The main transitions in LYSO:Ce,Ca are at \SI{307}{keV}, \SI{202}{keV}, and \SI{88}{keV}; they are usually detected in combinations~\cite{AlvaSanchez2018}. 

The auxiliary measurements are referred to as run 596 (reference) and run 597 (background). The DAQ settings during these runs were the same as for the beam runs (see~\cref{tab:DAQparameters}), each of the runs lasted \SI{300}{s}.
\chapter{Data analysis}
\section{Overview}
\begin{figure}[!htb]
\centering
\includegraphics[width = \textwidth]{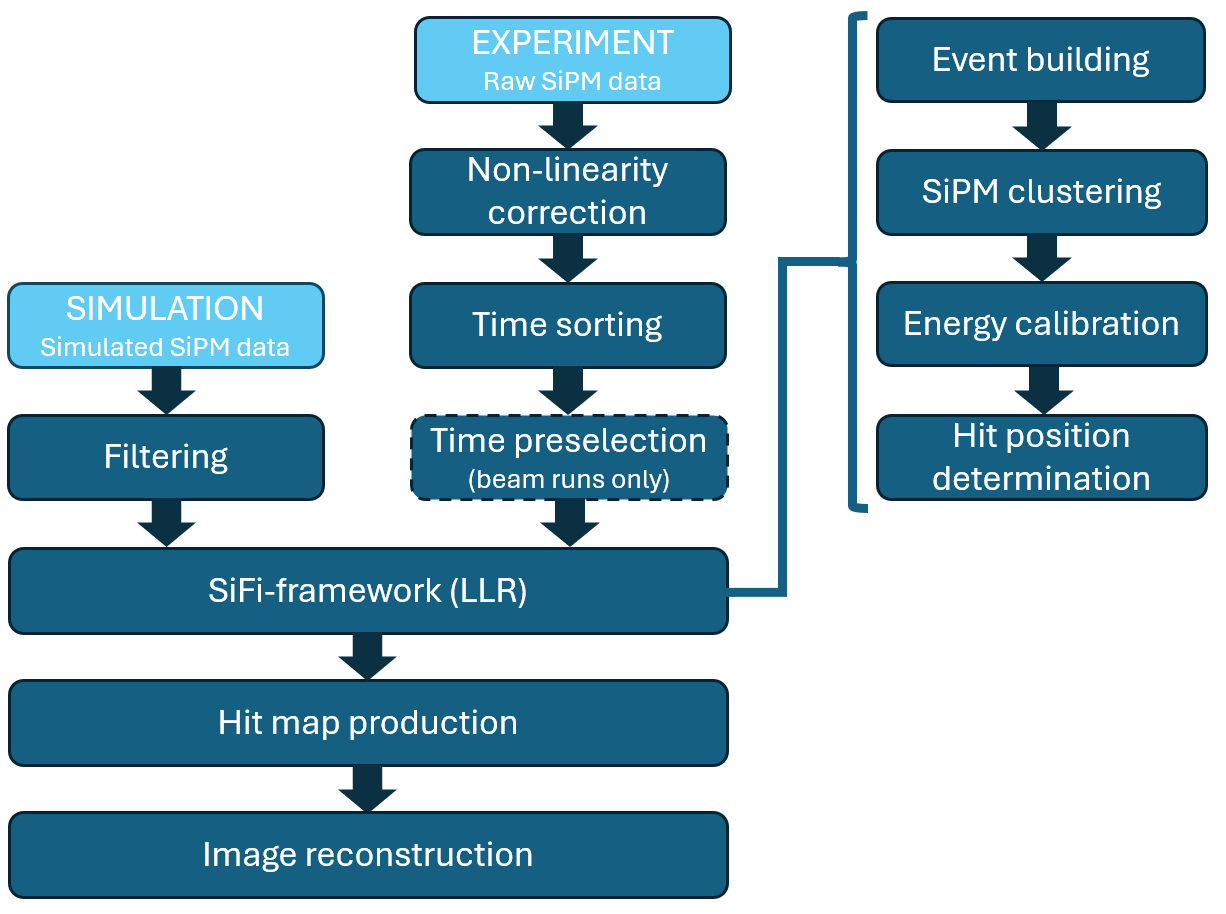}
\caption{The general processing scheme of the data from the \gls{1dcm} measurements and simulations.}
\label{fig:sifiArchitecture}
\end{figure}

In this chapter, all the steps needed to obtain prompt-gamma depth profiles from raw experimental data are described in detail, along with the method of correlating those profiles with the beam range. The processing and use of data from auxiliary measurements is also presented.  \Cref{fig:sifiArchitecture} shows the general data processing scheme for both experimental and simulation data. Its individual steps are discussed in detail in the following sections.
\section{Preprocessing}\label{sec:preprocessing}

\subsection{Conversion of raw data to root trees}
The \gls{daq} software provided by the  producer as part of the TOFPET2c \gls{daq} system~\cite{originalTOFPETSoftware} was adapted to our needs\footnote{The software was adapted by M.~L.~Wong.}. The version of the code used to process the data described here is available in~\cite{githubDAQ}. The software includes scripts and programmes that allow one to manage the data acquisition, save the data in a raw format, and process them to a desired format, e.g. a root tree. Conversion of the raw data is fast (in comparison with the processing time of the next analysis steps) due to the implementation of parallel processing. In our case, the raw data in the \verb|.rawf| format are processed into \verb|.root| files, and we refer to them as \textit{TOFPET trees} in the following sections.

\subsection{Nonlinearity correction}\label{sec:nonLinCorr}
As indicated in the TOFPET2c Software User Guide~\cite{PetSYSSoftwareUserGuide}, the response of the \gls{asic} \gls{qdc} preamplifier is nonlinear. Thus, the authors of the manual suggest either to perform one's own calibration with radioactive sources, or to linearise the dependence of energy $E$ versus measured charge $Q$ with the use of \cref{eq_nonLinCorr}: 
\begin{equation}
    E=P_0\cdot P_1^{Q^{P_2}}+ P_3\cdot Q - P_0, \label{eq_nonLinCorr}
\end{equation}
where $(P_0, P_1, P_2, P_3) = (8, 1.04676, 1.02734, 0.31909)$ - see Equation~(3.1), p. 25 in \cite{PetSYSSoftwareUserGuide}. The authors claim that the formula allows for nonlinearity correction up to a level of 2-3\% of the 511~keV peak energy resolution. We incorporated the correction into the modified \gls{daq} software \cite{githubDAQ}.

\subsection{Data splitting and sorting}
For the data files above $10^8$~entries there were memory issues in the \gls{llr}, so they were split into smaller batches of $10^8$~entries. They were merged again in the analysis steps downstream of \gls{llr}. Another technical issue was that the entries in the TOFPET trees were in some cases not ordered by time. This occurred only in groups of entries that were very close in time. However, incorrect ordering, even when the entries' times do not differ much, can result in incorrect event building, and thus in loss of data. To avoid this problem, we sorted all TOFPET trees by time.

\subsection{Time preselection}\label{sec:timePreselection}
\begin{figure}[!htb]
\centering
\includegraphics[width=\textwidth]{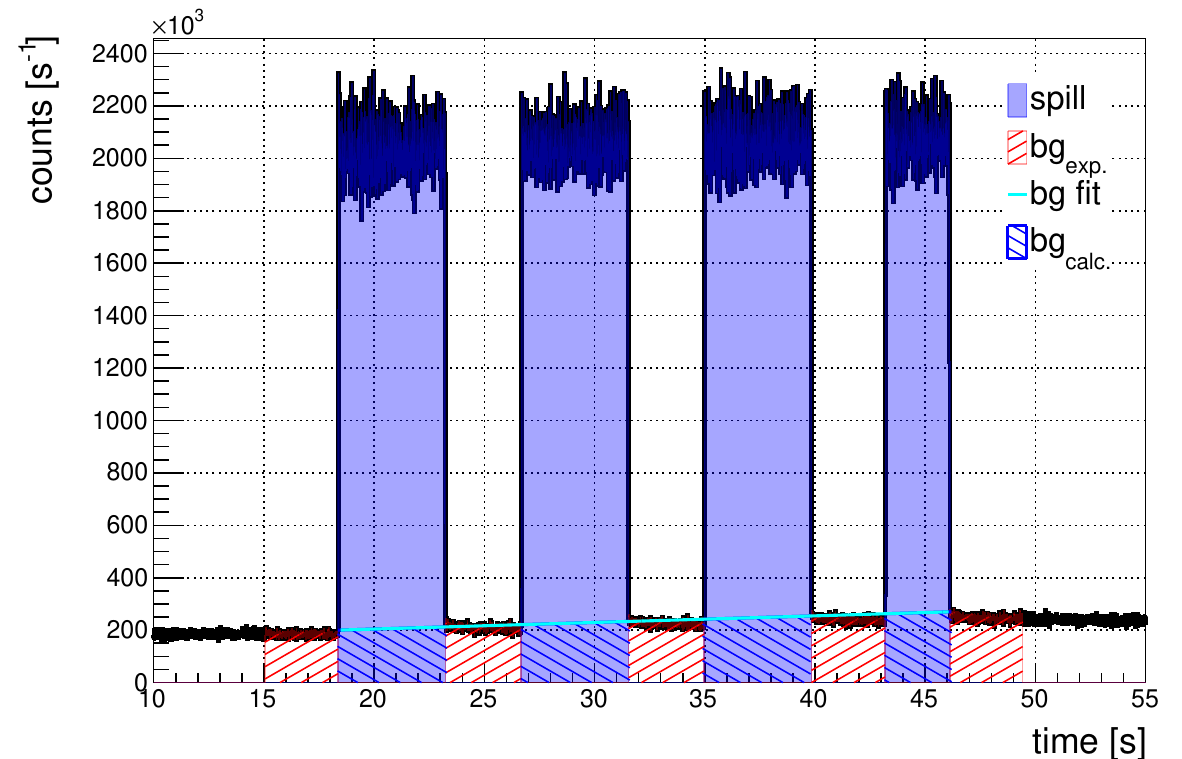}
      \caption{Example time distributions of events in the detector. The black line denotes the time structure of everything that was measured, spills are marked in blue, the region hatched in red is the experimental background, the one hatched in blue - the calculated background. The cyan line is the line of fit to the background parts.}
      \label{fig:timeStructureEdited}
\end{figure}

A synchrotron beam is not continuous in time, but is delivered in spills, which can be observed on a time distribution of events in the detector (see the blue regions of higher counts in~\cref{fig:timeStructureEdited}) that reflects the time structure of the beam. The number of spills depends on the total number of protons and the beam intensity.
The raw data from the measurements involving proton beam contain not only the information about interactions of \gls{pgh} quanta originating from the beam interaction with the phantom, but also the background before, after, and in between the spills (see the red hatched regions in~\cref{fig:timeStructureEdited}). In order to separate the spills from the background, time preselection was performed for all beam runs\footnote{The time preselection and determination of background scaling factor were conducted by A.~Wrońska and A.~Fenger.}: first, the start and stop times of all spills in each beam run were determined. Then, the data between the start and stop times were labelled spills and saved as root trees. The inter-spill regions, as well as \SI{3.3}{s} before the first and after the last spill, were labelled as background and also saved as root trees. 

\begin{figure}[!htb]
         \centering
         \includegraphics[width=\textwidth]{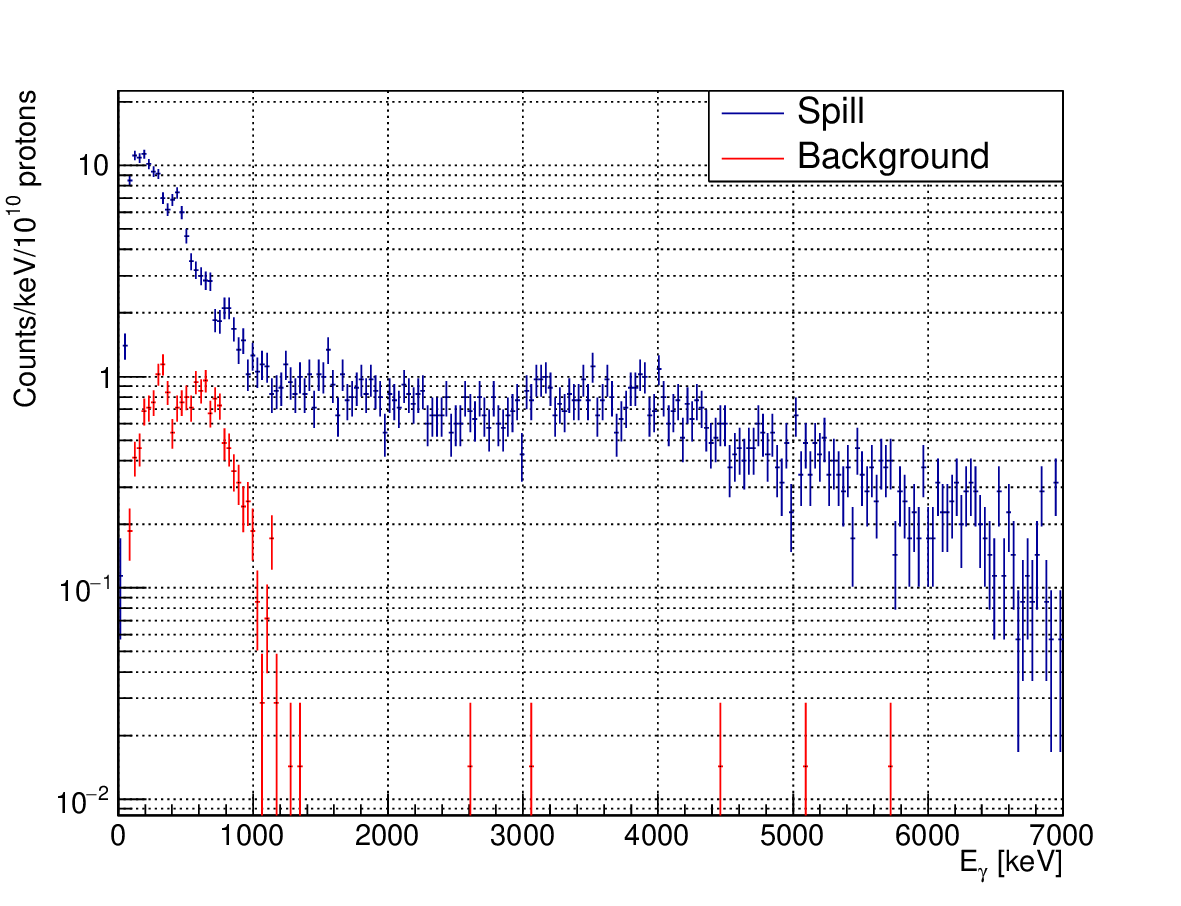}
         \caption{Energy spectrum for a selected fibre from run 569, divided into spill (blue) and background (red) parts based on the time information.}         \label{fig:spillsAndBackgroundQDC}
 \end{figure}

The energy spectra for an example run 569, divided into spills and background, can be seen in~\cref{fig:spillsAndBackgroundQDC}. 
There, the background was normalised to match the spills with a scaling factor, determined the following way: firstly, the average counts of the background over \SI{3.3}{s} before the first beam spill and the average counts over the same time after the last beam spill were calculated, making the $y$ coordinates of 2 points in~\cref{fig:timeStructureEdited}. Then, the $x$ coordinates of these points were assumed as the start time of the first spill and the stop time of the last spill, respectively.  The line connecting these two points ("bg fit" line in~\cref{fig:timeStructureEdited}) was used to estimate the background (blue hatched lines in~\cref{fig:timeStructureEdited}). Let us denote number of counts in the experimental background as $N_{\mathrm{background~exp.}}$ and number of counts in the calculated background (during spills) as $N_{\mathrm{background~calc.}}$. Then, the scaling factor $c$ used to normalise the background to match the spill time is (\cref{eq:bgScalingFactor}):
\begin{equation}\label{eq:bgScalingFactor}
    c=\frac{N_{\mathrm{background~calc.}}}{N_{\mathrm{background~exp.}}}
\end{equation}
The normalised background is later considered in the image reconstruction procedure (see~\cref{sec:imageReco}), to obtain a clearer image.

In~\cref{sec:experiments_rateCapability}, the spill counts ($N_{\mathrm{spill}}$) were determined using the scaling factor $c$, with~\cref{eq:determineNspills}; there, $K$ denotes number of spills, $N_{i~\mathrm{total}}$ - total counts in $i$-th spill, and $N_{\mathrm{background~exp.}}$ - counts in the experimental background:
\begin{equation}\label{eq:determineNspills}
    N_{\mathrm{spill}} = \Sigma_i N_{i~\mathrm{total}} - c N_{\mathrm{background~exp.}}, \quad i\in (1,...,K).
\end{equation}

\subsection{Simulation filtering}\label{simulationFiltering}
A realistic simulation (see~\cref{sec:simulationDescription}) was developed within the \gls{sificc} group\footnote{The simulation was developed by R.~Hetzel, M.~Kercz and L.~Mielke.}, reproducing the experiment in the \gls{1dcm} mode. The simulation assumes that all the \glspl{sipm} work properly. However, in the experimental data, there were acceptance gaps caused by faulty \glspl{sipm} (see~\cref{sec:deadSiPMsIssue}). To be able to directly compare the simulated and experimental results, we applied an additional preprocessing step to all the simulation data: entries associated with the \glspl{sipm} that were not responding during the experiment were excluded. The list of dead \glspl{sipm} can be found in~\cref{tab:deadSiPMs}. 

\section{Low-level reconstruction}\label{sifi_framework_step}
The \verb|sifi-framework| is a custom data analysis software developed by the \gls{sificc} group\footnote{The core part of the software was developed by R.~Lalik.}~\cite{githubSiFi}, used for \gls{llr}. The software (in the version for the \gls{1dcm} modality) takes the TOFPET trees containing \gls{sipm} data as input and outputs trees containing fibre hit position information. The author of this thesis developed the following parts of this software that are relevant to the analysis of the \gls{1dcm} data: classes representing data objects (called categories) at different processing stages, data importer, \gls{sipm} clustering (in parts), fibre grouping, and the hit position determination~\cite{githubSiFi}.

\subsection{Structure and operation}
\begin{figure}[!htb]
         \centering
         \includegraphics[width=\textwidth]{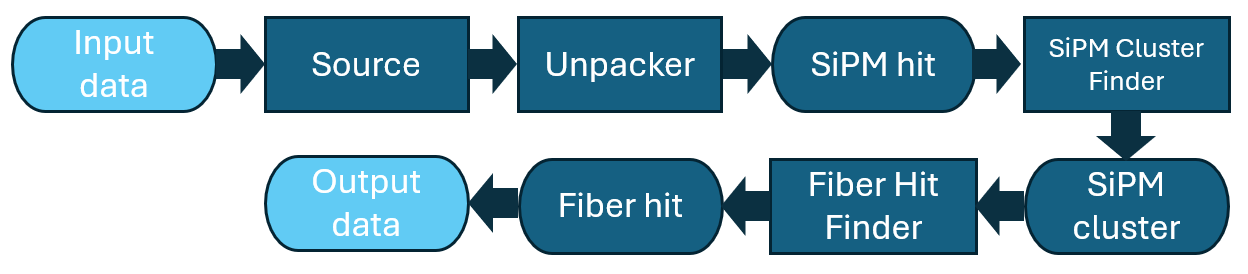}
         \caption{A scheme of the \gls{llr} software for the \gls{1dcm} modality.}
         \label{fig:sifi-framework-scheme}
 \end{figure}

The software is built out of three main types of classes: tasks, categories, and containers. The tasks process the data and write the result to categories, i.e. objects representing the data after a given processing stage. The containers store setup-specific data (\gls{sipm}-to-fibre mapping, calibration parameters, detector geometry details, etc.) and are used by the task classes. The detailed scheme of the \verb|sifi-framework| (for the \gls{1dcm} modality) is presented in~\cref{fig:sifi-framework-scheme}, where the tasks are represented by rectangular blocks, the categories by the blocks with rounded corners, and the input and output data are marked with a different colour.

\subsection{Reading in the data}
The \verb|Source| task reads in the data from a TOFPET tree in the form of entries, and the events are built: any entry belonging to a fixed trigger window of $T_\mathrm{TW} =\SI{15}{ns}$, is assigned to the same event. 

$T_\mathrm{TW}$ was chosen based on an earlier study\footnote{The study was performed by K.~Rusiecka.}, in which a wide trigger window was applied (\SI{50}{ns}) and the events were built. Then a histogram was filled with time differences between the first and the last entry within an event. The tail of the distribution of these differences ended at about \SI{15}{ns}. Hence, we chose this value to be the trigger window in the analysis of all measurements in the \gls{1dcm} modality.

Each entry in the TOFPET tree contains:
\begin{enumerate}
    \item ID of a \gls{daq} channel, 
    \item time of signal detection, elapsed from the start of measurement, in picoseconds,
    \item signal charge, expressed in arbitrary QDC units.
\end{enumerate}
 The time of each entry is rescaled from ps to ns. If there is a \gls{daq}-related offset in the \gls{daq} channel IDs (which is related to the \gls{pcie} board port ID used by TOFPET2c), it is subtracted so that the \gls{daq} channel numbering starts at 0. 
 
 The data in this form are passed to \verb|Unpacker|, where two containers are read in: for mapping and for energy calibration. First, the \gls{daq} channel IDs are translated to \gls{sipm} IDs using the mapping container, then the charge is rescaled to energy units (see~\cref{sec:energyCalibrationPerSiPM}). Finally, the data in the modified form are saved to \verb|SiPM hit| category. Any invalid hits (e.g. with a negative charge) are discarded at this stage.

\subsection{SiPM clustering}
The \gls{sipm} hits need to be clustered to account for the situations in which the energy deposit extends over several neighbouring fibres and consequently, clusters of neighbouring \glspl{sipm} respond on both detector sides. The clustering is performed by the \verb|SiPM Cluster| \verb|Finder|. The task reads in the data from the \verb|SiPM hit| category, performs clustering on its content, and writes the output to the \verb|SiPM cluster| category. The clustering is performed on a set of \gls{sipm} hits as follows: an empty cluster is created, and the first \gls{sipm} hit is appended to it. We check this \gls{sipm} hit against all remaining \gls{sipm} hits and check if there are any neighbours of it (\glspl{sipm} are considered neighbours if they have a common edge or corner). If there are neighbours, we add them to the same cluster and continue checking all \glspl{sipm} in the cluster against all remaining \glspl{sipm}, if they are neighbours. If there are no more new neighbours after iterating through a whole cluster, a new cluster is created and the first unassigned \gls{sipm} hit is appended to it. Then the procedure of neighbour search is repeated as for the first cluster. The whole procedure is repeated until no unassigned \glspl{sipm} hits are left. Then, all the cluster characteristics are set: \begin{enumerate}
    \item cluster ID (starting from 0 and incrementing by 1 until all clusters are assigned an ID),
    \item time: time of the earliest \gls{sipm} hit within the cluster,
    \item module and side - rewritten from the first \gls{sipm} hit,
    \item charge as the sum of all the charges from the \gls{sipm} hits constituting the cluster,
    \item position, calculated as the centre of gravity (the address weighted by charges associated with participating \gls{sipm} hits). 
\end{enumerate} 
The category \verb|SiPM Cluster| is filled with those characteristics.

\subsection{Finding fibre hits}\label{sec:fibreHits}
The next task takes \gls{sipm} clusters from the \verb|SiPM Cluster| category, identifies fibres connected to them, builds fibre groups, assigns event types, and finds the fibre hit positions. Then, the fibre groups are written to \verb|Fibre hit| category, which is the output of the \gls{llr} saved to a root tree. The \verb|Fibre hit| category contains the following variables: module, layer, fibre, energy information on both fibre sides, time on both fibre sides, event type. The procedure for finding the fibre hits is as follows (performed for each event):
\begin{enumerate}
    \item For each top-bottom cluster pair, common fibres are found. It is done using a parameter container that associates each \gls{sipm} ID to fibre IDs.
    \item We check if the common fibres for each cluster pair are grouped together spatially. A fibre belongs to a group if it has at least one common side or one common corner with any other fibre from the group. Any events with more than one fibre group are excluded from further analysis. 
    \item Based on the number of \gls{sipm} clusters on both detector sides ($c_{top}$, $c_{bottom}$) and fibre multiplicity ($m_f$) per cluster pair (i.e., how many fibres are assigned to a given cluster pair), the events are classified as belonging to one of five classes (see~\cref{fig:eventClasses}):
    \begin{figure}
    \centering
    \includegraphics[width = \textwidth]{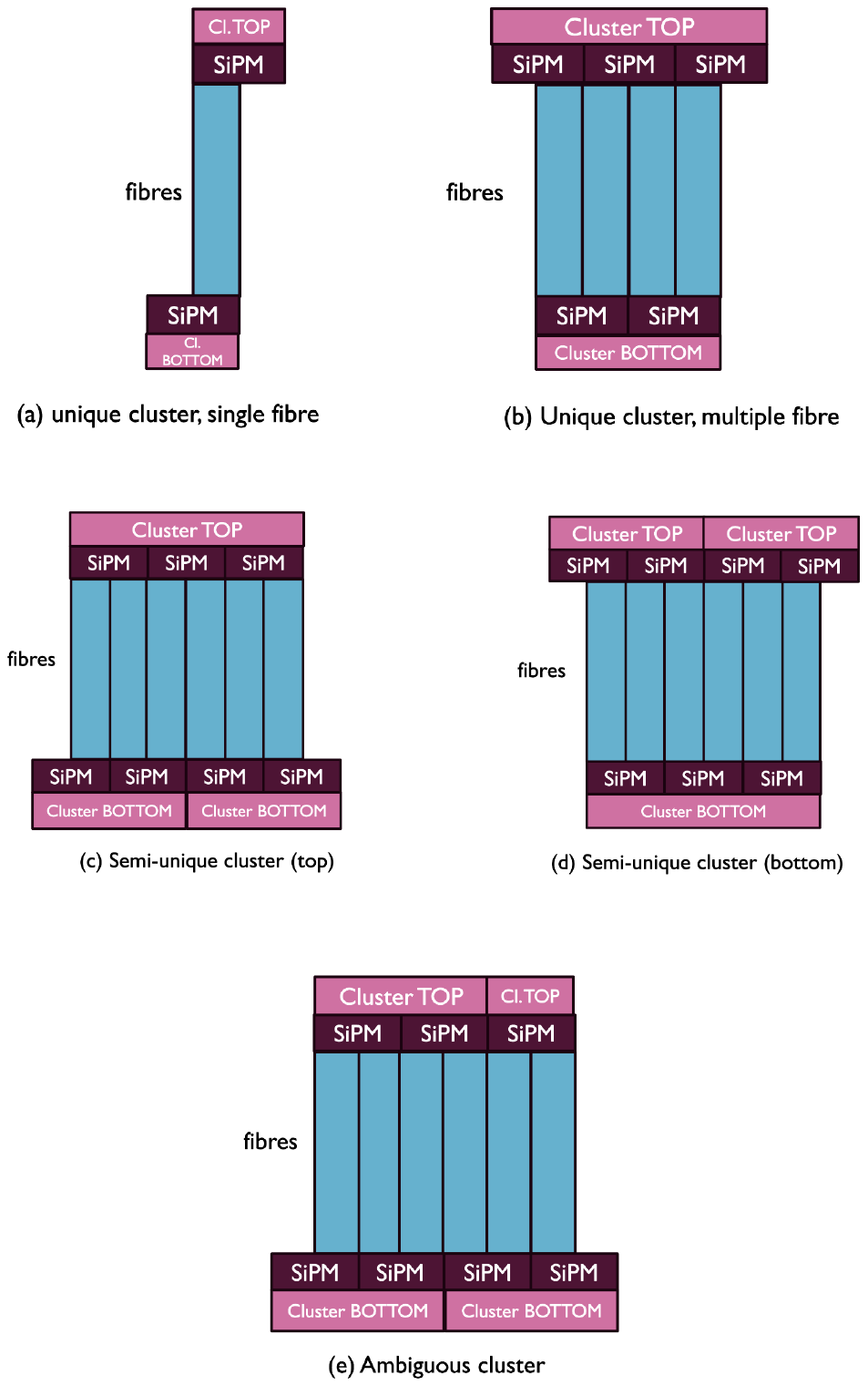}
    \caption{Classes of events in the \gls{llr}. This is a symbolic representation, as the actual clusters are formed in two dimensions.}
    \label{fig:eventClasses}
    \end{figure}
    \begin{enumerate}
        \item $c_{top}=1$, $c_{bottom}=1$, $m_f = 1$: unique clusters, unique fibre.
        \item $c_{top}=1$, $c_{bottom}=1$, $m_f > 1$: unique clusters, multiple fibres. 
        \item $c_{top}=2$, $c_{bottom}=1$, $m_f > 1$: top semi-unique cluster 
        \item $c_{top}=1$, $c_{bottom}=2$, $m_f > 1$: bottom semi-unique cluster
        \item $c_{top}>1$, $c_{bottom}>1$, $m_f > 1$: ambiguous clusters
    \end{enumerate}
    \item Based on the assigned class, the hit positions in $x$ and $z$ dimensions and energy deposits are found and assigned:
    \begin{itemize}
        \item Type a: the hit position is simply the position of the only fibre, and the energy deposits on the top and bottom fibre ends are the deposits from the top and bottom clusters. 
        \item Type b: The hit position along the layer ($x$) is defined as a weighted mean of active fibre IDs. The weight is defined as a geometric average of energy deposits in the \glspl{sipm} ($E_{top}$, $E_{bottom}$) at both fibre ends: $\sqrt{E_{top}E_{bottom}}$. The hit position across the layers ($z$) is assigned as the frontmost detector layer containing an active fibre. Then, the hit position ($x$, $z$) is rounded to the ID of the closest fibre. The energy deposits are determined as the energy deposits in top and bottom cluster, respectively. Note that the whole energy deposit in a given event is assigned to a single "main" fibre.
        \item Types c and d: the merged cluster energy deposit has first to be divided in two, according to the ratio of energy deposits in clusters on the other side. After the clusters separation, the procedure is the same as for b-type events.
        \item Type e: these events are excluded from further analysis, as there is currently no method implemented to determine the hit positions and energy deposits for them.
    \end{itemize}
    The time of the fibre hit is defined for all event classes as the time of the earliest \gls{sipm} hit in a given event.
    \item The \verb|Fiber hit| category is filled and saved in the tree. The data saved in this category form the output of \gls{llr}.
\end{enumerate}

\section{Hit maps production}\label{sec:hitmapsProduction}
A convenient representation of the data after \gls{llr} is a hit map, i.e. a spatial hit distribution over the detector module. It is represented by a 2D histogram, integrated over the $y$ coordinate. On the $x$ axis, there is the fibre number within one layer (0-54), and on the $y$ axis, there is the layer number (0-6). The colour scale denotes either the number of hits in each fibre, or the total energy deposit, depending on the map type. Example hit maps are presented in~\cref{fig:allHitmaps}, acceptance gaps due to several faulty \glspl{sipm} are visible. In the \gls{1dcm} modality, one fibre is one detector pixel (forming one bin of the hit map), and the hit position along the fibre is not needed. Thus, the hit map contains complete information about the detector response and can be input to the image reconstruction (see~\cref{sec:imageReco}). To create the hit maps, we iterate over the output trees from the \gls{llr} event by event, and fill the hit maps if appropriate criteria are met. In this analysis step, 16 hit maps were produced for each run, varying the following conditions:
\begin{itemize}
    \item 4 lower thresholds on energy deposits: 0, 500, \num{1000}, \SI{1500}{keV} (or the corresponding thresholds in the photon counts units in the simulation case; see~\cref{tab:PhotonCountsVsEnergy} in Appendix),
    \item 2 sets of event classes considered: unique clusters only~(\cref{fig:eventClasses}a,b) or both unique and semi-unique ones~(\cref{fig:eventClasses}a-d),
    \item 2 filling options: if there was a hit in a given detector pixel, the corresponding bin is incremented by 1 (map of hits) or by the energy value (energy deposition map). 
\end{itemize}
The best variant was then selected in the image reconstruction optimisation step. The upper threshold for hits included when filling all types of hit maps was fixed to \SI{7000}{keV}; the reason for choosing this value is explained in~\cref{sec:fibreEnergySpectraInspection}.

\section{Auxiliary studies}
In this section, various auxiliary studies are presented that provided insight into the \gls{1dcm} data. They are presented in this section along with their results to keep the logical order, since the conclusions from some of them influenced the main analysis chain.

\subsection{Fibre calibration parameter spread per SiPM}
A fitting procedure similar to the one in~\cref{sec:energyCalibrationPerSiPM} was also performed on charge spectra per fibre, after selecting only events with a single-fibre response (event type a, see~\cref{fig:eventClasses}a). This method provides the calibration parameters with slightly lower uncertainty, compared to the \gls{sipm} gain alignment factors. However, the obtained fibre calibration parameters were not used in the main data analysis chain because they are suitable only for the single-fibre events (where a fibre which was hit can be uniquely identified), while is it favourable to apply the same method of gain alignment across all the event classes.

\begin{figure}[!htb]
\centering
\begin{subfigure}[t]{0.48\textwidth}
\centering
\includegraphics[width=\textwidth]{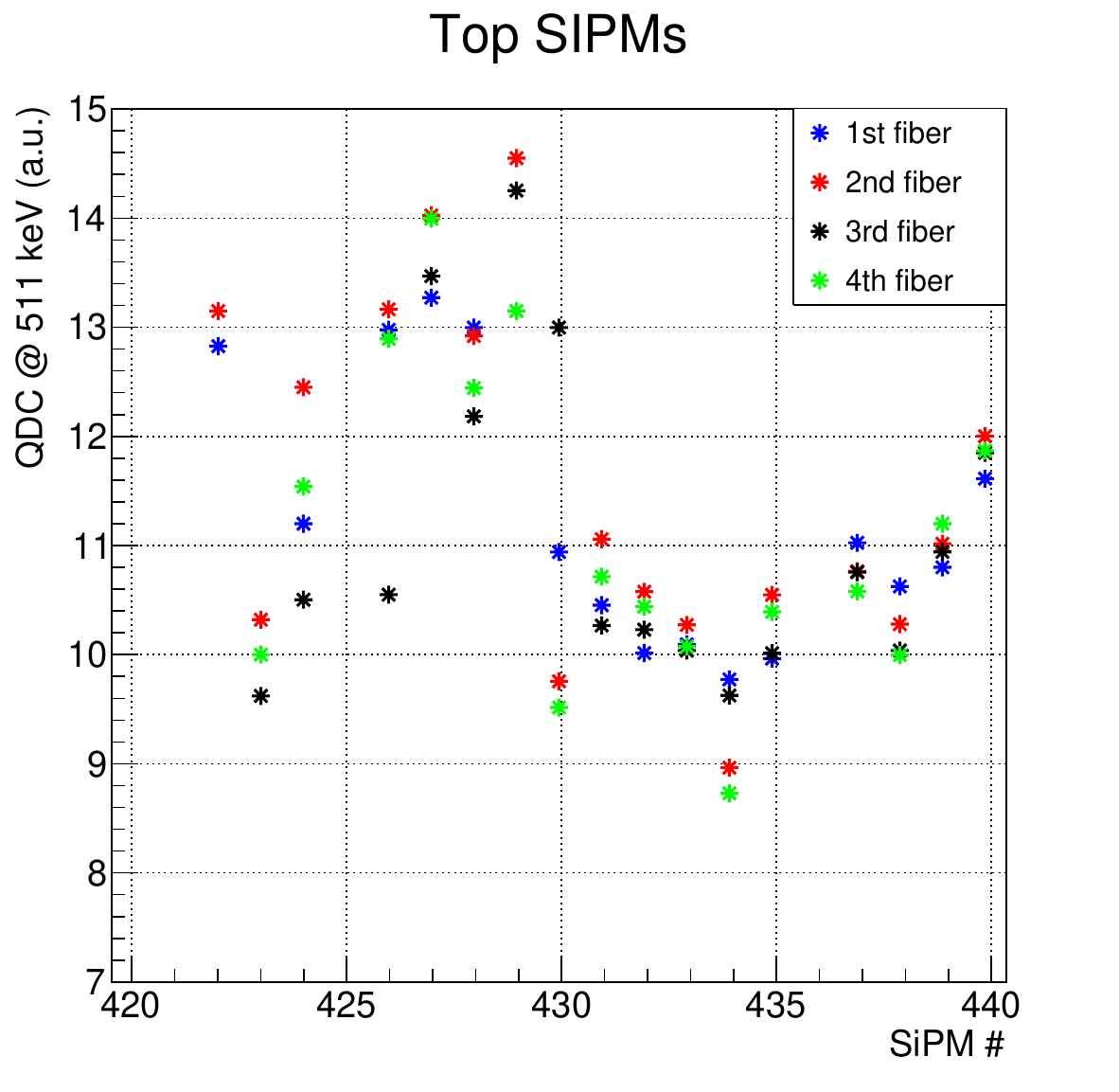} 
\caption{Selected top \glspl{sipm}.} \label{fig:TopSiPMsV2}
\end{subfigure}
\begin{subfigure}[t]{0.48\textwidth}
\centering
\includegraphics[width=\textwidth]{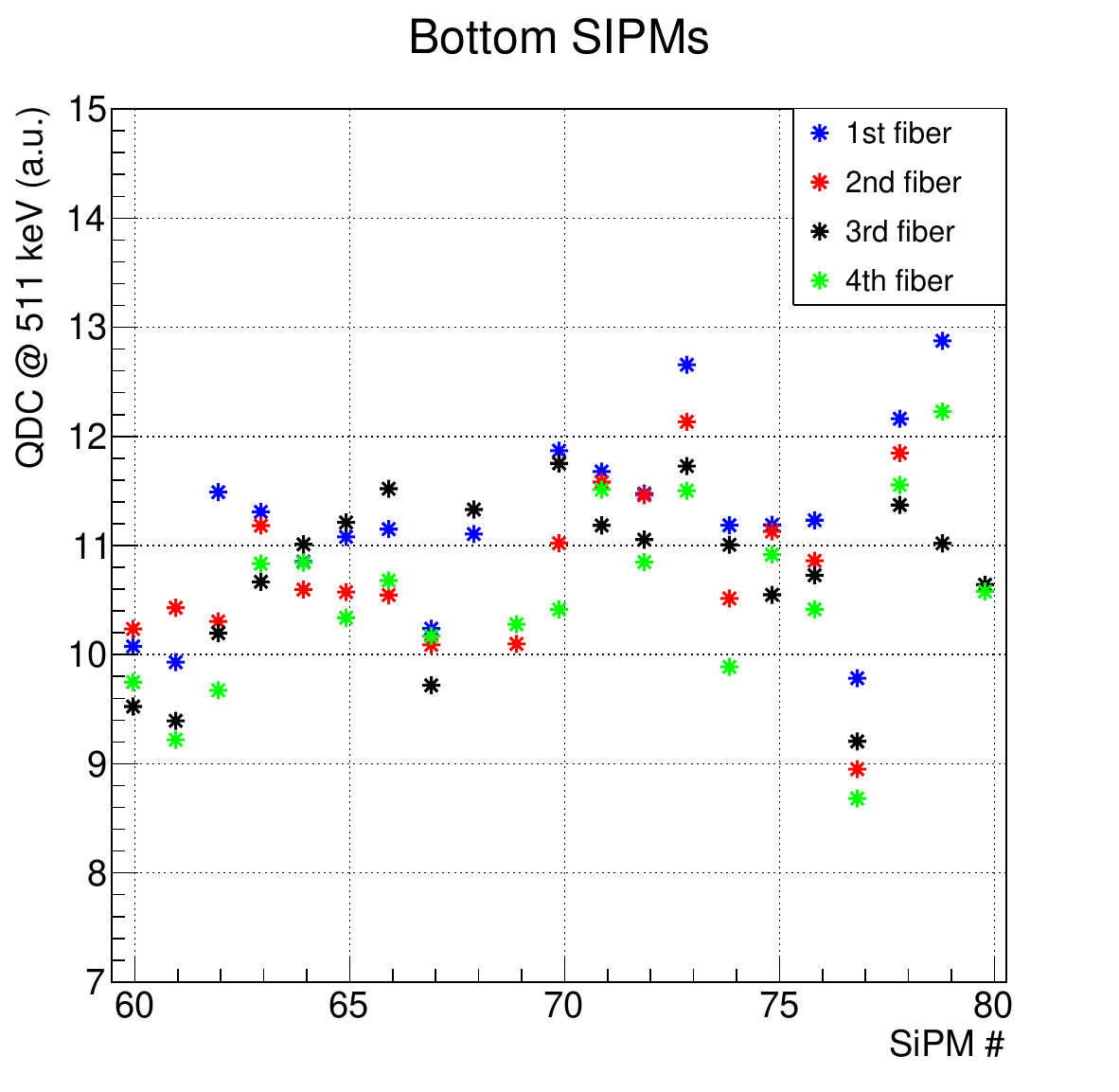} 
\caption{Selected bottom \glspl{sipm}.} \label{fig:BottomSiPMsV2}
\end{subfigure}
\caption{The spread of calibration parameters per fibre for selected \glspl{sipm}.}
\label{fig:gainSpreadPerFiber}
\end{figure}

The calibration parameters obtained per fibre served for another auxiliary study of the spread of gains in the detector. Generally, there are two main sources of calibration parameters spread: structural imperfections in fibres and uneven gain of the \glspl{sipm}. To check which effect is stronger, we calculated the average calibration parameter spread for fibres coupled to the same \gls{sipm} and compared it with the calibration parameter spread for all fibres. The plots for several selected \glspl{sipm} are presented in~\cref{fig:gainSpreadPerFiber}. 
The results are summarised in~\cref{tab:fiberGainSpread}, separately for top and bottom \gls{sipm} boards. There, one can see that the fibre calibration parameter spread within one \gls{sipm} is about 50\% of the total fibre calibration parameter spread, which implies that the remaining part of the effect can be attributed to differences between \glspl{sipm}. Therefore, the calibration parameter spread caused by fibre variability and the gain spread caused by \gls{sipm} variability are of the same order. The absolute spread is 28\% for the top detector side and 24\% for the bottom side; we did not study the origin of the spread difference between the sides. Overall, we find the magnitude of the spread effect to be acceptable for our purposes. In the data analysis, the effect is diminished by aligning the \gls{sipm} gains (see~\cref{sec:energyCalibrationPerSiPM}). 
\begin{table}[!ht]
    \centering
    \caption{Calibration parameter spread for fibres.}
    \begin{tabular}{cccc}\toprule
         ~ & Side & $\sigma$ of correction factor & Relative spread \\ \midrule \rowcolor[gray]{.95}
         fibres coupled to single \gls{sipm} & TOP & 0.67 & 12\% \\ 
         all fibres & TOP & 1.59 & 28\% \\ \midrule \rowcolor[gray]{.95}
         fibres coupled to single \gls{sipm} & BOTTOM & 0.76 & 13\% \\ 
        all fibres & BOTTOM & 1.37 & 24\% \\ \bottomrule
    \end{tabular}
    \label{tab:fiberGainSpread}
\end{table}

\subsection{Photon counts calibration}
\begin{figure}[!htb]
\centering
\includegraphics[width = \textwidth]{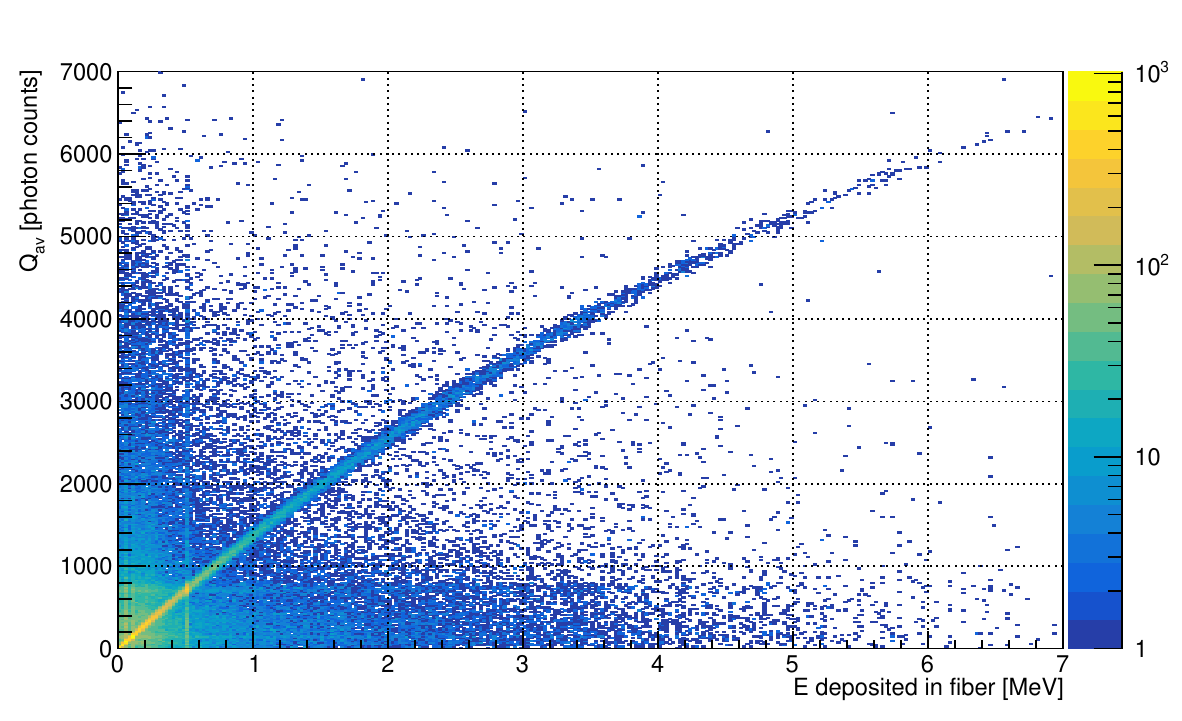}
\caption{Geometric mean of charges collected by the top and bottom \gls{sipm} ($Q_\mathrm{av}$),
versus energy deposited in the fibre (MC truth data).}
\label{fig:calibration}
\end{figure}

In the simulation data, the charge collected by the \glspl{sipm} is expressed in photon counts. To be able to apply energy thresholds on simulation hit maps, we need to know the corresponding photon count values. For this purpose, we plot a 2D histogram of $Q_{\mathrm{av}}$, which is a geometric mean of charges collected by the bottom and top \gls{sipm}, vs. energy deposited in the fibre, which is the MC truth data. In such a histogram (see~\cref{fig:calibration}), a bright line is visible that marks the dependence between the \gls{sipm} charge and the true energy deposited in the fibre. The $Q_{\mathrm{av}}$ values for the needed energy thresholds were read out from the histogram, they are listed in~\cref{tab:PhotonCountsVsEnergy}. These thresholds were applied to the simulation hit maps, so that they matched the energy thresholds on the experimental hit maps.

\begin{table}[]
\centering
\caption{Energy thresholds applied to the hit maps from the experiment, and the corresponding values in photon count units, applied to the hit maps from simulation. \label{tab:PhotonCountsVsEnergy} }
\begin{tabular}{cc} \toprule
Energy {[}keV{]} & SiPM photon counts  \\ \midrule \rowcolor[gray]{.95}
0                & 0(20)         \\
500              & 710(20)    \\ \rowcolor[gray]{.95}
1000             & 1362(20) \\
1500             & 2013(20)    \\ \bottomrule
\end{tabular}
\end{table}
\subsection{Inspection of fibre energy spectra}\label{sec:fibreEnergySpectraInspection}
\begin{figure}[!htb]
         \centering
         \includegraphics[width=\textwidth]{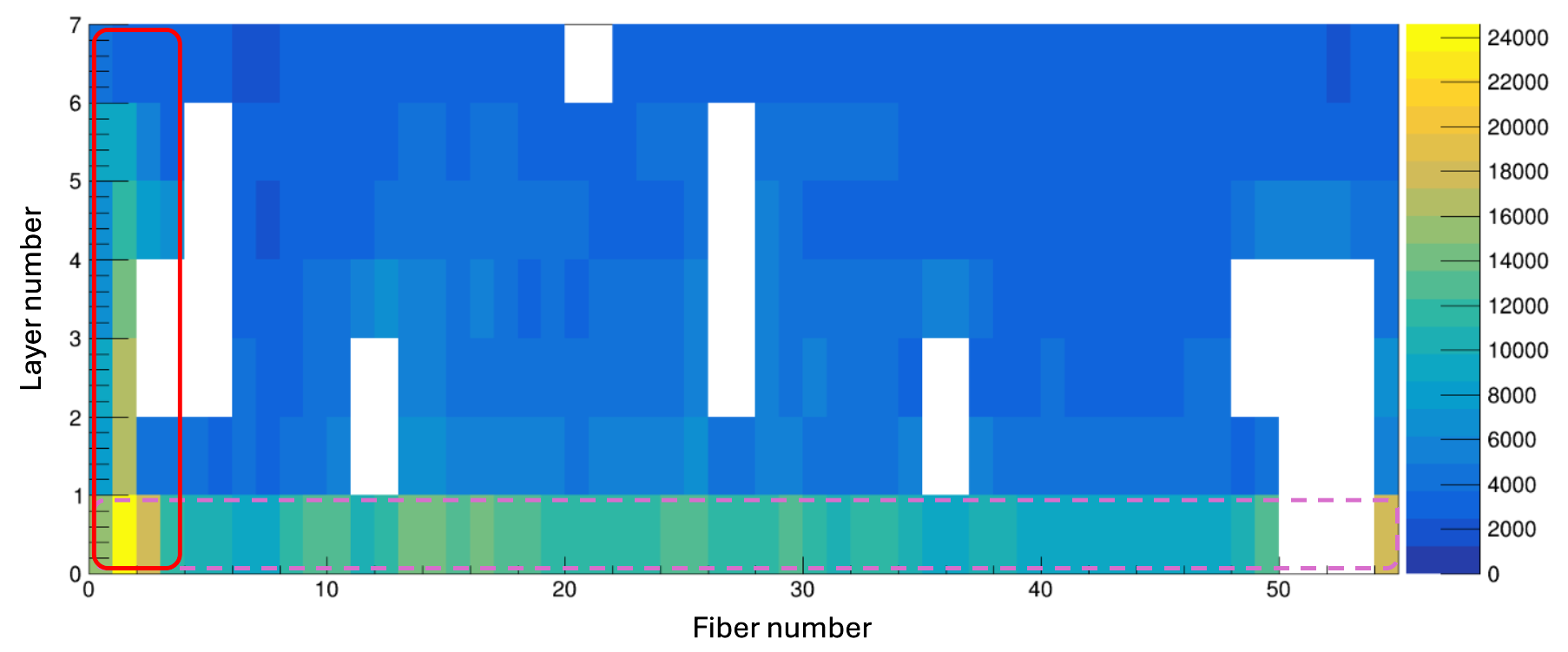}
         \caption{Hit map with no upper or lower energy threshold for an example run 569 (beam spot S1). Regions with unexpectedly high counts are marked: red contour, several first columns (region 1); dashed pink contour, first layer (region 2).}
         \label{fig:hitmapUncut569RegionsMarked}
 \end{figure}

\begin{figure}
\centering
\begin{subfigure}[t]{0.3\textwidth}
\centering
\includegraphics[width=\textwidth]{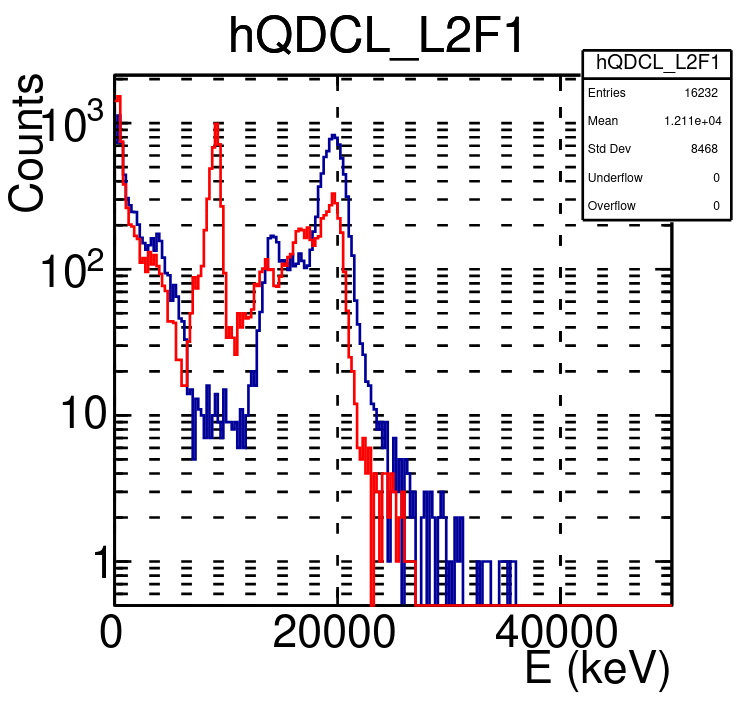} 
\caption{Layer 2, fibre 1.} \label{fig:QDC_L2F1}
\end{subfigure}
\begin{subfigure}[t]{0.3\textwidth}
\centering
\includegraphics[width=\textwidth]{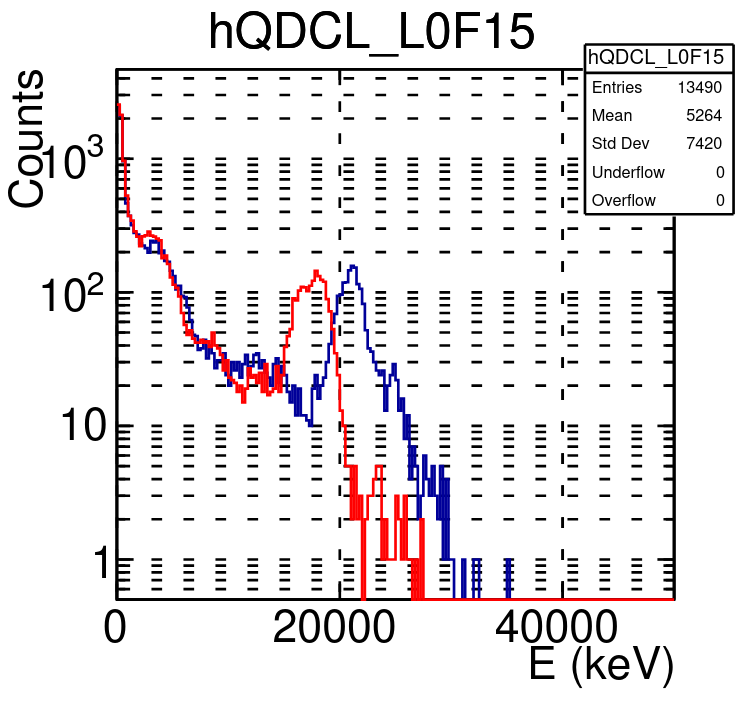} 
\caption{Layer 0, fibre 15.} \label{fig:QDC_L0F15}
\end{subfigure}
\begin{subfigure}[t]{0.3\textwidth}
\centering
\includegraphics[width=\textwidth]{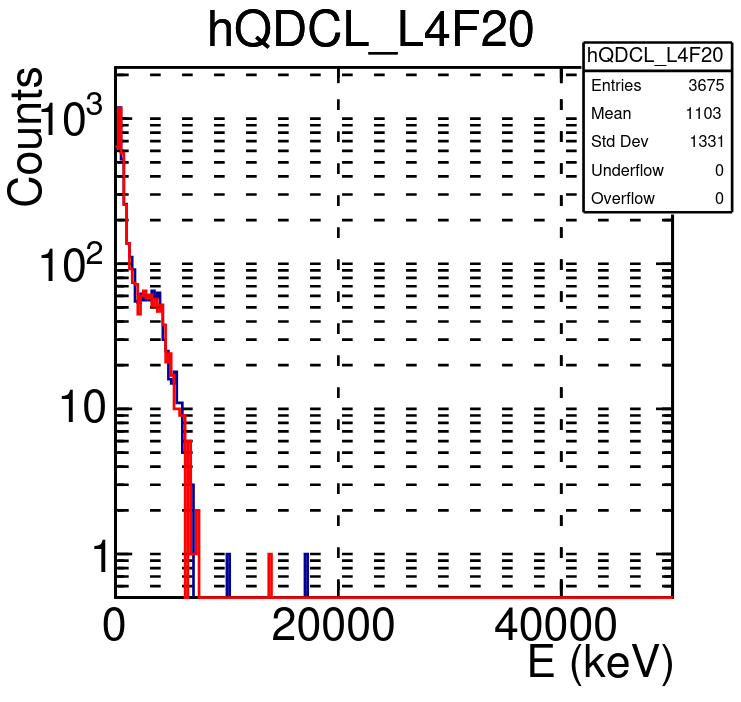} 
\caption{Layer 4, fibre 20.} \label{fig:QDC_L4F20}
\end{subfigure}
\caption{\gls{qdc} spectra for selected fibres: region 1 (a), region 2 (b) and reference from the middle of the detector (c). The red spectrum is from the \gls{sipm} on the top side, the blue one from the \gls{sipm} on the bottom side.}
\label{fig:FibreExampleSpectra}
\end{figure}

Looking at the hit maps created for the beam measurements without any energy cuts, we noticed unexpectedly high counts in some of the first fibre columns (region 1; see~\cref{fig:hitmapUncut569RegionsMarked}) and in the first layer (region 2). To explain these effects, we investigated the energy spectra of individual fibres. Example fibre spectra from these regions can be found in~\cref{fig:FibreExampleSpectra}:  (a) one of the first columns (region 1), (b) first layer (region 2), and (c) middle part of the detector (outside regions 1 and 2, for reference). In regions 1 and 2 we observed high-energy contributions in the spectra, up to a few tens of~MeV, which was not the case for the middle part of the detector. Since these energies were too high to be deposited by \glspl{pgh}, there must have been another source of these contributions. The initial assumptions were that the effect in the first few columns was caused by beam protons scattered in the exit nozzle, whereas the effect in the first detector layer was caused by protons scattered in the phantom. 

To verify the assumption for region 1, we checked how deep the particles penetrate the detector for the minimum and maximum beam energies used (S1 - \SI{70}{MeV}; S7 - \SI{108}{MeV}). For S1, the additional contribution in the spectrum was present up to the fourth column, which is 8~mm of range. For S7, it was seen up to the ninth layer, corresponding to 18~mm of range. Using SRIM~\cite{SRIM}, we calculated the proton energies that would correspond to such ranges in the scintillation material. We obtained the following values: $\sim$\SI{65}{MeV} for S1 and $\sim$\SI{100}{MeV} for S7, which are close to the beam energies used in these measurements. This is consistent with the assumption, that the effect of increased counts in the first few columns is caused by primary protons. As these additional counts are predominantly deposited in the high-energy region, which is not of our interest as we focus on \gls{pgh} detection, we can reduce this contribution by applying an upper threshold on energy deposits. Thus, an upper threshold of \SI{7}{MeV} was applied to all measurements with the beam before putting them into the image reconstruction.

We repeated a similar study to explain the effect of increased counts in the first detector layer (region 2), but the expected range of secondary protons (up to \SI{2}{mm} at the energy of up to \SI{30}{MeV}, as only the first layer was affected) does not match the expected secondary proton energies: from the kinematics of the two-body scattering, any secondary protons reaching the detector would have a lower energy ($\sim$\SI{1}{MeV}), which corresponds to \SI{9.19}{\micro m} of range. This energy is much lower than what we observe in the energy spectra, which span up to about \SI{40}{MeV}. For now, the origin of the high-energy contribution in the first detector layer is unclear, but since it is also mostly present at higher energies, it can be eliminated by application of the same upper threshold on fibre hit energy deposits.

\subsection{Hit maps in different beam spots}
\begin{figure}[H]
\centering
\includegraphics[width = \textwidth]{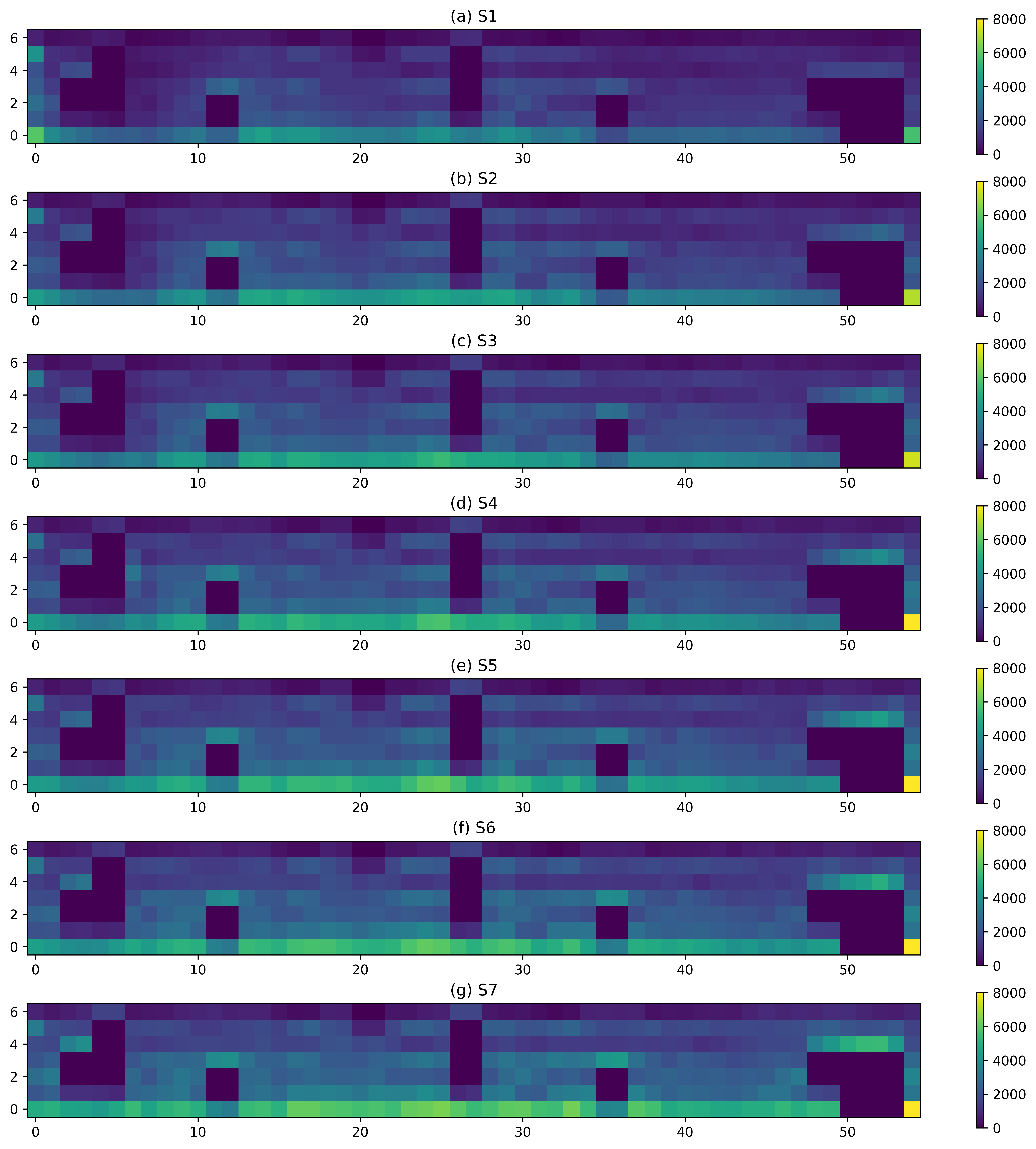}
\caption{Hit maps (detector response) for beam spots S1-S7. Cuts on energy deposits: 1-\SI{7}{MeV}, colour scale of fixed range denotes the number of hits per detector pixel. Horizontal axis - number of fibre within a layer, vertical axis - layer number.}
\label{fig:hitmapsInDifferentBeamSpots}
\end{figure}

The purpose of this study was to preliminarily assess the quality of the hit maps before inputting them to the image reconstruction. We investigated how the detector response changes with the beam penetration depth (beam spots S1-S7). The hit maps for these beam spots are presented in~\cref{fig:hitmapsInDifferentBeamSpots}a-g. There, one can see the number of entries increasing with penetration depth and that the further columns of the detector register more hits, which are the expected effects. On top of this, the irregular pattern is visible, which we attribute to the mask shape. Several other effects can be observed: 
\begin{enumerate}
    \item increased counts in the bottom right pixel and increased counts in several pixels of the first few columns (especially pronounced for S1) - the reasons for this effect were studied in~\cref{sec:fibreEnergySpectraInspection}.
    \item brighter spots behind the acceptance gaps (in the deeper layers) and darker spots in front of them (in the shallower layers). 
    \end{enumerate}
These effects are unfavourable, because they do not originate from the \glspl{pgh} that are of interest, but are either background effects or artefacts of the analysis due to the acceptance gaps. The first effect is already mitigated by applying an upper threshold on the hit maps but is not completely eliminated. This is another reason for excluding the regions with increased counts from further analysis (see~\cref{sec:imageRecoOptimization}). The other effect is a distortion of the hit map due to the applied algorithm that assigns \gls{sipm} hits to fibres and the presence of acceptance gaps; we did not study it further, but instead attempted to account for it (see~\cref{simulationFiltering}).

\subsection{System stability over time}
For beam spots S2-S4, measurements were repeated roughly 30 minutes after the main measurement series with statistics increased 10 times (i.e., to $10^{11}$ protons). We refer to them as S2'-S4'. The pairs of measurements (Sn,Sn') were used to assess the stability of the detection system over time. The hit maps for measurements S2'-S4' were first normalised to match the statistics of measurements S2-S4. Then, we created maps of the differences in entries per pixel ($N_\mathrm{Sn} - N_\mathrm{Sn'}$) and checked if they are uniformly distributed or show irregular patterns. Then, we looked at 1D histograms of the said differences. We expect the differences to form a Gaussian shape around 0. A non-Gaussian shape or a large shift with respect to 0 could be a sign of some unfavourable effects in the experimental setup, e.g. a noise contribution growing with time due to \gls{sipm} heating. 

\begin{figure}[!htb]
\centering
\includegraphics[width = \textwidth]{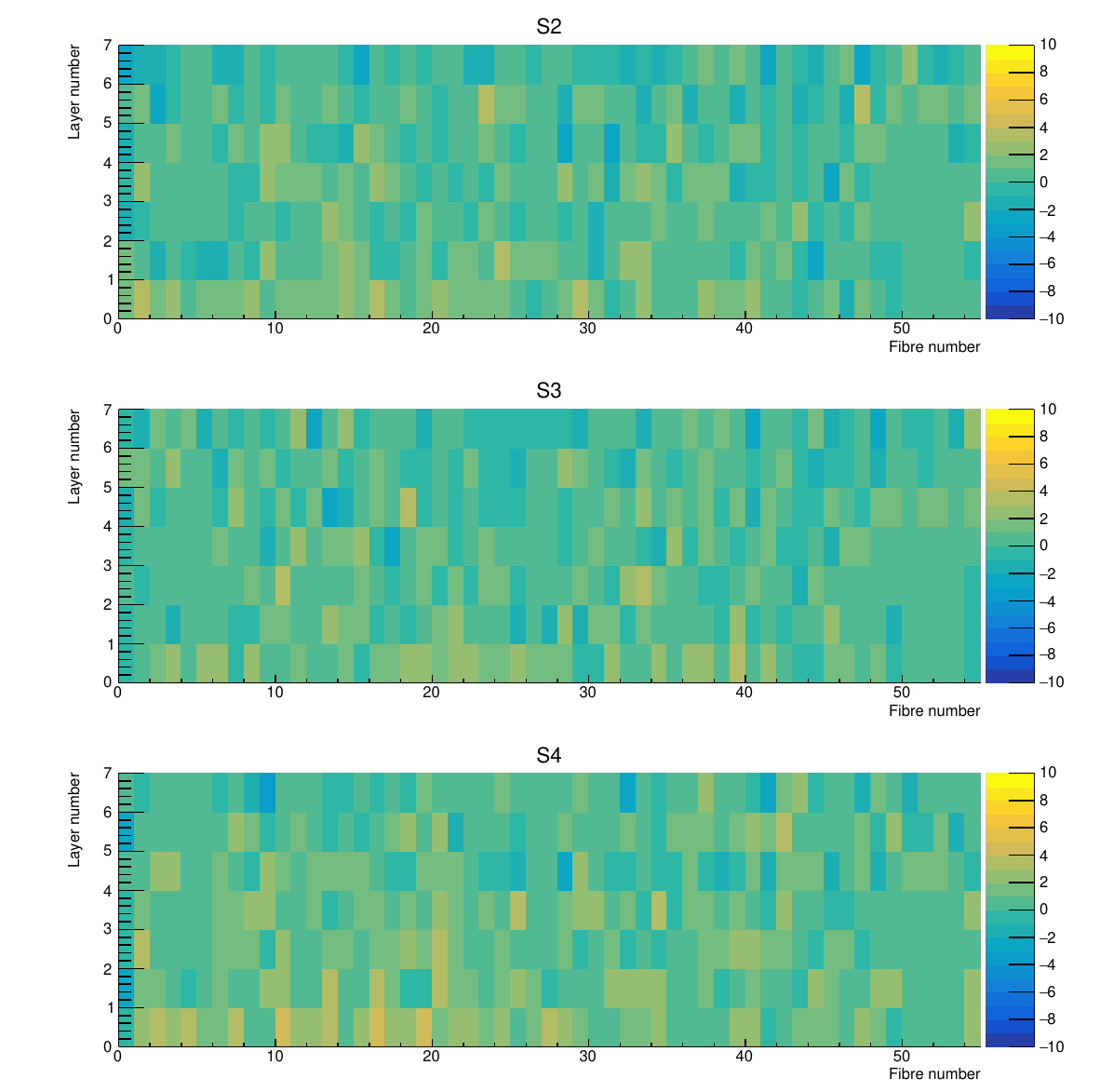}
\caption{Maps of differences between numbers of events per pixel in two different measurements for the same beam spot, in units of standard deviation, normalised.}
\label{fig:hitmaps_residuals}
\end{figure}

\begin{figure}[!htb]
\centering
\includegraphics[width = \textwidth]{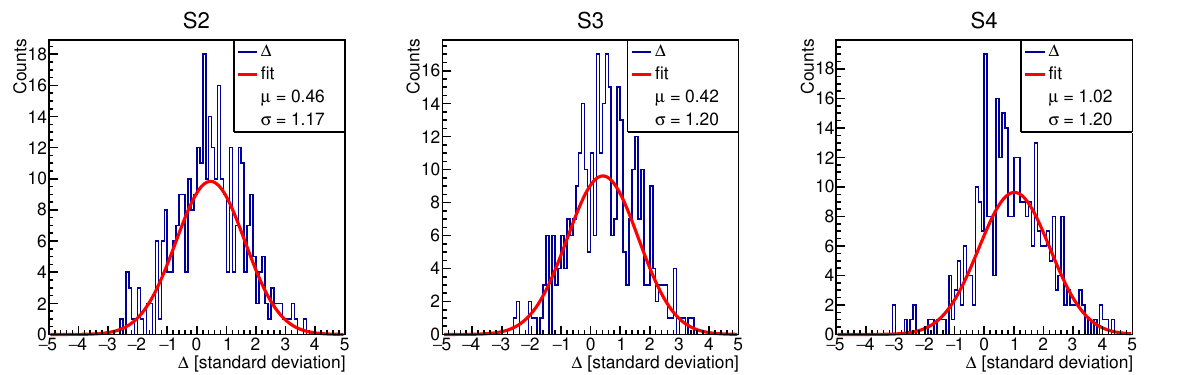}
\caption{Distributions of differences ($\Delta$) between numbers of hits per pixel in two different measurements for the same beam spot, in units of standard deviation. The red line denotes a Gaussian fit to the distribution.}
\label{fig:1D_withoutEmptyBins}
\end{figure}

The maps of the differences ($N_\mathrm{Sn} - N_\mathrm{Sn'}$) are presented in~\cref{fig:hitmaps_residuals}. 
There, one can see that the differences are uniformly scattered for all beam spots, as expected. The histograms of the differences in entries per pixel ($N_\mathrm{Sn} - N_\mathrm{Sn'}$) are presented in~\cref{fig:1D_withoutEmptyBins}, in units of standard deviation. The dead fibres were excluded. There, one can see that for all three investigated beam spots, the distribution of differences is Gaussian-like, as expected. It is not centred around 0, but instead shifted towards the positive values. Nevertheless, for all the examined beam spots, the shift ($\mu$ in the histogram legend) is up to about 1 standard deviation, which we consider a small shift without significant influence on the main analysis. There is no clear trend in the shift values, but the largest shift occurs for the deepest beam spot examined (S4).

\subsection{Background studies}
\begin{figure}[!htb]
\centering
\includegraphics[width = 0.9\textwidth]{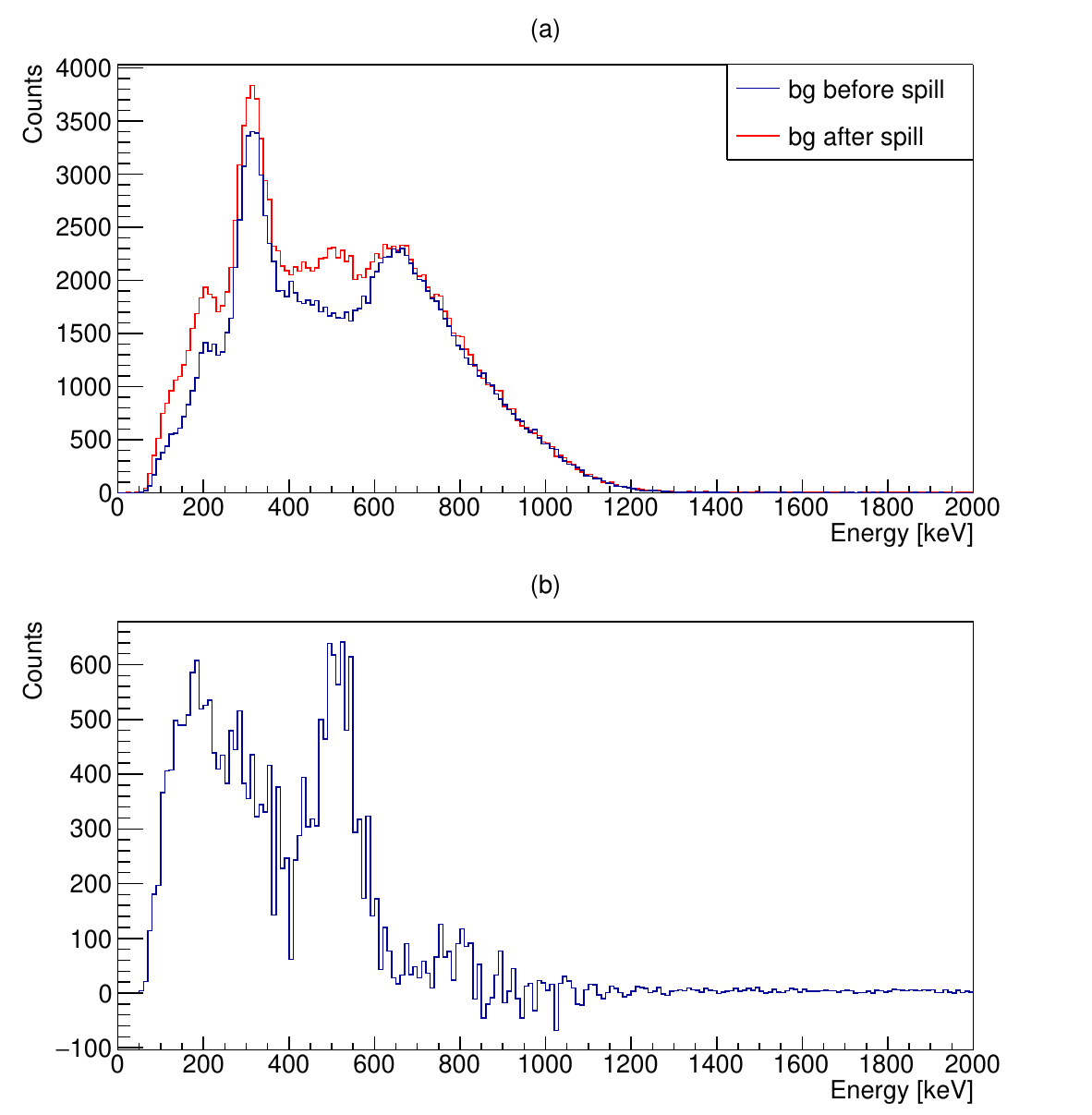}
\caption{(a) Energy spectra of the background before (blue) and after (red) the beam spill for beam spot S4; (b) Difference of spectra from the plot in (a) -  beam-induced background.}
\label{fig:bg_before_vs_after_run00567}
\end{figure}

The shape of the energy spectrum hints at the background origins. In~\cref{fig:bg_before_vs_after_run00567}a, a comparison is presented of the background energy spectra before and after the beam spill. Both measurements lasted \SI{3.3}{s}. There, one can see that the spectrum after the spill has higher intensity in the region up to \SI{700}{keV}. Several peaks are visible in both spectra, corresponding to the peaks in the LYSO:Ce,Ca activity spectrum (see, for example, the simulation results in~\cite{AlvaSanchez2018}). In the second spectrum, an additional peak of energy close to \SI{511}{keV} can be observed, along with an increased contribution to the spectrum at energies below the peak energy. The difference of the two spectra (the spectrum before the spill was subtracted from the spectrum after the spill) is presented in~\cref{fig:bg_before_vs_after_run00567}b. There, one can clearly see the annihilation peak at \SI{511}{keV}, and the associated Compton continuum at lower energies. This confirms that production of $\beta^+$ emitters is the main beam-generated background source to \gls{pgh} detection.

\subsection{Contributions of event classes in LLR}
We examined the contributions of event types a-e (see~\cref{sec:fibreHits}) in the experimental data. The distribution of the event types is similar for both the spill and background part of run S4: the single-fibre events (type a) are about 41\% of all events, the multi-fibre events with unique clusters (type b) are almost 50\%, 10\% are the semi-unique cluster events (types c and d), and the ambiguous events are up to 0.2\% of the data sample. For run 596 with a radioactive source, there is a larger fraction of single-fibre events in the data sample: 62\%, while the multi-fibre events form 32\% of all events. Semi-unique events are 7\% in total, while ambiguous events are only 0.05\%. The exact values of the event type distributions are summarised in~\cref{tab:typesOfEvents}. 

\begin{table}[!ht]
\caption{Multiplicity of event types (a-e) in various run types.\newline \label{tab:typesOfEvents}}
    \centering
    \begin{tabular}{cccccc}\toprule
        ~ & Type a [\%] & Type b [\%] & Type c [\%] & Type d [\%] & Type e [\%] \\ \midrule  \rowcolor[gray]{.95}
        S4 (spills) & 40.86 & 49.28 & 5.58 & 4.07 & 0.21  \\ 
        S4 (background) & 40.61 & 47.89 & 5.90 & 5.67 & 0.10  \\   \rowcolor[gray]{.95}
        Source run (596) & 61.87 & 31.28 & 3.32 & 3.49 & 0.05 \\ \bottomrule
    \end{tabular}
\end{table}
Based on the distribution of event types, we decided that the semi-unique cluster events (types c and d) form a significant part of the data sample, so we need to include them in the analysis. As the ambiguous events constitute only a small fraction of the data sample and including them would involve developing a dedicated procedure to separate the clusters, we decided to disregard them in the analysis.

\section{Analysis of Monte Carlo data}\label{sec:simulationDescription}
In order to reproduce the experimental conditions of the beam tests of the detector in \gls{1dcm} modality, a realistic Geant4~\cite{Geant4} simulation was set up\footnote{The simulation was developed by R.~Hetzel, M.~Kercz and Linn.~Mielke.}. The simulation was based on the previous developments by J.~Kasper, described in~\cite{KasperPhD}. The version of the Geant4 code was 10.4.2, the physics list used was QGSP\_BIC\_HP\_EMZ (see~\cite{PhysicsListGuide}), an additional package was used to manage the properties of scintillation crystals and \glspl{sipm}~\cite{GODDeSS}. The simulation consisted of the following steps:
\begin{enumerate}
    \item Interaction of the beam with the phantom: a simulated proton beam of fixed energy hits the phantom. The secondaries from beam-phantom interactions propagate in the phantom up to its borders. There, their data (i.e., direction, energy, and time) are saved as phase-space files.
    \item Interactions of the secondary particles from the phantom with the detector: the secondary particles (saved as phase-space files in the previous step) are emitted outside the phantom and interact with the detector. The simulation stops when the \glspl{sipm} register the optical photons generated in the scintillation fibres. The output files from the simulation contain the data registered by the \glspl{sipm}, the information about the particles that triggered the detector, and the Monte Carlo truth information about the fibre hits.
\end{enumerate}
The simulation data are adjusted to the experimental data by introducing the acceptance gaps that were present in the experiment (see~\cref{simulationFiltering}), and then processed with the same analysis chain as the experimental data. In this way, we can directly compare the simulation and the experimental results. Additionally, the full simulation without the acceptance gaps is also processed and compared with the adjusted one, to estimate the influence of the acceptance gaps on the detector performance in terms of the image reconstruction quality.

\section{Image reconstruction}
\subsection{MLEM algorithm}\label{sec:imageReco}
In order to reconstruct the \gls{pgh} image (which is in our case a 1D \gls{pgh} depth profile) from the detector response in the form of hit maps (described in~\cref{sec:hitmapsProduction}), we use the \gls{mlem}~\cite{Shepp1982, Lange1984} algorithm, implemented in a custom software\footnote{The software for image reconstruction in the \gls{1dcm} modality was written by V.~Urbanevych, and it was later modified and extended by the author of this thesis.}. The \gls{mlem} algorithm is provided in~\cref{MLEMEquation}:
\begin{equation}\label{MLEMEquation}
    \bm{f}^{(k+1)} = \frac{\bm{f}^{(k)}}{\bm{S}} \textbf{A}^T \frac{ \bm{y}}{\textbf{A}\bm{f}^{(k)}+\bm{b}},
\end{equation}
where the system matrix is denoted with $\textbf{A}$; $\bm{S}$ is the sensitivity; $\bm{y}$ denotes the measured detector response; $\bm{b}$ stands for background; vectors $\bm{f}^{(k)}$ and $\bm{f}^{(k+1)}$ are the reconstructed \gls{pgh} depth profiles (projections) after iterations $k$ and $k+1$, respectively. The vector element $f^{(k)}_j$ is the reconstructed number of \glspl{pgh} emitted from the \gls{fov} pixel $j$, the length of this vector $J=100$ equals to the number of \gls{fov} pixels. Elements of $\bm{f}^{(0)}$ are initialised with ones.

The components of the formula are explained in the following section.

\subsection{Input to the image reconstruction}\label{sec:analysis_inputToImageReco}

\subsubsection{System matrix}
The \gls{sm} is a fundamental concept in statistical iterative image reconstruction algorithms. It relates the image space to the projection space. An \gls{sm} element $a_{ij}$ expresses the probability that a particle emitted from the source pixel $j$ is registered in the detector pixel $i$~\cite{Rafecas2003}. 

\begin{figure}[!htb]
\centering
\includegraphics[width = 0.7\textwidth]{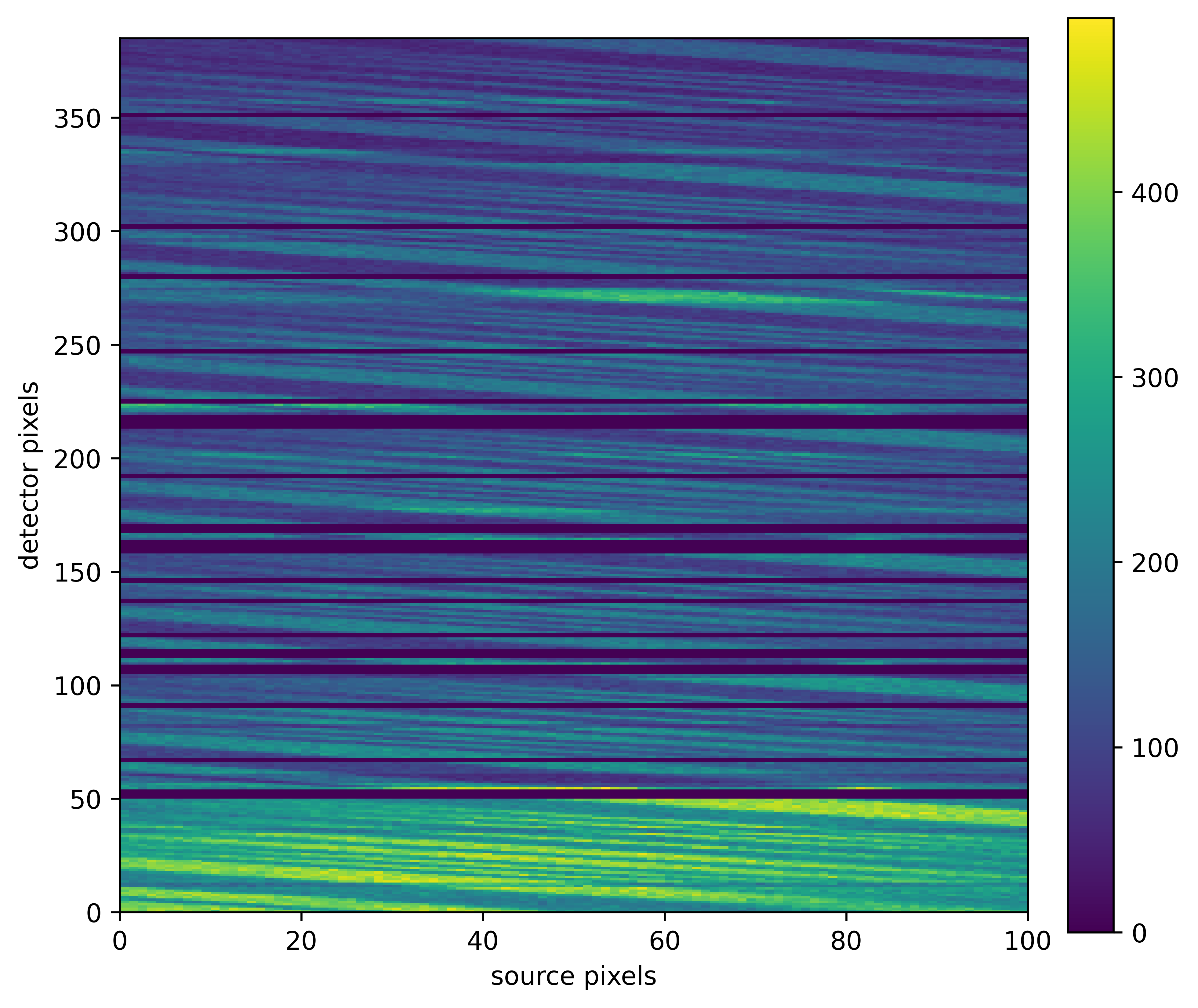}
\caption{The system matrix (before normalisation and removal of elements corresponding to dead fibres), used in the reconstruction of images from the \gls{1dcm} modality detector tests.}
\label{fig:systemMatrix}
\end{figure}

The \gls{sm} for our purpose was generated\footnote{The \gls{sm} was generated by M.~Kercz.} using the simulations described in \cref{sec:simulationDescription}. The first stage of the simulation was modified though: instead of a phantom interacting with the proton beam and generating secondary particles including \glspl{pgh}, a point-like gamma source was used as an event generator. The \gls{fov} of the detector (at $y=0$ and $z=0$) was divided into 100 bins (pixels) of equal length. Then, a gamma source was placed in the centre of the first pixel and the detector response was recorded. The procedure was repeated for source positions in all the \gls{fov} pixels. For each source position, the full detector response was saved, the hits from dead pixels were filtered out, the data were processed with the usual analysis chain and the hit maps were created. Then, the hit maps of dimensions 7 layers $\times$ 55 fibres were flattened\footnote{Here and in the further parts of this thesis, flattening a hit map means creating a vector by appending the hit map rows to it, one after another.} into vectors of length 385. Each of such vectors formed one column of the \gls{sm}, which can be seen in~\cref{fig:systemMatrix}. There, \gls{sm} is not normalised, each bin contains the number of hits for a given detector pixel. To be interpreted as probability, \gls{sm} needs to be normalised to the total number of emitted gammas into the full solid angle (which was 37.798$\times10^6$). This is an ideal \gls{sm} for uniform efficiency, $\tilde{a}_{ij}$. Before putting the \gls{sm} in the \gls{mlem} formula, it was corrected for efficiency: $a_{ij} = \epsilon_i \tilde{a}_{ij}$ (see the efficiency calculation in~\cref{sec:efficiencyCorrection}); moreover, the matrix elements corresponding to the response in dead fibres were excluded, so the resulting \gls{sm} has dimensions of $100\times303$.
\subsubsection{Sensitivity}
The sensitivity is defined as the probability that a \gls{pgh} emitted from a given \gls{fov} pixel is registered by the detector. We calculate the sensitivity by summing up the system matrix elements column-wise over the detector pixels. The resulting vector has 100 elements, which is the number of \gls{fov} pixels.

\subsubsection{Spills and background separation}
\begin{figure}[!htb]
\centering
\includegraphics[width = \textwidth]{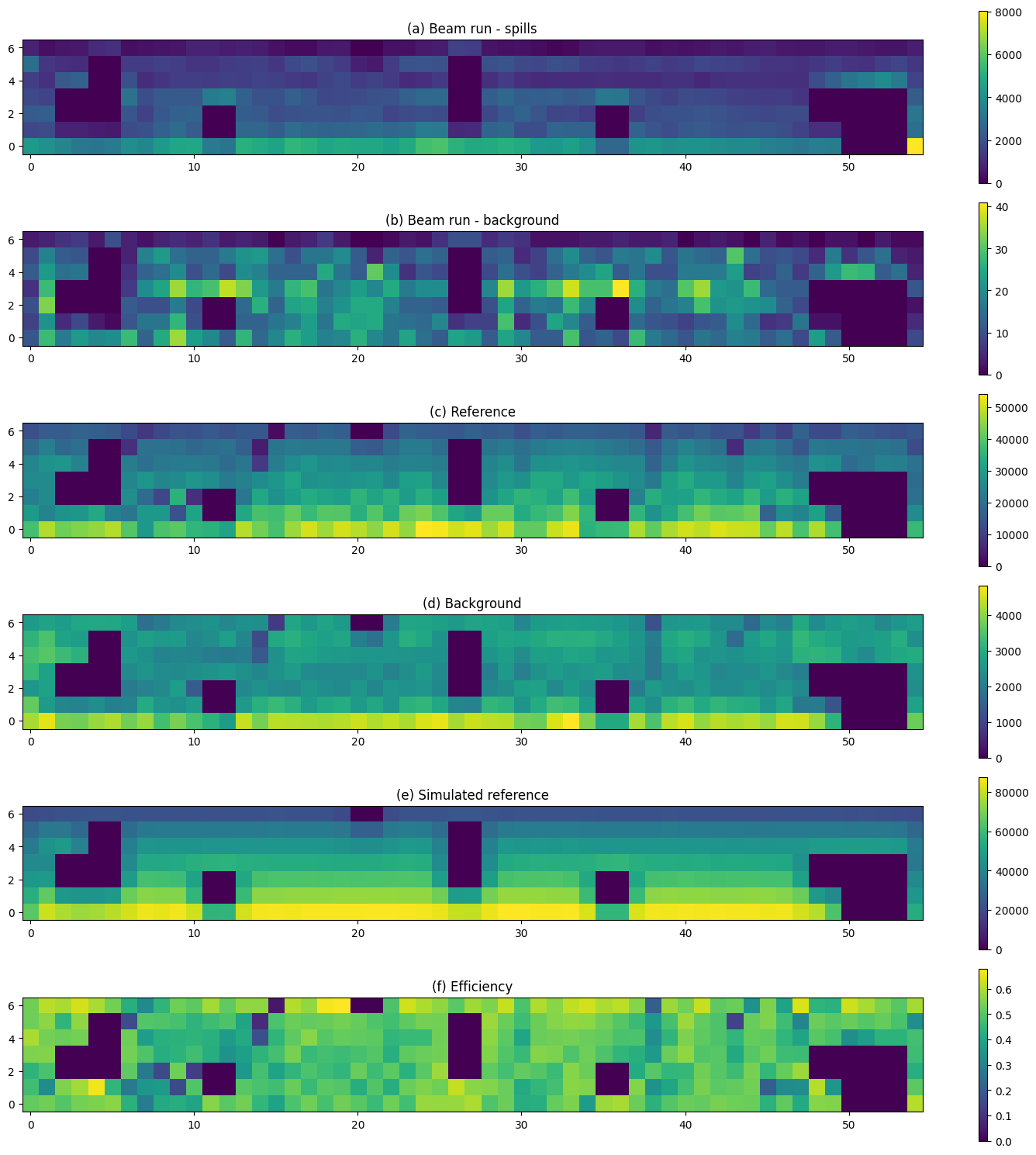}
\caption{Example hit maps; detailed description in the text. Horizontal axis - number of fibre within a layer, vertical axis - layer number.}
\label{fig:allHitmaps}
\end{figure}

After the time preselection (see~\cref{sec:timePreselection}), there are two types of files from the runs with the proton beam: spills and background. The flattened hit maps produced based on spill data are inputted into the MLEM formula (\cref{MLEMEquation}) as the measured detector response $\bm{y}$, while the flattened hit maps obtained for background, normalised according to the procedure described in~\cref{sec:timePreselection}, are put into the formula as background $\bm{b}$. Both $\bm{y}$ and $\bm{b}$ vectors have 303 elements, i.e., the number of all detector pixels (385) minus the dead pixels. The hit maps for the two parts of an example run 568 (beam spot S4) are presented in~\cref{fig:allHitmaps}a for spills and in~\cref{fig:allHitmaps}b for background.
\subsection{Efficiency correction}\label{sec:efficiencyCorrection}
Detection efficiency can vary for each detector pixel due to various hardware imperfections, e.g., non-uniformities in the fibre structure, different thicknesses of the coupling layer, or some dead cells in the \glspl{sipm}. To correct for these effects, it was necessary to perform an auxiliary measurement with uniform detector irradiation (reference), a simulation reproducing conditions of such a measurement (simulated reference), and a measurement of background. Both measurements were described in~\cref{sec:experiments_auxiliaryMeasurements}. The hit map for the reference measurement is presented in~\cref{fig:allHitmaps}c, the background hit map can be found in~\cref{fig:allHitmaps}d.

A simulation reproducing the reference measurement was performed\footnote{The simulation was prepared by R.~Hetzel.}. The simulated detection setup was identical to the experimental one, the number of simulated particles was $4\times10^8$ and the particle emission angle was limited to $35^{\circ}$, so that particles from all regions of the source contribute to the data. The simulated reference hit map is presented in~\cref{fig:allHitmaps}e.

All three data sets (reference, simulated reference and background) were processed in the same way with the \gls{llr} reconstruction software and hit maps were created, with selected hit energies within 2$\sigma$ around the \SI{511}{keV} peak - the main peak in the registered energy spectra from the used radioactive source. The hit maps were flattened into vectors. As a result, we obtained three vectors of 303 elements each: $\bm{R}$ - reference measurement, $\bm{SR}$ - simulated reference measurement, and $\bm{B}$ - background. For the detector element $i$, the efficiency $\epsilon_i$ was calculated according to~\cref{eq:efficiencyFormula}:
\begin{equation}\label{eq:efficiencyFormula}
    \epsilon_i = \frac{R_i-B_i}{SR_i}.
\end{equation}
The efficiency map obtained, normalised to unity in its maximum, can be found in~\cref{fig:allHitmaps}f. In the calculations, it was flattened to one dimension, but it is presented as a 2D hit map for visualisation purposes, with the dead pixels manually set to 0. The efficiency is included in the system matrix.

\subsection{Distal falloff position determination}
The MLEM formula was applied to the data for the chosen number of iterations, and the \gls{pgh} depth profiles were obtained for each of the beam spots (S1-S7). A \gls{pgh} depth profile represents the distribution of the \gls{pgh} emission vertices from the phantom, we use the standard normalisation to unity in the profile maximum. The distal falloff of a \gls{pgh} depth profile is strongly correlated with the proton beam range~\cite{Min2006}, so it is the crucial part of the profile. The determination of \gls{dfp} is done according to the procedure described in~\cite{Gueth2013}. A spline is fit to the \gls{pgh} depth profile, the distal falloff is the part of such a profile between the global maximum of the profile and a first following downstream minimum (in our representation, this corresponds to the profile part to the right of the maximum). We then define the \gls{dfp} as the depth ($x$ position) of the point with a half value between the maximum and the minimum. This value is compared with the proton range, calculated with the use of  PSTAR~\cite{PSTAR}, for PMMA of the \SI{1.19}{\g/\cm\tothe{3}} density, and the proton beam energies used in the experiment. 

\subsection{Optimisation of image reconstruction}\label{sec:imageRecoOptimization}
The components of the \gls{mlem} formula presented in~\cref{sec:analysis_inputToImageReco} have previously been optimised according to the principles described in this section.
\subsubsection{Optimisation targets}
In the course of finding the optimal set of input parameters of image reconstruction, we tested combinations of the following options: 
\begin{enumerate}
    \item number of iterations: 1-1000,
    \item threshold of energy deposit in maps: 
    \begin{itemize}
        \item lower threshold: 0, 500, 1000, 1500~keV,
        \item upper threshold: 7~MeV or none,
    \end{itemize}
    
    \item post-processing - \gls{pgh} depth profile smoothing function (from the \verb|scipy.ndimage| package~\cite{GaussianFilterPython}): 
    \begin{itemize}
        \item smoothing function: Gaussian, median or none,
        \item standard deviation of the kernel (corresponding to degree of smoothing): 1-10 pixels,
    \end{itemize}
    \item excluded detector regions:
    \begin{itemize}
        \item dead fibres, listed in~\cref{tab:deadFibers} in Appendix,
        \item empty fibres: if there were any pixels with no entries (apart from the dead fibres), they were excluded,
        \item $n$ first and last columns, $n\in(0,7)$,
        \item $m$ layers (starting from the one most distant from the source or phantom), $m\in (0,6)$,
    \end{itemize}
    \item classes of events included: only unique (types a and b) or unique and semi-unique (types a-d),
    \item type of the detector response map: hit maps or maps of energy deposits, 
    \item summing up the maps layer-wise or not,
    \item application of the efficiency correction: to the measured data or to \gls{sm},
    \item accounting for the presence of background: subtraction on the input level prior to the image reconstruction or inclusion of the background term in the MLEM formula.
\end{enumerate}
The quality of the parameter sets was assessed based on the values of the performance metrics (see~\cref{sec:performanceMetrics}), calculated for each image reconstruction variant.

For some of the parameter options listed above (e.g. number of iterations, energy thresholds, smoothing method), it is understandable that they needed to be tested and optimised; others may require some explanation of the reason why they were included in the optimisation. 

We started parameter optimisation with the options that worked well in a previous study involving CM image reconstruction~\cite{Hetzel2023}: summing the data across layers, subtracting the background from the measured data before feeding them to the MLEM algorithm, and applying the efficiency correction to the measured data (and not to the \gls{sm}). However, these options were ruled out in the optimisation procedure, as other approaches tested were found to give better results (see~\cref{recoImages}). We decided to try to exclude lateral columns, as due to the lack of detector shielding we observed signals generated by primary protons in several upstream detector columns (see~\cref{sec:fibreEnergySpectraInspection}).  

The number of iterations was first roughly optimised independently of the other parameters, yielding a range of values to be considered. Next, the other parameters were compared in direct parameter scans: several parameter sets were chosen, image reconstructions were done looping over the parameter values, and the obtained performance metrics were compared. Several options, e.g. subtracting the background before applying the MLEM formula or thresholds 0 and \SI{1500}{keV}, yielded consistently worse results across all parameter scans, so they were excluded from the list of possible parameter values. Two parameters: classes of events included and type of detector response map, did not have significant influence on the image reconstruction quality. Thus, we decided to include more event classes (types a-d), as it is generally preferred to use all available information, and chose hit maps over maps of energy deposits. The hit maps content is governed by the Poissonian counting statistics, which is one of the assumptions on which the derivation of the \gls{mlem} formula is based, thus this choice is better justified from the mathematical point of view. Finally, a wide scan of the number of iterations was performed again, based on which this parameter was fixed.

\subsubsection{Performance metrics}\label{sec:performanceMetrics}
The following metrics were used to assess the performance of image reconstruction:
\begin{enumerate}
    \item Pearson correlation coefficient ($PCC$) between the determined \glspl{dfp} and calculated proton ranges (PRs) for beam spots S1-S7,
    \item Coefficient of determination ($R^2$) between \glspl{dfp} and PRs for beam spots S1-S7,
    \item Root mean squared error ($RMSE$): 
    \begin{equation}
       \mathrm{RMSE} = \sqrt{\frac{\Sigma_{i,j}(\Delta_{\mathrm{PR}}-\Delta_{\mathrm{DFP}})^2}{2n}},    \end{equation}
        \begin{equation}
        \Delta_{\mathrm{PR}} = \mathrm{PR}_{i}-\mathrm{PR}_{j}
        \end{equation}
        \begin{equation}
        \Delta_{\mathrm{DFP}} = \mathrm{DFP}_{i}-\mathrm{DFP}_{j}
    \end{equation}
    where $n=21$, number of all pairs of measurements, and $i,j\in [1,7]$, number of beam spots,
    \item Standard deviation ($SD$) of the \glspl{dfp} of the spots with the same $x$ coordinate ($S4$, $S4'$, S4a-d),
    \item Slope of the linear fit to \glspl{dfp} vs. PRs for all spots ($s$).
\end{enumerate}

The $PCC$ quantifies the level of correlation between two variables. It takes values from a range of (-1,1). A strong positive correlation between the \gls{dfp} and the Bragg peak position was demonstrated experimentally, so it is preferred in our study to have $PCC$ as close to 1 as possible. 

The $R^2$ takes values from a range of (0,1). The closer it is to 1, the better the assumed model replicates the measured data. Thus, it is also preferred to be as close to 1 as possible.

The standard deviation $SD$ of the profiles of the same depth is a preliminary measure of the image reconstruction reproducibility on different data sets, so it is preferred to be as low as possible. 

The $RMSE$ can be interpreted as the accuracy of the image reconstruction; thus it is the crucial metric and it was our priority to minimise it. 

Finally, the slope $s$ of the fit line should be close to 1, different slope would mean that an additional calibration would be needed to translate the measured DFP shifts into PR shifts. We did not impose any limit on the intercept, because the \gls{dfp} is known to be shifted with respect to the \gls{bp} position, and there could also be a physical shift in the setup.

To select the parameter sets that provided the best image reconstructions, we applied the following cuts on the performance metrics:
\begin{itemize}
    \item $C_C> 0.99$,
\item $C_D>0.8$,
\item $SD<1$,
\item $RMSE<1.88$,
\item $0.98<s<1.02$.
\end{itemize}
After selecting the parameter sets that fulfil the above criteria, we chose the best ones, looking at the shape of the \gls{pgh} depth profiles and the stability of the performance metrics (if a slight change in the parameter value, e.g. incrementing the iteration number by 1, does not cause a drop of performance). 

\section{Analysis of statistical precision}\label{sec:analysis_statisticalPrecision}
The measurements for the beam spots S1-S7 were performed for $10^{10}$~impinging protons, which is much more than administered to a typical spot during irradiation in proton therapy ($10^8$, see~\cref{sec:introduction_requirementsForPTMonitoring}). Therefore, there was a need to check the image reconstruction performance also for lower proton statistics. Such a study was carried out on subsets of data corresponding to the following statistics: (20, 10, 4, 2, 1)$\times 10^8$~protons. For each of these statistics, 100 subsets (in the form of \gls{llr} output trees) were created with the use of bootstrapping~\cite{Efron1992}. Then, image reconstruction was performed on each of the subsets, and the spread of the reconstructed \glspl{dfp} (quantified as a standard deviation) was considered a measure of the statistical precision in \gls{dfp} determination for a given subset statistics.

\chapter{Results}
In this chapter, we present the results of the \gls{1dcm} detector performance test in proton beam conditions. Then, we compare the experimental results with those obtained from the simulated detector response. Finally, we examine the performance of the detector without acceptance gaps based on simulated data.

\section{Rate capability}\label{sec:results_rateCapability}
Based on the measurements described in~\cref{sec:experiments_rateCapability}, average channel occupation for the maximum beam intensity used for patient treatment at HIT was calculated to be~\SI{26.4}{kcps} (assuming uniform channel occupation), while the limit declared by the \gls{daq} manufacturer is \SI{480}{kHz}~\cite{TOFPET2overview}. Therefore, the detector can potentially work with up to $18$ times higher beam intensities.

\section{Detection efficiency}
The map of the detection efficiency per pixel is presented in~\cref{fig:allHitmaps}f. The mean value of relative pixel efficiencies, after excluding the dead pixels, yielded 0.746. The relative standard deviation of pixel efficiency was 11.39\%.

\section{Dynamic range}

\begin{figure}[!htb]
\centering
\includegraphics[width = \textwidth]{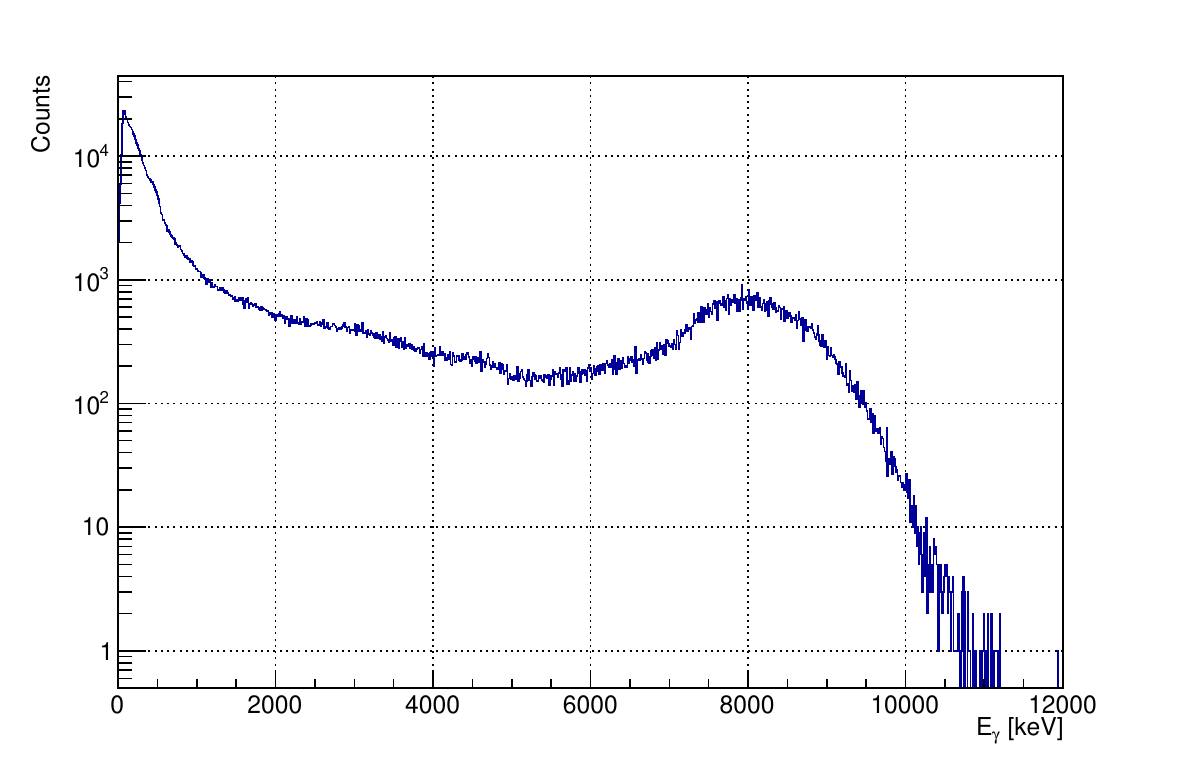}
\caption{Charge spectrum from \gls{sipm} 11 of the first layer on the top side of the detector, rescaled to energy units.}
\label{fig:SiPM_0_0_11_582_energySpectrum}
\end{figure}
In~\cref{fig:SiPM_0_0_11_582_energySpectrum}, we present an example of a \gls{sipm} energy spectrum from an example beam spot S4, for the statistics of $10^{11}$ protons. The spectrum was calibrated according to the procedure described in~\cref{sec:energyCalibrationPerSiPM} and  corrected for \gls{daq} non-linearity (see~\cref{sec:nonLinCorr}). In the spectrum, we can see a hint of the characteristic PG peak at the energy of~\SI{4.4}{MeV}. There is no hint of the \SI{6.1}{MeV} peak though, which could be attributed to too low statistics. The registered energy deposits do not exceed \SI{11}{MeV} and there is an increase in counts starting from about \SI{6}{MeV}, which is a sign of saturation. Therefore, we assume that above this value, the energy information obtained could not be considered fully reliable. Nevertheless, the image reconstruction presented in this thesis is not affected by saturation, since it does not exploit the energy information. However, the region of interest for detecting the \gls{pgh} characteristic peaks is not significantly affected by the saturation effect. Hence, we conclude that the detector has an appropriate dynamic range to register \gls{pgh} spectra in proton therapy. 

\section{Reconstructed images}\label{recoImages}
The image reconstruction was optimised, as described in~\cref{sec:imageRecoOptimization}. The following parameters provided the best image reconstruction performance, as well as its stability and good-quality \gls{pgh} depth profiles:
\begin{enumerate}
    \item number of iterations: $23$,
    \item threshold of energy deposit in hit maps: 
    \begin{itemize}
        \item lower threshold: $1000$~keV,
        \item upper threshold: $7$~MeV,
    \end{itemize}
    
    \item Post-processing: smoothing with a Gaussian filter with kernel standard deviation of 3 pixels,
    \item excluded detector regions:
    \begin{itemize}
        \item dead fibres,
        \item empty fibres,
        \item $n=3$ first and last columns, 
        \item $m=0$ further layers,
    \end{itemize}
    \item classes of events included: unique and semi-unique (types a-d),
    \item input type: hit maps - filled with entries per fibre, 
    \item not summing up the hit maps layer-wise,
    \item efficiency correction applied to: the \gls{sm},
    \item background considered in: the MLEM formula.
\end{enumerate}

\begin{figure}[!htb]
\centering
\includegraphics[width = \textwidth]{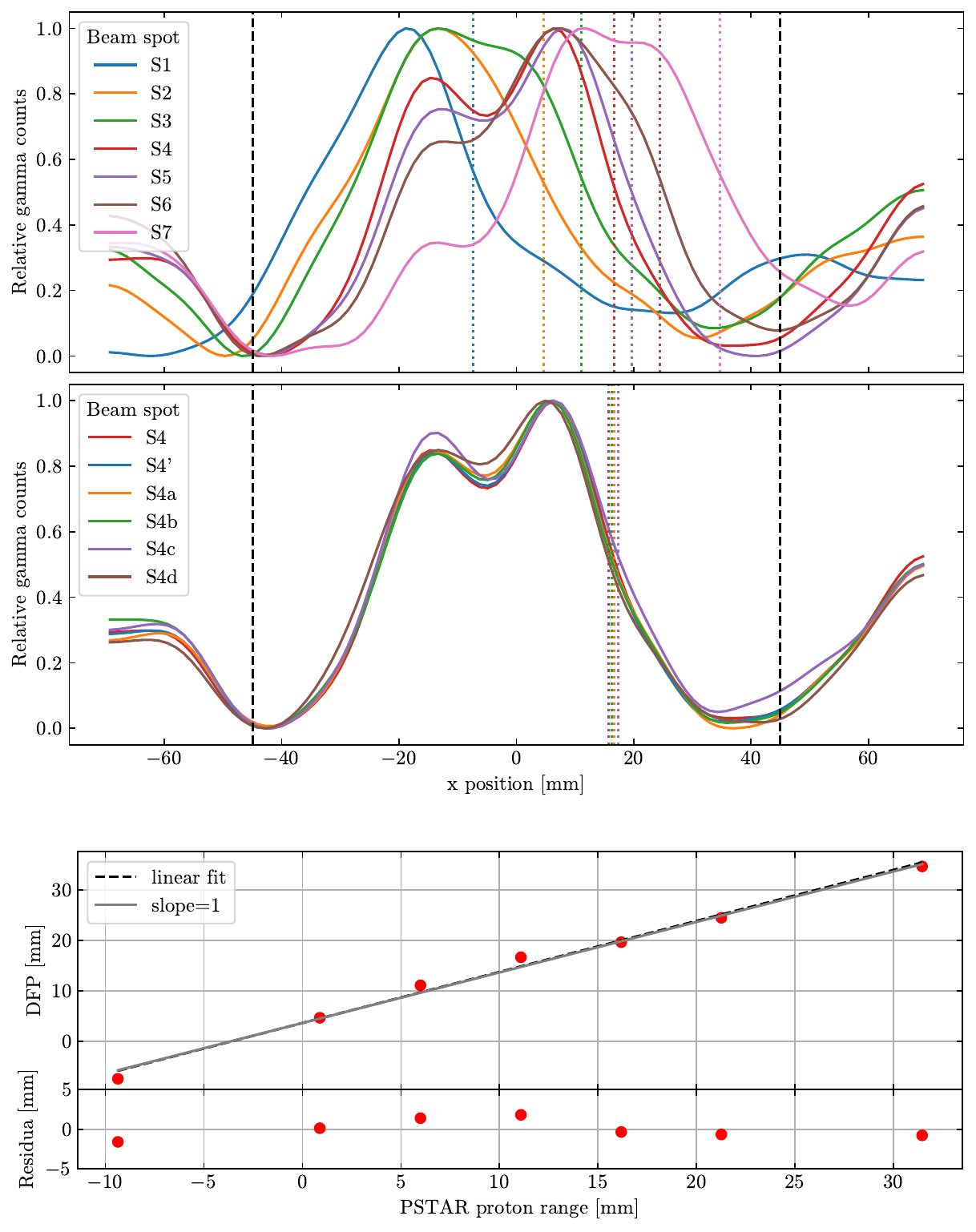}
\caption{Top: \gls{pgh} depth profiles for S1-S7, black dashed line marks the phantom borders and the dotted colour lines correspond to the \gls{dfp} positions determined for the corresponding profiles; middle: \gls{pgh} depth profiles for S4, S4', S4a-d; bottom: \gls{dfp} vs. PSTAR proton range, with a linear fit (dashed line) and a reference line with a slope equal to~1 (solid line).}
\label{fig:profiles_thr1000}
\end{figure}

The values of the performance metrics obtained for this configuration of the image reconstruction are summarised in~\cref{tab:performanceMetrics}. 
\begin{table}[!htb]
\centering
\caption{The values of the performance metrics, obtained for the optimal set of image reconstruction parameters. Full simulation is the simulation with all \glspl{sipm} active, i.e. without the acceptance gaps (see~\cref{sec:unfiltered_simulation}). The number of iterations of image reconstruction is given in the last row.}
\label{tab:performanceMetrics}   
    \begin{tabular}{cccc}\toprule
        ~ & Experiment & Simulation & Full simulation \\ \midrule \rowcolor[gray]{.95}
        $PCC$ & $0.996$ & $0.9981$ & $0.9998$ \\ 
        $R^2$ & $0.902$ & $0.992$ & $0.992$ \\ \rowcolor[gray]{.95}
        $RMSE$~[mm] & $1.7$ & $1.6$ & $0.4$ \\ 
        $SD$~[mm] & $0.48$ & - & - \\ \rowcolor[gray]{.95}
        Slope $s$ & $1.0102(16)$ & $0.94351(68)$ & $0.999981(92)$ \\
        Intercept~[mm] & $3.63(45)$ & $1.01(19)$ & $-1.070(25)$ \\ \hline 
        Number of iterations & 23 & 23 & 600 \\ \bottomrule
    \end{tabular}
\end{table}
The profiles for the beam spots S1-S7 are presented in~\cref{fig:profiles_thr1000}~(top). There, one can see that the distal edge shifts with the beam energy, as expected. The dependence between the distal falloff positions and the corresponding proton ranges is illustrated in~\cref{fig:profiles_thr1000}~(bottom), along with a linear fit and a residual plot. In the middle plot, six profiles for the same depth coordinate are presented. One can observe that the shape of the profiles is consistent, the reproducibility of the profile shape and lack of its dependence from the beam lateral position within the range of $\pm$\SI{1}{cm}. The variation of the obtained profiles here is quantified by the $SD$ parameter (described in~\cref{sec:performanceMetrics}). Apart from the desired behaviour of the distal edge, the rest of the profile presents some unfavourable effects. Although the profile widens with beam energy, as expected, additional wide peaks appear in the profile structure. Moreover, the reconstructed gamma intensity is too small in the entrance region: we would expect to see the rising edge of the profile around the phantom border and then a slow increase in the middle region of the profile (see, e.g., the profile in~\cref{fig:PromptGammaVsProtonRange}). There are also some artefacts in the regions outside the phantom: the relative gamma count increases towards the image edges, even though we do not expect any events there. Despite all the described flaws of the profiles, we conclude that the image reconstruction performs very well, as indicated by the parameter metrics: the $RMSE$ yields \SI{1.7}{\mm}. This metric defines the image reconstruction accuracy. The reconstructed \glspl{dfp} are strongly correlated with the calculated proton ranges, as indicated by high values of both the $PCC=0.996$ and $R^2=0.902$. The slope of the \gls{dfp}-proton range dependence is close to unity: 1.0102(16), the intercept is positive: \SI{3.63(45)}{\mm}. The standard deviation of the same-depth profiles' \gls{dfp} equals \SI{0.48}{\mm}. 

\section{Analysis of statistical precision}
\begin{figure}[!htb]
\centering
\includegraphics[width = \textwidth]{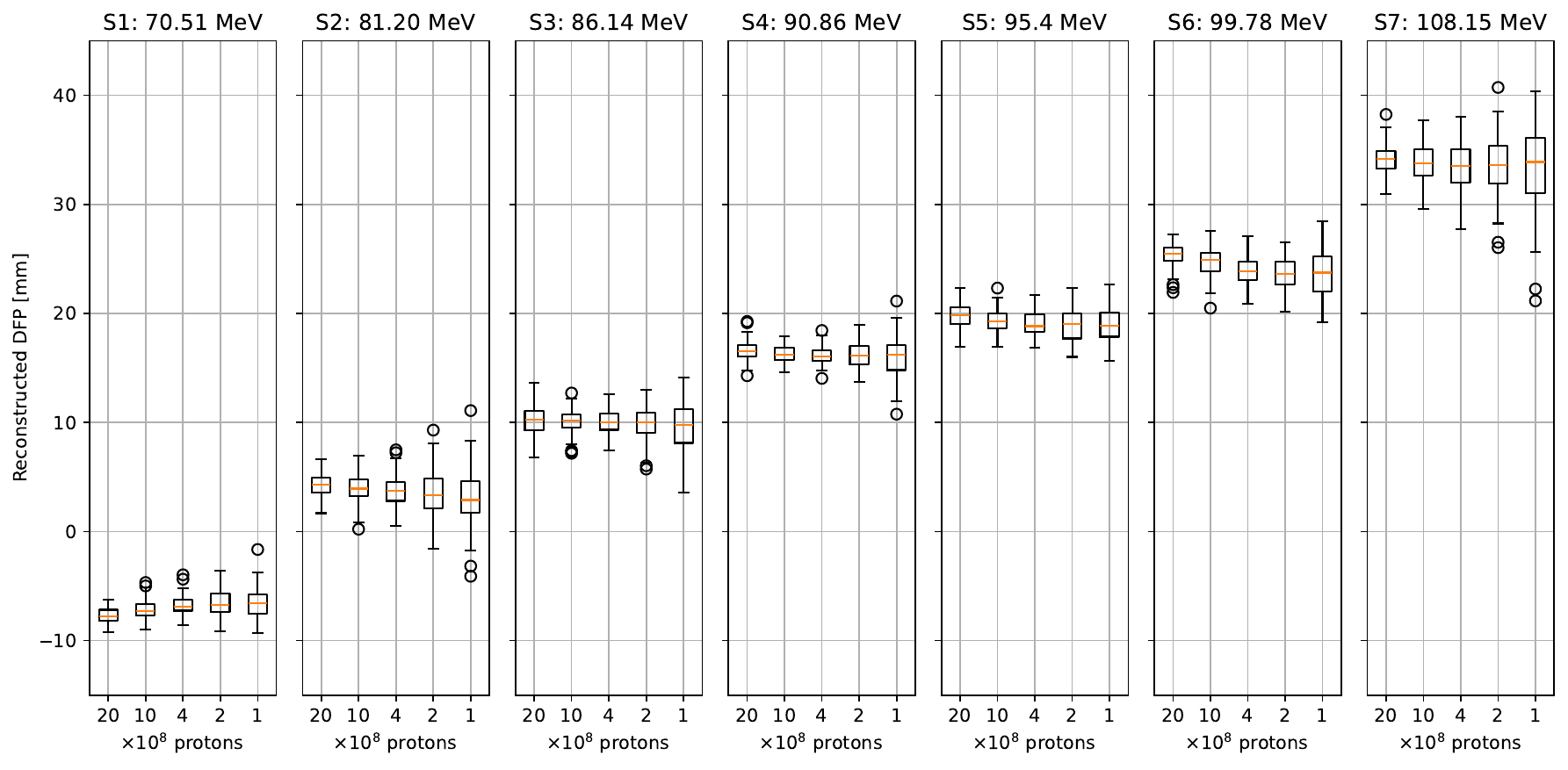}
\caption{Reconstructed \gls{dfp} vs. sample size for beam spots S1-S7.}
\label{fig:statisticalAnalysis}
\end{figure}

For each of 100 subsets of the experimental data (created for each investigated sample size, according to the procedure described in~\cref{sec:analysis_statisticalPrecision}), image reconstruction was performed. The configuration of the image reconstruction was the same as for the image reconstruction of the full-size data samples. We summarise the results of these image reconstructions in a box plot, presented in~\cref{fig:statisticalAnalysis} for beam spots S1-S7 (arranged left to right). In each panel, the distribution of the reconstructed \gls{dfp} values is presented for each sample size. The rectangle (box) marks the interquartile range (IQR) between the quartiles Q1
and Q3. The orange line on the rectangle denotes the median. The whiskers extend to the furthest data
point that falls into a range of $1.5\times$IQR, starting from the box edge. Outliers are marked with empty circles.
There, one can see that the trend is, with some exceptions, that the smaller the data sample, the larger the IQR, which is expected, since the image reconstruction suffers from the statistical fluctuations in the input data. The IQR for beam spots S1-S6 is below \SI{3}{mm}, while for the deepest spot S7 it is larger, up to about \SI{5}{mm} for the lowest investigated statistics. Another observed effect is the drift of median with sample size, while we would expect it to be constant. The magnitude of the effect is between \SI{0.46}{mm} (S3) and \SI{1.8}{mm} (S6). No clear pattern is observed regarding the decrease or increase of the median value with sample size. 

\begin{figure}[!htb]
\centering
\includegraphics[width =0.7\textwidth]{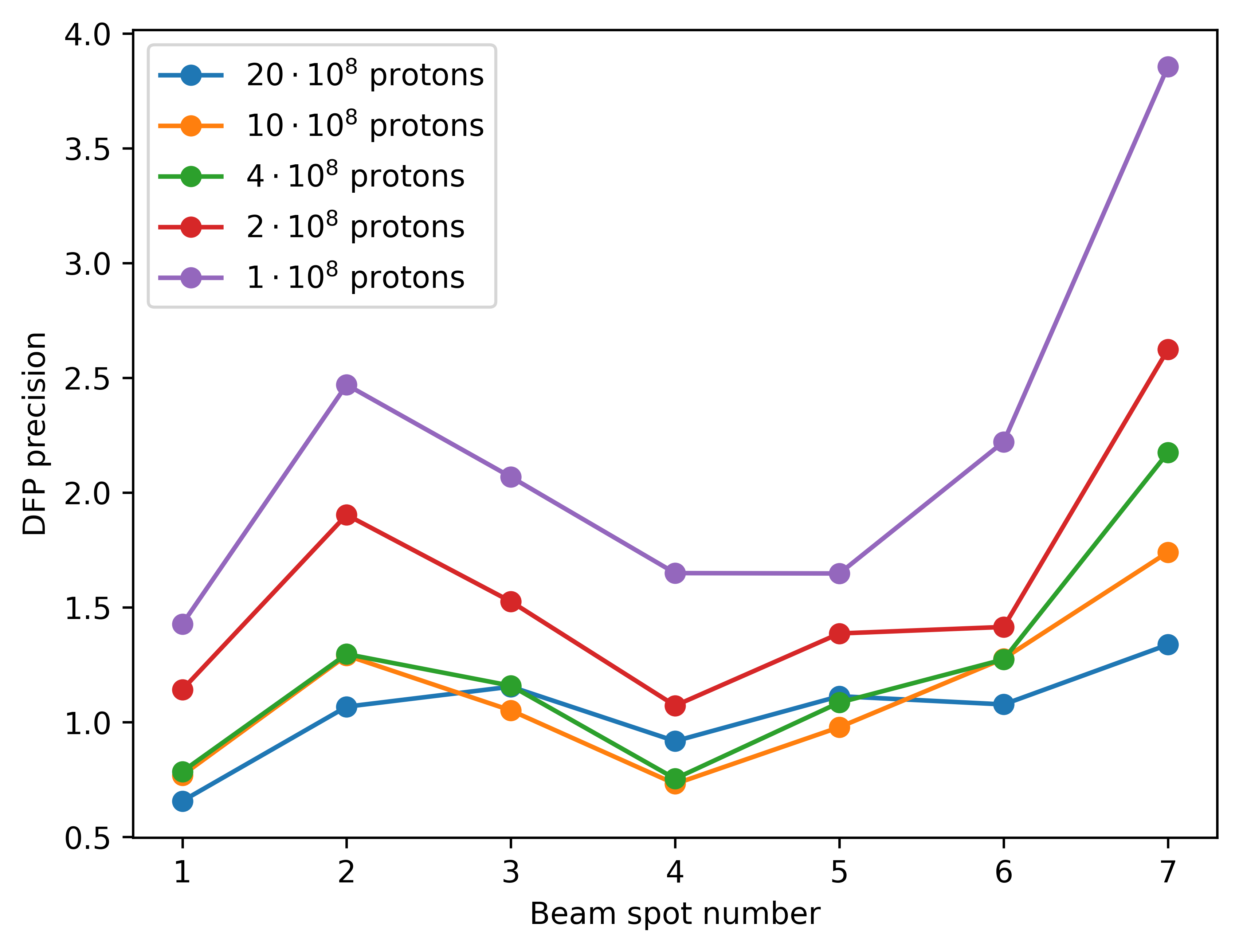}
\caption{\gls{dfp} precision ($\sigma_\mathrm{DFP}$) vs. number of the beam spot. Different line colours correspond to different statistics of data used as input (see legend).}
\label{fig:DFPP}
\end{figure}

Another, more popular in the literature, measure of the image reconstruction quality is the standard deviation of the obtained \glspl{dfp}: $\sigma_\mathrm{DFP}$. The standard deviations for all beam spots and examined sample sizes are presented in~\cref{fig:DFPP}. There, one can see that $\sigma_\mathrm{DFP}$ depends not only on the sample statistics, but also on the Bragg peak position in the phantom. One can observe a consistent pattern across all the sample sizes: the $\sigma_\mathrm{DFP}$ is the lowest in spots S1 and S4; in spots S2, S3, S5 and S6, the $\sigma_\mathrm{DFP}$ is higher, in the range of $1-$\SI{2.5}{mm}, and the highest $\sigma_\mathrm{DFP}$ is observed for spot S7 (the deepest one). A possible explanation for such behaviour can be the irregularity of the collimator pattern. The broader the slits in certain collimator region, the worse the imaging resolution would be in the corresponding detector region. This is the possible cause for the worse performance in the last spot. Moreover, the neutron background is larger at higher energies, which could also contribute to the deterioration of the achieved \gls{dfp} precision. 

\section{Simulation vs. experiment}\label{sec:simulationVsExperiment}
\begin{figure}[!htb]
\centering
\includegraphics[width = \textwidth]{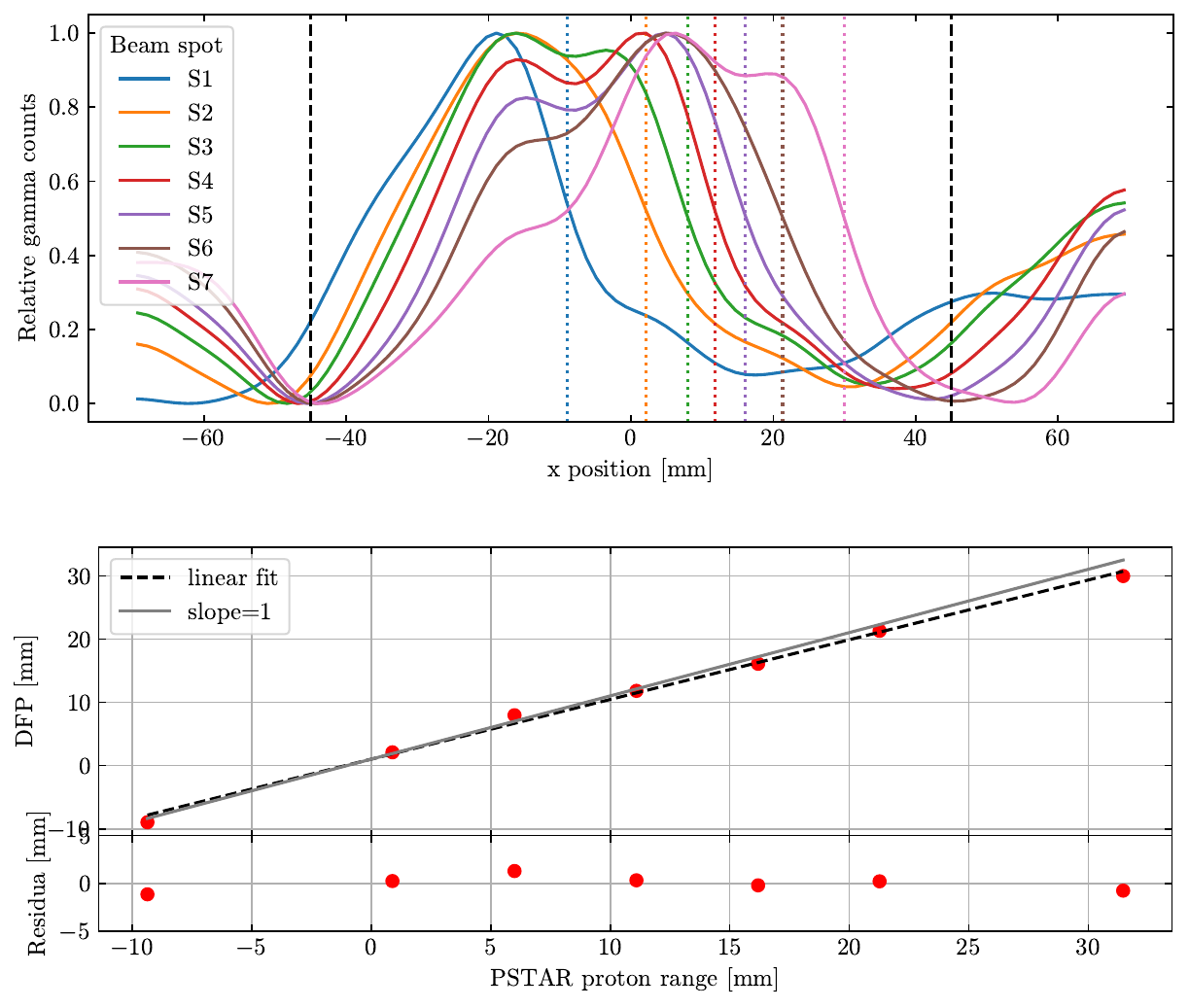}
\caption{Top: Simulated \gls{pgh} depth profiles for S1-S7, colour coding the same as in~\cref{fig:profiles_thr1000}; bottom: simulated \gls{dfp} vs. PSTAR proton range,  with a linear fit (dashed line) and a reference line with a slope equal to~1 (solid line).}
\label{fig:profiles_sim}
\end{figure}

To verify the results of the experimental detector tests, data from a realistic simulation reproducing the experimental setup were prepared. The data files from the simulation were processed in the same way as the experimental ones and the image reconstruction was performed on both data sets with the same parameters (found in a process of optimisation on experimental data). The resulting \gls{pgh} depth profiles are presented in~\cref{fig:profiles_sim}, and the obtained values of performance metrics are summarised in~\cref{tab:performanceMetrics}. The $RMSE$ for the simulation data is \SI{1.6}{\mm}, slightly better than for the experimental data (\SI{1.7}{mm}). Both the $PCC$ and $R^2$ were also better (0.9981 and 0.992, respectively), but the slope was further away from 1. The intercept lowered with respect to the experimental data to \SI{1.01(19)}{mm}. The standard deviation of the same beam spot profiles was not determined, as only one data set was simulated per beam spot. The simulated profiles presented in~\cref{fig:profiles_sim} are very similar to the experimental ones, all the effects present in the experimental results are also observed in the simulation: the change of the profile shape with beam energy, the wide peaks in the profiles, the artefacts in the regions outside of the phantom. Based on these observations and on the similarity of the performance metrics, we can conclude that the simulation is accurate and correctly predicts all major effects observed in the analysis of experimental data.

\section{Non-filtered simulation}\label{sec:unfiltered_simulation}
\begin{figure}[!htb]
\centering
\includegraphics[width = \textwidth]{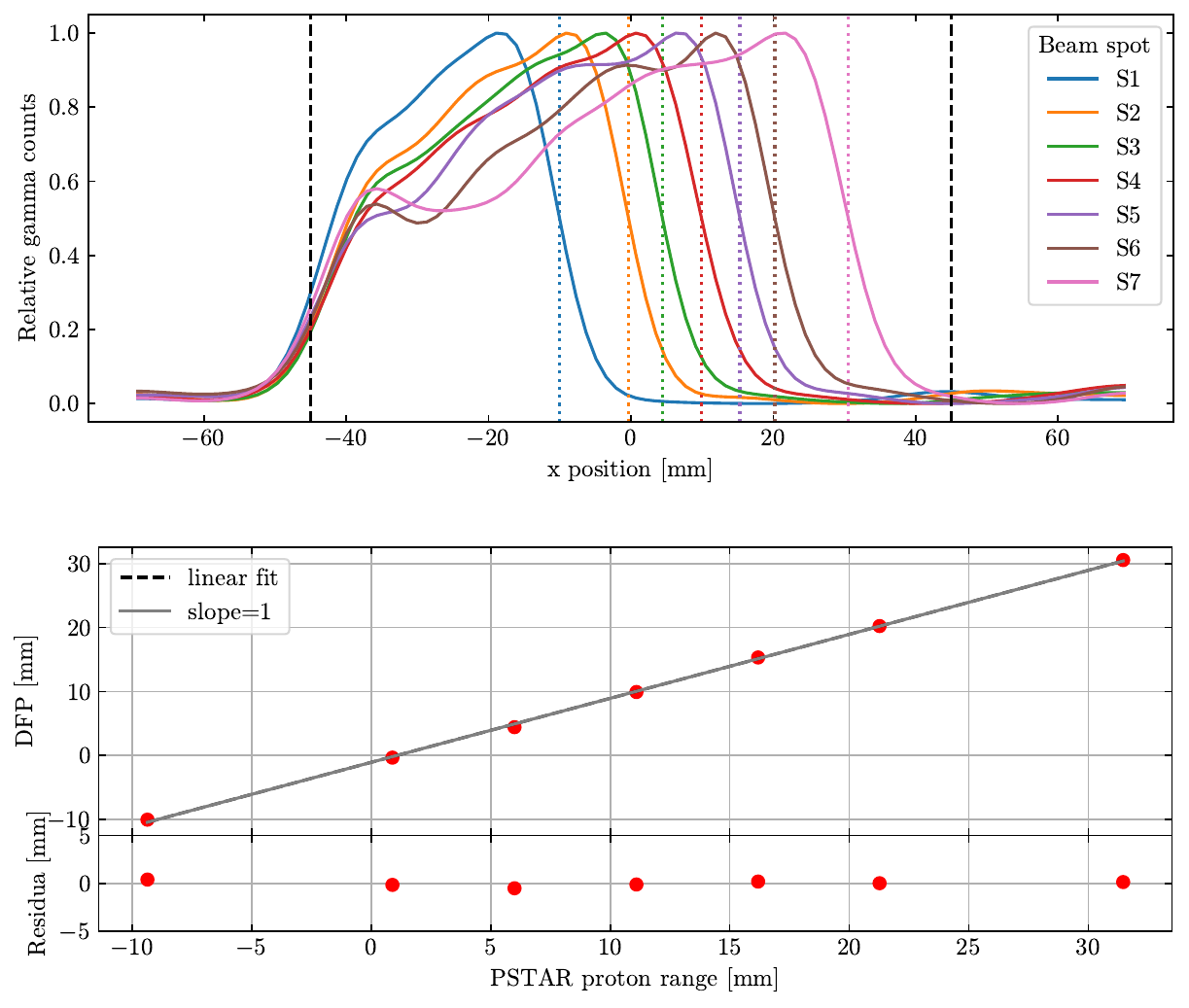}
\caption{Same as in~\cref{fig:profiles_sim}, but the simulation results were not filtered.}
\label{fig:profiles_sim_nonFiltered}
\end{figure}

An additional study was performed to assess the influence of the acceptance gaps on the image reconstruction performance: the full simulated data set without accounting for acceptance gaps (referred to as non-filtered) was processed in the same way as the experimental data, with the same software configuration. Only the number of iterations was re-optimised and the new optimum was found to be much higher: 600 iterations. The resulting \gls{pgh} depth profiles are presented in~\cref{fig:profiles_sim_nonFiltered}. The profiles are of much better shape than the ones reconstructed from data with acceptance gaps, both simulated and experimental. Firstly, the profiles start earlier: the rising edges are located at the phantom proximal border, as expected, and do not shift with the beam energy, which is also physically correct, as the gammas are emitted along the whole proton path, regardless of the range.
No artefacts are present in the regions outside the phantom. Moreover, the performance metrics are better than for the filtered case (see~\cref{tab:performanceMetrics}), with $RMSE$ of \SI{0.4}{\mm}, $PCC$ of 0.9998, slope $s$ of 0.999981(92) (equal to 1 within uncertainty) and slightly negative intercept of \SI{-1.070(25)}{\mm}.

The increase of the iteration number for the non-filtered simulation was only possible, because there were no distortions or wide peaks in the profiles. These distortions were present in both the experimental data and the filtered simulation, and were amplified with each iteration, leading to the deterioration of the image reconstruction performance. For completeness, in the appendix~\ref{appendix} we present the image reconstruction after 600 iterations for the experimental data in~\cref{fig:profiles_iter600} and for simulated data with acceptance gaps in~\cref{fig:profiles_sim_iter600}. In both cases, one can see the amplified artefacts in the \gls{pgh} depth profiles. Moreover, one can observe that the residua associated with the linear fit to the dependence of \glspl{dfp} vs. proton ranges are larger than in~\cref{fig:profiles_thr1000} (where the number of iterations was optimal), which is a sign of a poorer image reconstruction performance. The values of performance metrics associated with the presented reconstructed images were also consistently poorer than for the optimal number of iterations.
We also present the full simulation profile after 23 iterations (the number optimal for data with acceptance gaps) in~\cref{fig:profiles_unfiltered_iter23}. There, one can see that the profiles are smooth, but their shape is not yet optimal, definitely more iterations are needed.

The comparison of the image reconstruction performance for filtered and non-filtered data leads us to the conclusion that all the unfavourable effects in the experimental profiles, along with the poorer values of the performance metrics, are likely to have their origin in the presence of the acceptance gaps. 

\chapter{Discussion and outlook}
\section{Detector performance}
Based on the rate capability study described in~\cref{sec:results_rateCapability}, we conclude that the detector is operable rate-wise under synchrotron conditions. The channel occupation at maximum possible intensity of the synchrotron beam was 18 times lower than the \gls{daq} limit. Therefore, it can be assumed that the detector can operate also in cyclotron conditions, where the beam intensities, and thus also the detector rates, are about 10 times higher than at a synchrotron~\cite{Wronska2021}.

In the analysis of the \glspl{dfp} of the profiles obtained for different lateral positions of beam spots (S4, S4', S4a-d, see~\cref{fig:profiles_thr1000}, middle panel), we obtained the \glspl{dfp} standard deviation of \SI{0.48}{mm}, which is below the $\sigma_{\mathrm{DFP}}$ obtained for the nearest investigated statistics corresponding to \SI{e9}{protons}. Thus, we confirm that the detector is not sensitive to changes in the lateral spot position of the beam by~$\pm$\SI{1}{cm}.

The main flaw in the experimental setup, which limited the detector performance, was the presence of acceptance gaps due to several non-working \glspl{sipm} (the issue was described in~\cref{sec:deadSiPMsIssue}). When we repeated the analysis and image reconstruction on a simulated data set with all \glspl{sipm} working (\cref{sec:unfiltered_simulation}), we obtained much better profile shapes, as well as the performance metrics. Notably, $RMSE$ decreased from \SI{1.6}{\mm} (simulation with acceptance gaps) to \SI{0.4}{\mm}, suggesting that eliminating acceptance gaps could improve detector precision in terms of range shift determination up to 4 times. This number has to be taken with caution, because we reason based on the simulated data. We conclude that in the next detector generation particular care must be taken to ensure that all detector pixels are active, as it turned out to be crucial for the detector performance.

\section{PG depth profiles - experimental and simulated}
The simulated profiles are very similar to the experimental ones, both qualitatively, by looking at the profiles' shapes, and quantitatively, by comparing the performance metrics. Particularly, similar $RMSE$ values (\SI{1.7}{\mm} for the experimental data and \SI{1.6}{\mm} for the simulated data) were obtained. We conclude that our simulation realistically and accurately reproduces the \gls{1dcm} experiment. 

\section{Clinical feasibility, significance}
The study described in this thesis is the first experimental test of a coded-mask gamma camera used for beam range verification in proton therapy in clinical conditions. Prior to this test, two articles were published on the coded-mask approach to proton beam monitoring:~\cite{Sun2020a} (presenting simulation data) and~\cite{Hetzel2023} (presenting simulation data and the results of laboratory tests), the latter published by our group. The achieved precisions declared in these articles for the statistics of $10^8$ protons were, respectively, \SI{2.1}{mm} (for the lowest tested beam energy of \SI{122.7}{MeV}) and \SI{0.72}{mm} (averaged over several beam spots in the energy range of 85.9-\SI{107.9}{MeV}). 

There are also several solutions comparable to ours that were investigated experimentally or by means of simulations: knife-edge slit (KES) systems~\cite{Xie2017,Berthold2021}, and multi-parallel slit (MPS) systems~\cite{Pinto2014,Ku2023}. Of these two approaches, the KES systems are the most mature, with a record of in-human tests~\cite{Berthold2021,Bertschi2023}. We compare the beam range reconstruction precision ($\sigma_\mathrm{DFP}$) obtained by these systems in the conditions as close as possible to the conditions of our experiment. In~\cite{Xie2017}, a precision of \SI{2}{mm} at a beam energy of \SI{160}{MeV} was obtained (there, the statistics was slightly larger: $1.4\times 10^8$ protons). The other clinical solution~\cite{Berthold2021} provides also \SI{2}{mm} precision for aggregated spots, with an energy range of 100-\SI{160}{MeV}, for $10^8$ protons. In a simulation study~\cite{Pinto2014}, the precision is 1.30-\SI{1.66}{mm}, depending on the camera geometry settings. This was verified for the energy \SI{160}{MeV} and the standard statistics of $10^8$ protons. The experimental study of a MPS setup by another group~\cite{Ku2023} yielded a precision of \SI{1.5}{mm} at \SI{99.68}{MeV} with $10^8$ protons. Our result of \SI{1.7}{mm} precision with $10^8$ protons and \SI{90.86}{MeV} is comparable to the precisions obtained by the presented detectors. However, removing the acceptance gaps is expected to significantly improve our detector's performance, yielding better precision than any of the presented systems.

\section{Prospects of moving to 2D~CM imaging}
The setup can be extended to enable 2D imaging, provided that the method of determining the position in the dimension along the fibre is improved. Currently, it is not sufficient, due to reasons described in~\cref{sec:Experiments_Calibration}. The planned modification of the active detector part in the next detector iteration is increasing the attenuation along the fibre, which is expected to enhance the position resolution. 

\chapter{Summary and conclusions}
In this thesis, a process of construction of a coded-mask gamma camera and the results of its first experimental test under proton beam conditions are presented. 

Chapter 1 provides the theoretical background and the motivation for building a coded-mask gamma camera as a solution for proton beam range verification in proton therapy. 

In Chapter 2, we characterise the SiFi-CC project and discuss how it addresses the issues observed in prompt-gamma-based monitoring of proton therapy; we also describe the process of optimising setup components, particularly the 
\gls{daq} system. We conclude that the optimal \gls{daq} system among the tested options is the TOFPET2c and should therefore be used in the experimental setup. We also outline the previous steps that led to the development of the current version of the detector, including a small-scale prototype and a pilot CM study with radioactive sources. 

In Chapter 3, the detector setup and experimental conditions for studying the CM gamma camera with a proton beam are presented. The test was carried out under clinical conditions at HIT. During the test, a \gls{pmma} phantom was irradiated with proton beams of energies ranging from \SI{70.15}{MeV} to \SI{108.15}{MeV} and of three beam intensities of $8\times10^7$, $6\times10^8$, and $3.2\times10^9$~protons/s. The \glspl{pgh} emitted from the phantom were registered with the scintillation-fibre-based CM gamma camera operating in 1D-imaging modality.

Chapter 4 contains the description of data analysis methods and algorithms used (e.g. \gls{llr}, MLEM, \gls{dfp} determination), along with a summary of several auxiliary studies aimed to better understand the data and the detector low-level performance. Also, the configuration of the simulation application is provided in this part.

In Chapter 5, the results obtained based on both experimental and simulation data are confronted, followed by their detailed discussion in Chapter 6. The CM gamma camera in the 1D modality achieved the precision of the \gls{dfp} determination of \SI{1.7}{\mm} with statistics of $10^8$ protons, which is comparable to the typical statistics per distal beam spot in clinical irradiation. The beam energy for this irradiation was \SI{90.68}{MeV}. The experimental results were shown to be consistent with those obtained for simulation data: the shape of the reconstructed profiles and the obtained performance metrics were very similar in both cases. In the experiment, the detector had acceptance gaps, which were also accounted for in the simulation by filtering out the dead pixels. However, the same simulation without the filtering (i.e., with all detector pixels active) proved to achieve 4 times better performance in terms of the range shift determination. Therefore, there is still space for improving the detector performance. In terms of hit rate, the setup proved to be compatible with synchrotron conditions, with the prospect of working also under the more demanding cyclotron conditions. The performance of the setup is already comparable to that of similar available solutions for \gls{pgh} proton therapy monitoring (KES, MPS), and it can be further improved if the conclusions from the present work are applied.

The main objective of this thesis was to demonstrate the feasibility of using a coded-mask gamma camera for proton therapy monitoring. The presented research constitutes a proof-of-principle for this proposed solution and thus provides a basis for a novel method of monitoring the beam range with potential for clinical application.

\appendix
\cleardoublepage
\chapter{Appendix}
\label{appendix}

\begin{table}[!ht]
\caption{List of dead SiPMs. \newline \label{tab:deadSiPMs}}

    \centering
    \begin{tabular}{cc}\toprule
        Dead SiPM IDs & Side \\ \midrule \rowcolor[gray]{.95}
        $34$ & bottom \\ 
        $46$ & ~ \\ \rowcolor[gray]{.95} \hline
        $462$ & top \\ 
        $426$ & ~ \\ \rowcolor[gray]{.95}
        $437$ & ~ \\ 
        $397$ & ~ \\ \rowcolor[gray]{.95}
        $398$ & ~ \\ 
        $409$ & ~ \\ \rowcolor[gray]{.95}
        $420$ & ~ \\ 
        $421$ & ~ \\ \rowcolor[gray]{.95}
        $422$ & ~ \\ 
        $393$ & ~ \\ \bottomrule 
    \end{tabular}
\end{table}

\begin{table}[!ht]
\caption{List of dead fibers, i.e. the fibers that were connected to dead SiPMs.\newline \label{tab:deadFibers}}
    \centering
    \begin{tabular}{cc|cc}\toprule
    Layer ID & Fiber ID & Layer ID & Fiber ID\\ \midrule \rowcolor[gray]{.95}
       $0$	&	$50$	&	$2$	&	$51$   \\  
       $0$	&	$51$	&	$2$	&	$52$   \\ \rowcolor[gray]{.95}
       $0$	&	$52$	&	$2$	&	$53$   \\  
       $0$	&	$53$	&	$3$	&	$2$    \\ \rowcolor[gray]{.95}
       $1$	&	$11$	&	$3$	&	$3$    \\ 
       $1$	&	$12$	&	$3$	&	$4$    \\  \rowcolor[gray]{.95}
       $1$	&	$35$	&	$3$	&	$5$    \\ 
       $1$	&	$36$	&	$3$	&	$26$   \\ \rowcolor[gray]{.95}
       $1$	&	$50$	&	$3$	&	$27$   \\ 
       $1$	&	$51$	&	$3$	&	$48$   \\ \rowcolor[gray]{.95}
       $1$	&	$52$	&	$3$	&	$49$   \\ 
       $1$	&	$53$	&	$3$	&	$50$   \\ \rowcolor[gray]{.95}
       $2$	&	$2$	&	$3$	&	$51$   \\ 
       $2$	&	$3$	&	$3$	&	$52$   \\ \rowcolor[gray]{.95}
       $2$	&	$4$	&	$3$	&	$53$   \\ 
       $2$	&	$5$	&	$4$	&	$4$    \\ \rowcolor[gray]{.95}
       $2$	&	$11$	&	$4$	&	$5$    \\ 
       $2$	&	$12$	&	$4$	&	$26$   \\ \rowcolor[gray]{.95}
       $2$	&	$26$	&	$4$	&	$27$   \\
       $2$	&	$27$	&	$5$	&	$4$    \\ \rowcolor[gray]{.95}
       $2$	&	$35$	&	$5$	&	$5$    \\ 
       $2$	&	$36$	&	$5$	&	$26$   \\ \rowcolor[gray]{.95}
       $2$	&	$48$	&	$5$	&	$27$   \\ 
       $2$	&	$49$	&	$6$	&	$20$   \\ \rowcolor[gray]{.95}
       $2$	&	$50$	&	$6$	&	$21$   \\ \bottomrule

    \end{tabular}
\end{table}

\begin{table}[!ht]
\caption{DAQ parameters.\newline  \label{tab:DAQparameters}}
    \centering
    \begin{tabular}{rl} \toprule
        General DAQ settings & ~ \\ \midrule \rowcolor[gray]{.95}
        Overvoltage & 8 V \\ 
        Preset time & 30 s \\ \rowcolor[gray]{.95}
        Pre BDV & 30 V \\
        BDV & 33 V \\ \rowcolor[gray]{.95}
        Threshold T1 & 20 \\ 
        Threshold T2 & 20 \\ \rowcolor[gray]{.95}
        Threshold E & 22 \\ \hline
        Hardware trigger settings & ~ \\ \rowcolor[gray]{.95} \midrule 
        type & builtin \\ 
        threshold & 1 \\ \rowcolor[gray]{.95}
        pre\_window & 3 \\
        post\_window & 15 \\ \rowcolor[gray]{.95}
        coincidence\_window & 3 \\ 
        single\_acceptance\_period & 1000 \\ \rowcolor[gray]{.95}
        single\_acceptance\_length & 0 \\ 
        global.disc\_lsb\_T1 & 60 \\ \bottomrule 
    \end{tabular}
\end{table}

\begin{table}[] 
\caption{List of runs for rate capability assessment. I - intensity; $N_{\mathrm{total}} = \Sigma_i N_{i~\mathrm{total}}$ - total counts in the spill regions; $N_{\mathrm{background~calc.}}$ - background contribution to the spill regions, calculated according to~\cref{eq:determineNspills} in~\cref{sec:timePreselection}; $N_{\mathrm{spills}}=N_{\mathrm{total}}-N_{\mathrm{background~calc.}}$ - counts associated with beam spills. \newline \label{Tab:rateCapability_listOfRuns}}
\centering
\begin{tabular}{cccccc}\toprule
Run ID & I [protons/s] & E [MeV] & $\Sigma_i N_{i~\mathrm{total}}$ & $N_{\mathrm{background~calc.}}$ &  $N_{\mathrm{spills}}$\\ \midrule \rowcolor[gray]{.95}
$496$   & $3.2\times10^9$             & 108.15          & \num{30201200}     & \num{844945}     & \num{29356300}$\pm$\num{5500}                   \\
495   & $6\times10^8$              & 108.15          & \num{36139700}      & \num{3865110}    & \num{32274600}$\pm$\num{6300}                   \\ \rowcolor[gray]{.95}
494   & $8\times10^7$              & 108.15          & \num{58006100}      & \num{26205400}   & \num{31800700}$\pm$\num{9600}                  \\
\hline
493   & $3.2\times10^9$             & 90.86          & \num{23217000}      & \num{730090}     & \num{22486900}$\pm$\num{4900}                   \\\rowcolor[gray]{.95}
492   & $6\times10^8$              & 90.86          & \num{27696600}      & \num{3461950}    & \num{24234700}$\pm$\num{5600}                   \\
$491$   & $8\times10^7$              & 90.86          & \num{48503600}      & \num{24504200}   & \num{23999400}$\pm$\num{9000}       \\ \bottomrule
\end{tabular}

\end{table}

\begin{table}[!ht]
\caption{List of runs for rate overvoltage scan.\newline \label{Tab:overvoltageScan_listOfRuns}}
    \centering
    \begin{tabular}{cc}\toprule
        Run ID & Overvoltage [V] \\ \midrule \rowcolor[gray]{.95}
        $552$ & $4$ \\ 
        $554$ & $6$ \\ \rowcolor[gray]{.95}
        $551$ & $8$ \\ 
        $553$ & $12$ \\ \rowcolor[gray]{.95}
        $556$ & $14$ \\  \bottomrule 
    \end{tabular}
    
\end{table}

\begin{figure}[!htb]
\centering
\includegraphics[width = \textwidth]{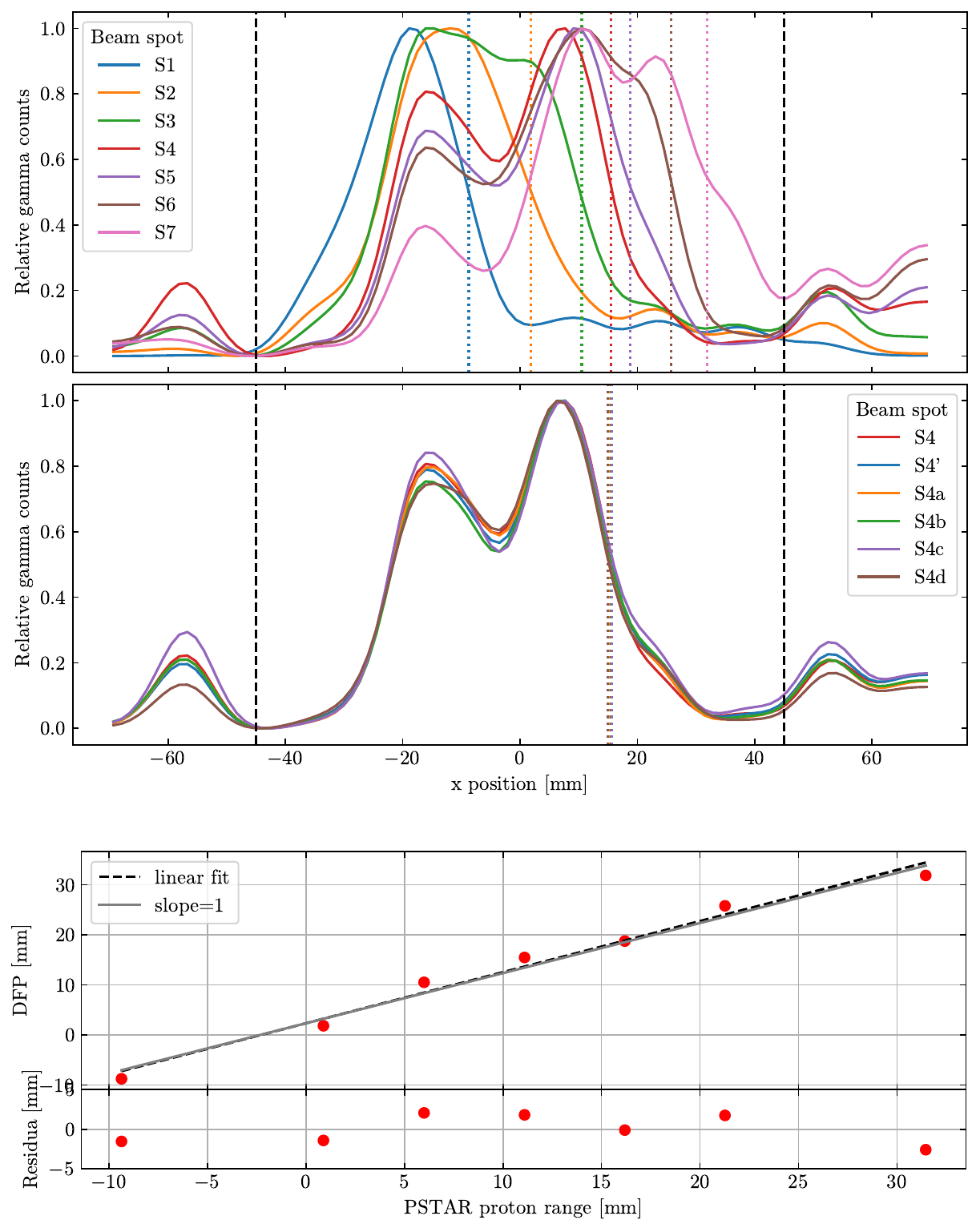}
\caption{Same as in~\cref{fig:profiles_thr1000}, but the number of iterations is 600.}
\label{fig:profiles_iter600}
\end{figure}

\begin{figure}[!htb]
\centering
\includegraphics[width = \textwidth]{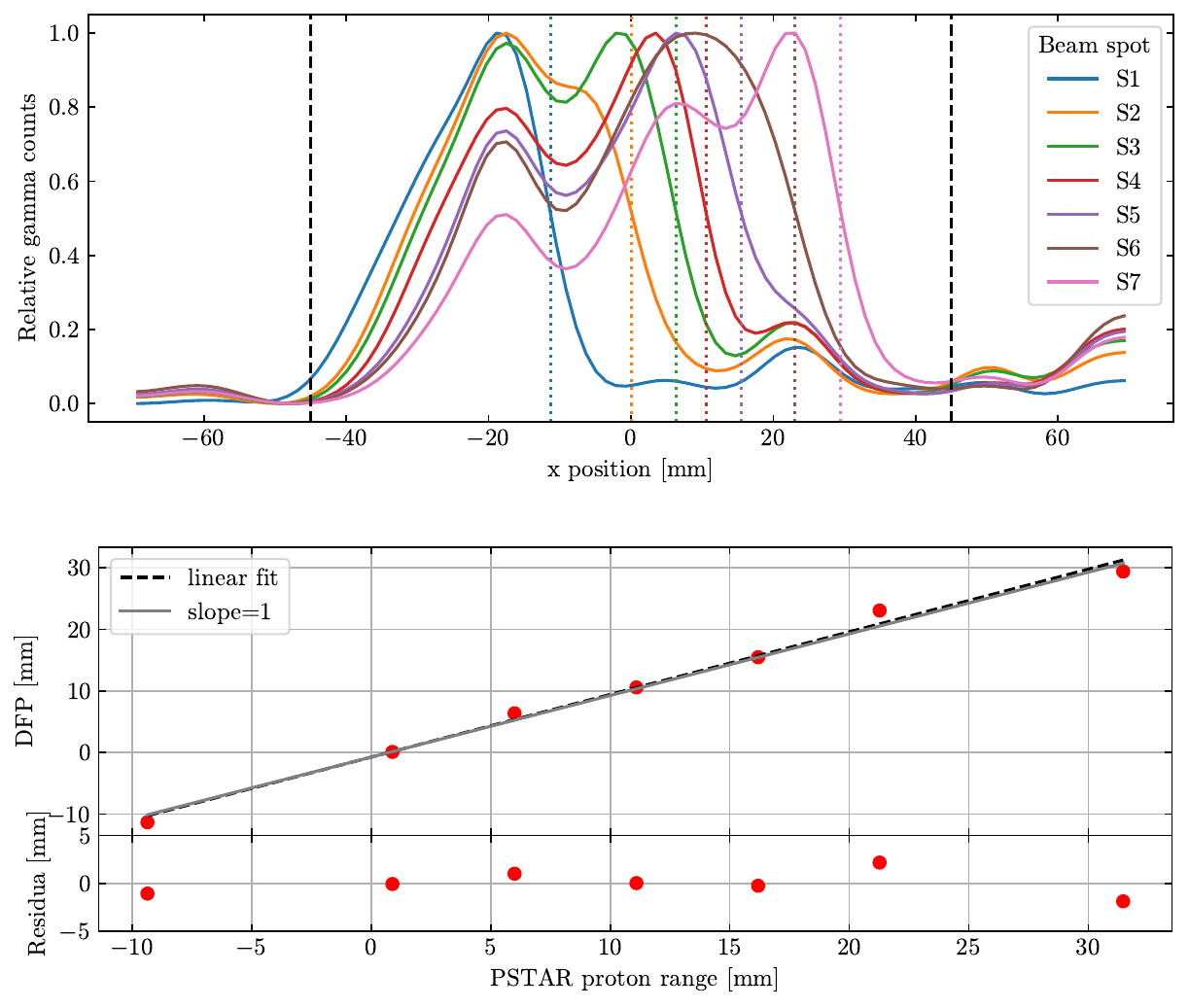}
\caption{Same as in~\cref{fig:profiles_sim}, but the number of iterations is 600.}
\label{fig:profiles_sim_iter600}
\end{figure}

\begin{figure}[!htb]
\centering
\includegraphics[width = \textwidth]{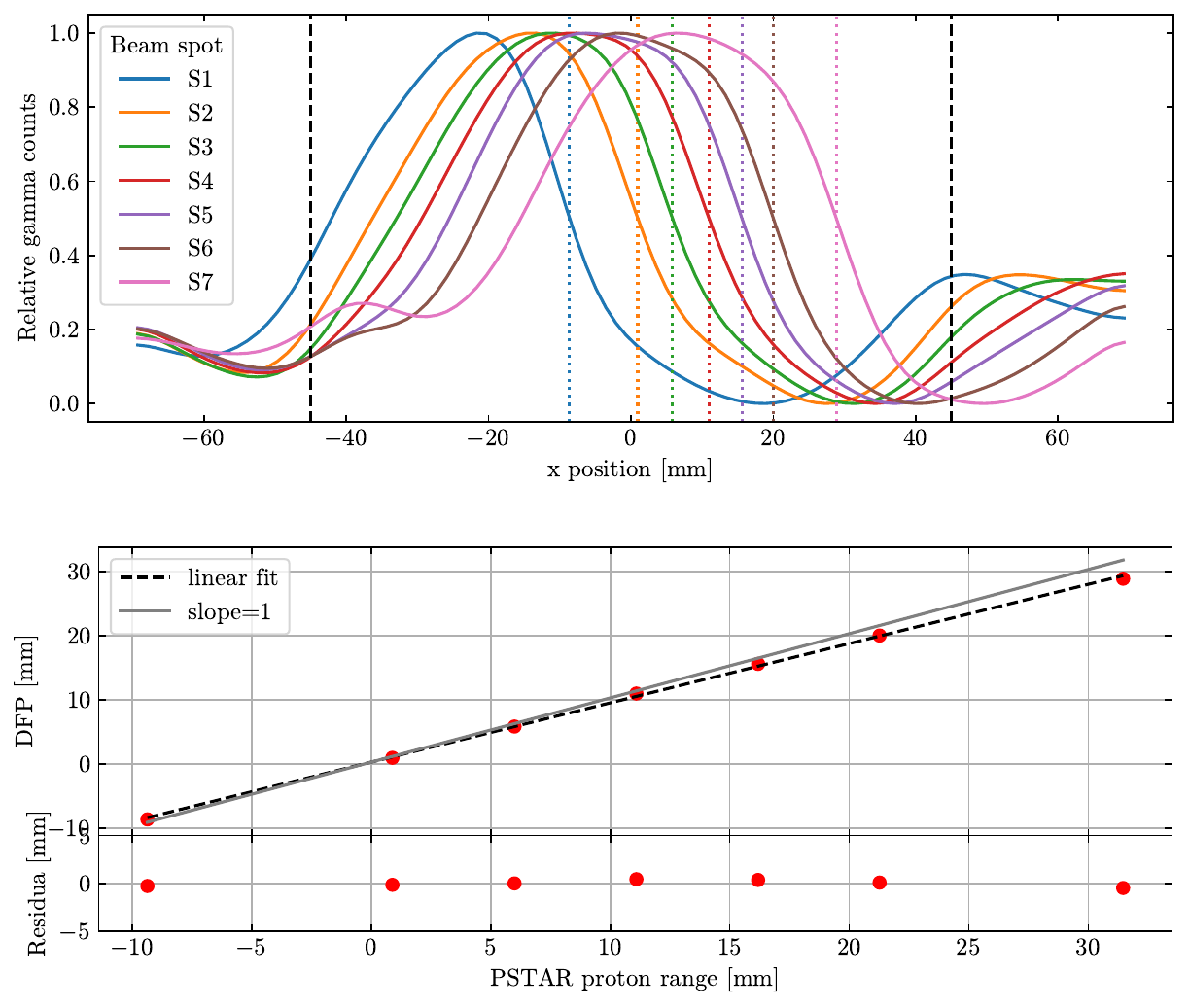}
\caption{Same as in~\cref{fig:profiles_sim_nonFiltered}, but the number of iterations is 23.}
\label{fig:profiles_unfiltered_iter23}
\end{figure}
\cleardoublepage
\manualmark
\markboth{\spacedlowsmallcaps{\bibname}}{\spacedlowsmallcaps{\bibname}} 
\refstepcounter{dummy}
\addtocontents{toc}{\protect\vspace{\beforebibskip}} 
\addcontentsline{toc}{chapter}{\tocEntry{\bibname}}
\label{app:bibliography}


\begin{thebibliography}{100}

\bibitem{Eurostat2021}
Eurostat.
\newblock {Causes of death statistics}.
\newblock
  \url{https://ec.europa.eu/eurostat/statistics-explained/index.php?title=Causes_of_death_statistics}
  (Accessed: 2024-31-12).

\bibitem{Siegel2021}
Rebecca~L. Siegel, Kimberly~D. Miller, Hannah~E. Fuchs, and Ahmedin Jemal.
\newblock Cancer statistics, 2021.
\newblock {\em CA: A Cancer Journal for Clinicians}, 71(1):7–33, January
  2021.

\bibitem{MaLomax2012}
C.-M.C. Ma and T.~Lomax.
\newblock Proton and carbon ion therapy.
\newblock CRC Press, 2012, \url{http://dx.doi.org/10.1201/b13070}.

\bibitem{Jones2016}
Bleddyn Jones.
\newblock Why {RBE} must be a variable and not a constant in proton therapy.
\newblock {\em The British Journal of Radiology}, 89(1063):20160116, July 2016.

\bibitem{Lapen2023}
Kaitlyn Lapen and Yoshiya Yamada.
\newblock The development of modern radiation therapy.
\newblock {\em Current Physical Medicine and Rehabilitation Reports},
  11(2):131–138, April 2023.

\bibitem{Paganetti2012}
H.~Paganetti.
\newblock {Proton Therapy Physics}.
\newblock CRC Press, 2012, \url{http://dx.doi.org/10.1201/b22053}.

\bibitem{Newhauser2015}
Wayne~D Newhauser and Rui Zhang.
\newblock The physics of proton therapy.
\newblock {\em Physics in Medicine and Biology}, 60(8):R155–R209, 2015.

\bibitem{Bethe1930}
H.~Bethe.
\newblock Zur theorie des durchgangs schneller korpuskularstrahlen durch
  materie.
\newblock {\em Annalen der Physik}, 397(3):325–400, January 1930.

\bibitem{rpp2022}
{Particle Data Group}.
\newblock The review of particle physics (2022), passage of particles through
  matter.
\newblock
  \url{https://pdg.lbl.gov/2022/reviews/rpp2022-rev-passage-particles-matter.pdf}
  (Accessed: 2024-31-12).

\bibitem{Paganetti2014}
Harald Paganetti.
\newblock Relative biological effectiveness ({RBE}) values for proton beam
  therapy. {V}ariations as a function of biological endpoint, dose, and linear
  energy transfer.
\newblock {\em Physics in Medicine and Biology}, 59(22):R419–R472, October
  2014.

\bibitem{Paganetti2015}
Harald Paganetti.
\newblock Relating proton treatments to photon treatments via the relative
  biological effectiveness—should we revise current clinical practice?
\newblock {\em International Journal of Radiation Oncology*Biology*Physics},
  91(5):892–894, April 2015.

\bibitem{Paganetti2019}
Harald Paganetti, Eleanor Blakely, Alejandro Carabe‐Fernandez, David~J.
  Carlson, Indra~J. Das, Lei Dong, David Grosshans, Kathryn~D. Held, Radhe
  Mohan, Vitali Moiseenko, Andrzej Niemierko, Robert~D. Stewart, and Henning
  Willers.
\newblock {Report of the AAPM TG-256 on the relative biological effectiveness
  of proton beams in radiation therapy}.
\newblock {\em Medical Physics}, 46(3), February 2019.

\bibitem{Lhr2018}
Armin L\"{u}hr, Cl\"{a}re von Neubeck, Mechthild Krause, and Esther~G.C.
  Troost.
\newblock Relative biological effectiveness in proton beam therapy – current
  knowledge and future challenges.
\newblock {\em Clinical and Translational Radiation Oncology}, 9:35–41,
  February 2018.

\bibitem{McMahon2018}
Stephen~Joseph McMahon.
\newblock The linear quadratic model: usage, interpretation and challenges.
\newblock {\em Physics in Medicine and Biology}, 64(1):01TR01, 2018.

\bibitem{Wilson1946}
Robert~R. Wilson.
\newblock Radiological use of fast protons.
\newblock {\em Radiology}, 47(5):487–491, November 1946.

\bibitem{Lawrence1958}
J.~H. Lawrence and C.~A. Tobias.
\newblock Pituitary irradiation with high-energy proton beams: a preliminary
  report.
\newblock {\em Cancer Research}, 18(2):121--134, February 1958.

\bibitem{Larsson1958}
B\"{o}rje Larsson, Lars Leksell, Bror Rexed, Patrick Sourander, William Mair,
  and Bengt Andersson.
\newblock {The High-Energy Proton Beam as a Neurosurgical Tool}.
\newblock {\em Nature}, 182(4644):1222–1223, November 1958.

\bibitem{Kjellberg1962}
R~N Kjellberg, W~H Sweet, W~M Preston, and A~M Koehler.
\newblock {The Bragg peak of a proton beam in intracranial therapy of tumors}.
\newblock {\em Transactions of the American Neurological Association (U.S.)},
  87, 1 1962.

\bibitem{PTCOGStats}
{Particle Therapy Co-Operative Group}.
\newblock Particle therapy facilities in clinical operation.
\newblock \url{https://www.ptcog.site/index.php/facilities-in-operation-public}
  (Accessed: 2024-31-12).

\bibitem{Koehler1975}
A.~Koehler, R.~Schneider, and J.~Sisterson.
\newblock Range modulators for protons and heavy ions.
\newblock {\em Nuclear Instruments and Methods}, 131(3):437–440, December
  1975.

\bibitem{Koehler1977}
A.~M. Koehler, R.~J. Schneider, and J.~M. Sisterson.
\newblock Flattening of proton dose distributions for large‐field
  radiotherapy.
\newblock {\em Medical Physics}, 4(4):297–301, July 1977.

\bibitem{Kanai1980}
Tatsuaki Kanai, Kiyomitsu Kawachi, Yoshikazu Kumamoto, Hirotsugu Ogawa,
  Takanobu Yamada, Hideo Matsuzawa, and Tetsuo Inada.
\newblock Spot scanning system for proton radiotherapy.
\newblock {\em Medical Physics}, 7(4):365–369, July 1980.

\bibitem{Pedroni1995}
Eros Pedroni, Reinhard Bacher, Hans Blattmann, Terence B\"{o}hringer, Adolf
  Coray, Antony Lomax, Shixiong Lin, Gudrun Munkel, Stefan Scheib, Uwe
  Schneider, and Alexander Tourovsky.
\newblock {The 200‐MeV proton therapy project at the Paul Scherrer Institute:
  Conceptual design and practical realization}.
\newblock {\em Medical Physics}, 22(1):37–53, January 1995.

\bibitem{Lin2021}
Binwei Lin, Feng Gao, Yiwei Yang, Dai Wu, Yu~Zhang, Gang Feng, Tangzhi Dai, and
  Xiaobo Du.
\newblock {FLASH Radiotherapy: History and Future}.
\newblock {\em Frontiers in Oncology}, 11, May 2021.

\bibitem{Daugherty2022}
E.C. Daugherty, A.E. Mascia, M.G.B. Sertorio, Y.~Zhang, E.~Lee, Z.~Xiao,
  J.~Speth, J.~Woo, C.~McCann, K.~Russell, L.~Levine, R.~Sharma, D.~Khuntia,
  J.P. Perentesis, and J.C. Breneman.
\newblock {FAST-01: Results of the First-in-Human Study of Proton FLASH
  Radiotherapy}.
\newblock {\em International Journal of Radiation Oncology *Biology*Physics},
  114(3):S4, November 2022.

\bibitem{Mascia2023}
Anthony~E. Mascia, Emily~C. Daugherty, Yongbin Zhang, Eunsin Lee, Zhiyan Xiao,
  Mathieu Sertorio, Jennifer Woo, Lori~R. Backus, Julie~M. McDonald, Claire
  McCann, Kenneth Russell, Lisa Levine, Ricky~A. Sharma, Dee Khuntia,
  Jeffrey~D. Bradley, Charles~B. Simone, John~P. Perentesis, and John~C.
  Breneman.
\newblock {Proton FLASH Radiotherapy for the Treatment of Symptomatic Bone
  Metastases: The FAST-01 Nonrandomized Trial}.
\newblock {\em JAMA Oncology}, 9(1):62, January 2023.

\bibitem{Berthold2021}
Jonathan Berthold, Chirasak Khamfongkhruea, Johannes Petzoldt, Julia Thiele,
  Tobias Hölscher, Patrick Wohlfahrt, Nils Peters, Angelina Jost, Christian
  Hofmann, Guillaume Janssens, Julien Smeets, and Christian Richter.
\newblock {First-In-Human Validation of CT-Based Proton Range Prediction Using
  Prompt Gamma Imaging in Prostate Cancer Treatments}.
\newblock {\em International Journal of Radiation Oncology*Biology*Physics},
  111:1033--1043, 11 2021.

\bibitem{NUPECC2014}
{Nuclear Physics European Collaboration Committee}.
\newblock {Nuclear Physics For Medicine}.
\newblock \url{https://www.nupecc.org/pub/npmed2014.pdf} (Accessed:
  2024-31-12).

\bibitem{Son2018}
Jaeman Son, Se~Lee, Youngkyung Lim, Sung Park, Kwanho Cho, Myonggeun Yoon, and
  Dongho Shin.
\newblock {Development of Optical Fiber Based Measurement System for the
  Verification of Entrance Dose Map in Pencil Beam Scanning Proton Beam}.
\newblock {\em Sensors}, 18(1):227, January 2018.

\bibitem{Parodi2018}
Katia Parodi and Jerimy~C. Polf.
\newblock \textit{In vivo} range verification in particle therapy.
\newblock {\em Medical Physics}, 45(11), November 2018.

\bibitem{Fischetti2020}
M.~Fischetti, G.~Baroni, G.~Battistoni, G.~Bisogni, P.~Cerello, M.~Ciocca,
  P.~De~Maria, M.~De~Simoni, B.~Di~Lullo, M.~Donetti, Y.~Dong, A.~Embriaco,
  V.~Ferrero, E.~Fiorina, G.~Franciosini, F.~Galante, A.~Kraan, C.~Luongo,
  M.~Magi, C.~Mancini-Terracciano, M.~Marafini, E.~Malekzadeh, I.~Mattei,
  E.~Mazzoni, R.~Mirabelli, A.~Mirandola, M.~Morrocchi, S.~Muraro, V.~Patera,
  F.~Pennazio, A.~Schiavi, A.~Sciubba, E.~Solfaroli~Camillocci, G.~Sportelli,
  S.~Tampellini, M.~Toppi, G.~Traini, S.~M. Valle, B.~Vischioni, V.~Vitolo, and
  A.~Sarti.
\newblock {Inter-fractional monitoring of $^{12}$C ions treatments: results
  from a clinical trial at the CNAO facility}.
\newblock {\em Scientific Reports}, 10(1), November 2020.

\bibitem{FlixBautista2024}
Renato Félix-Bautista, Laura Ghesquière-Diérickx, Pamela Ochoa-Parra,
  Laurent Kelleter, Gernot Echner, J\"{u}rgen Debus, Oliver J\"{a}kel, Mária
  Martišíková, and Tim Gehrke.
\newblock {Inhomogeneity detection within a head-sized phantom using tracking
  of charged nuclear fragments in ion beam therapy}.
\newblock {\em Physics in Medicine and Biology}, 69(22):225003, November 2024.

\bibitem{Marafini2017}
M.~Marafini, L.~Gasparini, R.~Mirabelli, D.~Pinci, V.~Patera, A.~Sciubba,
  E.~Spiriti, D.~Stoppa, G.~Traini, and A.~Sarti.
\newblock {MONDO: a neutron tracker for particle therapy secondary emission
  characterisation}.
\newblock {\em Physics in Medicine and Biology}, 62(8):3299–3312, March 2017.

\bibitem{Schellhammer2023}
Sonja~M. Schellhammer, Ilker Meric, Steffen L\"{o}ck, and Toni K\"{o}gler.
\newblock Hybrid treatment verification based on prompt gamma rays and fast
  neutrons: multivariate modelling for proton range determination.
\newblock {\em Frontiers in Physics}, 11, December 2023.

\bibitem{Llosa2023}
Gabriela Llosá and Magdalena Rafecas.
\newblock {Hybrid PET/Compton-camera imaging: an imager for the next
  generation}.
\newblock {\em The European Physical Journal Plus}, 138(3), March 2023.

\bibitem{Bisogni2016}
Maria~Giuseppina Bisogni, Andrea Attili, Giuseppe Battistoni, Nicola Belcari,
  Niccolo’ Camarlinghi, Piergiorgio Cerello, Silvia Coli, Alberto Del~Guerra,
  Alfredo Ferrari, Veronica Ferrero, Elisa Fiorina, Giuseppe Giraudo,
  Eleftheria Kostara, Matteo Morrocchi, Francesco Pennazio, Cristiana Peroni,
  Maria~Antonietta Piliero, Giovanni Pirrone, Angelo Rivetti, Manuel~D. Rolo,
  Valeria Rosso, Paola Sala, Giancarlo Sportelli, and Richard Wheadon.
\newblock {INSIDE in-beam positron emission tomography system for particle
  range monitoring in hadrontherapy}.
\newblock {\em Journal of Medical Imaging}, 4(1):011005, December 2016.

\bibitem{Min2006}
Chul-Hee Min, Chan~Hyeong Kim, Min-Young Youn, and Jong-Won Kim.
\newblock Prompt gamma measurements for locating the dose falloff region in the
  proton therapy.
\newblock {\em Applied Physics Letters}, 89(18), October 2006.

\bibitem{Kelleter2017}
Laurent Kelleter, Aleksandra Wrońska, Judith Besuglow, Adam Konefał, Karim
  Laihem, Johannes Leidner, Andrzej Magiera, Katia Parodi, Katarzyna Rusiecka,
  Achim Stahl, and Thomas Tessonnier.
\newblock Spectroscopic study of prompt-gamma emission for range verification
  in proton therapy.
\newblock {\em Physica Medica}, 34:7–17, February 2017.

\bibitem{Richter2016}
Christian Richter, Guntram Pausch, Steffen Barczyk, Marlen Priegnitz, Isabell
  Keitz, Julia Thiele, Julien Smeets, Francois~Vander Stappen, Luca Bombelli,
  Carlo Fiorini, Lucian Hotoiu, Irene Perali, Damien Prieels, Wolfgang
  Enghardt, and Michael Baumann.
\newblock First clinical application of a prompt gamma based \textit{in vivo}
  proton range verification system.
\newblock {\em Radiotherapy and Oncology}, 118(2):232–237, February 2016.

\bibitem{Xie2017}
Yunhe Xie, El~Hassane Bentefour, Guillaume Janssens, Julien Smeets,
  Fran\c{c}ois Vander~Stappen, Lucian Hotoiu, Lingshu Yin, Derek Dolney,
  Stephen Avery, Fionnbarr O’Grady, Damien Prieels, James McDonough,
  Timothy~D. Solberg, Robert~A. Lustig, Alexander Lin, and Boon-Keng~K. Teo.
\newblock {Prompt Gamma Imaging for In Vivo Range Verification of Pencil Beam
  Scanning Proton Therapy}.
\newblock {\em International Journal of Radiation Oncology*Biology*Physics},
  99(1):210–218, September 2017.

\bibitem{Smeets2016}
Julien Smeets, Frauke Roellinghoff, Guillaume Janssens, Irene Perali, Andrea
  Celani, Carlo Fiorini, Nicolas Freud, Etienne Testa, and Damien Prieels.
\newblock {Experimental Comparison of Knife-Edge and Multi-Parallel Slit
  Collimators for Prompt Gamma Imaging of Proton Pencil Beams}.
\newblock {\em Frontiers in Oncology}, 6, June 2016.

\bibitem{Koide2018}
Ayako Koide, Jun Kataoka, Takamitsu Masuda, Saku Mochizuki, Takanori Taya, Koki
  Sueoka, Leo Tagawa, Kazuya Fujieda, Takuya Maruhashi, Takuya Kurihara, and
  Taku Inaniwa.
\newblock {Precision imaging of 4.4~MeV gamma rays using a 3-D position
  sensitive Compton camera}.
\newblock {\em Scientific Reports}, 8:8116, 12 2018.

\bibitem{Draeger2018}
E.~Draeger, D.~Mackin, S.~Peterson, H.~Chen, S.~Avery, S.~Beddar, and J.~C.
  Polf.
\newblock {3D prompt gamma imaging for proton beam range verification}.
\newblock {\em Physics in Medicine and Biology}, 63(3):035019, 2018.

\bibitem{Babiano2020}
V.~Babiano, J.~Balibrea, L.~Caballero, D.~Calvo, I.~Ladarescu, J.~Lerendegui,
  S.~{Mira Prats}, and C.~Domingo-Pardo.
\newblock {First i-TED demonstrator: A Compton imager with Dynamic Electronic
  Collimation}.
\newblock {\em Nuclear Instruments and Methods in Physics Research Section A:
  Accelerators, Spectrometers, Detectors and Associated Equipment}, 953:163228,
  2020.

\bibitem{Munoz2021}
Enrique Muñoz, Ana Ros~García, Marina Borja-Lloret, John Barrio, Peter
  Dendooven, Josep Oliver, Ikechi Ozoemelam, Jorge Roser, and Gabriela Llosá.
\newblock {Proton} range verification with {MACA-}{CO II} {Compton} camera
  enhanced by a neural network for event selection.
\newblock {\em Scientific Reports}, 11:9325, 04 2021.

\bibitem{Golnik2014}
Christian Golnik, Fernando Hueso-González, Andreas M\"{u}ller, Peter
  Dendooven, Wolfgang Enghardt, Fine Fiedler, Thomas Kormoll, Katja Roemer,
  Johannes Petzoldt, Andreas Wagner, and Guntram Pausch.
\newblock Range assessment in particle therapy based on prompt-gamma ray timing
  measurements.
\newblock {\em Physics in Medicine and Biology}, 59(18):5399–5422, August
  2014.

\bibitem{Verburg2014}
Joost~M Verburg and Joao Seco.
\newblock Proton range verification through prompt gamma-ray spectroscopy.
\newblock {\em Physics in Medicine and Biology}, 59(23):7089–7106, November
  2014.

\bibitem{HuesoGonzalez2018}
Fernando Hueso-González, Moritz Rabe, Thomas~A Ruggieri, Thomas Bortfeld, and
  Joost~M Verburg.
\newblock A full-scale clinical prototype for proton range verification using
  prompt gamma-ray spectroscopy.
\newblock {\em Physics in Medicine and Biology}, 63(18):185019, September 2018.

\bibitem{Krimmer2017}
J.~Krimmer, G.~Angellier, L.~Balleyguier, D.~Dauvergne, N.~Freud, J.~Hérault,
  J.~M. Letang, H.~Mathez, M.~Pinto, E.~Testa, and Y.~Zoccarato.
\newblock A cost-effective monitoring technique in particle therapy via
  uncollimated prompt gamma peak integration.
\newblock {\em Applied Physics Letters}, 110(15), April 2017.

\bibitem{Pinto2014}
M.~Pinto, D.~Dauvergne, N.~Freud, J.~Krimmer, J.~M. Letang, C.~Ray,
  F.~Roellinghoff, and E.~Testa.
\newblock {Design optimisation of a TOF-based collimated camera prototype for
  online hadrontherapy monitoring}.
\newblock {\em Physics in Medici- ne and Biology}, 59(24):7653–7674, November
  2014.

\bibitem{Ku2023}
Youngmo Ku, Sehoon Choi, Jaeho Cho, Sehyun Jang, Jong~Hwi Jeong, Sung~Hun Kim,
  Sungkoo Cho, and Chan~Hyeong Kim.
\newblock {Tackling range uncertainty in proton therapy: Development and
  evaluation of a new multi-slit prompt-gamma camera (MSPGC) system}.
\newblock {\em Nuclear Engineering and Technology}, 55:3140--3149, 9 2023.

\bibitem{Ready2016}
J.~Ready, V.~Negut, L.~Mihailescu, and K.~Vetter.
\newblock {MO-FG-CAMPUS-JeP1-01: Prompt Gamma Imaging with a Multi-Knife-Edge
  Slit Collimator: Evaluation for Use in Proton Beam Range Verification}.
\newblock {\em Medical Physics}, 43(6Part31):3717–3717, June 2016.

\bibitem{Sun2020a}
Shifeng Sun, Yang Liu, and Xiaoping Ouyang.
\newblock Design and performance evaluation of a coded aperture imaging system
  for real-time prompt gamma-ray monitoring during proton therapy.
\newblock {\em Radiation Physics and Chemistry}, 174:108891, September 2020.

\bibitem{Hetzel2023}
Ronja Hetzel, Vitalii Urbanevych, Andreas Bolke, Jonas Kasper, Monika Kercz,
  Magdalena Kołodziej, Andrzej Magiera, Florian Mueller, Sara M\"{u}ller,
  Magdalena Rafecas, Katarzyna Rusiecka, David Schug, Volkmar Schulz, Achim
  Stahl, Bjoern Weissler, Ming-Liang Wong, and Aleksandra Wrońska.
\newblock Near-field coded-mask technique and its potential for proton therapy
  monitoring.
\newblock {\em Physics in Medicine and Biology}, 68(24):245028, December 2023.

\bibitem{Wronska2021}
Aleksandra Wrońska and Denis Dauvergne.
\newblock Range verification by means of prompt-gamma detection in particle
  therapy.
\newblock In Krzysztof Iniewski and Jan Iwańczyk, editors, {\em Radiation
  Detection Systems, vol. 2}, Devices, Circuits, and Systems. CRC
  Press/Routledge, 2021.

\bibitem{Anger1958}
Hal~O. Anger.
\newblock Scintillation camera.
\newblock {\em Review of Scientific Instruments}, 29(1):27–33, January 1958.

\bibitem{Anger1963}
H.~O. Anger and A.~Gottschalk.
\newblock Localization of brain tumors with the position scintillation camera.
\newblock {\em Journal of Nuclear Medicine (U.S.)}, Vol: 4, 1963.

\bibitem{Leo1994}
William~R. Leo.
\newblock {\em Techniques for Nuclear and Particle Physics Experiments: A
  How-to Approach}.
\newblock Springer Berlin Heidelberg, 1994.

\bibitem{Tavernier2009}
Stefaan Tavernier.
\newblock {\em Experimental Techniques in Nuclear and Particle Physics}.
\newblock Springer Berlin Heidelberg, 2009.

\bibitem{Roush1964}
M.L. Roush, M.A. Wilson, and W.F. Hornyak.
\newblock Pulse shape discrimination.
\newblock {\em Nuclear Instruments and Methods}, 31(1):112–124, December
  1964.

\bibitem{SiPMHamamatsuWeb}
Hamamatsu.
\newblock {What is an SiPM and how does it work?}
\newblock
  \url{https://hub.hamamatsu.com/us/en/technical-notes/mppc-sipms/what-is-an-SiPM-and-how-does-it-work.html}
  (Accessed: 2024-31-12).

\bibitem{Georgel2022}
Rachel Georgel, Konstantin Grygoryev, Simon Sorensen, Huihui Lu, Stefan
  Andersson-Engels, Ray Burke, and Daniel O’Hare.
\newblock {Silicon Photomultiplier—A High Dynamic Range, High Sensitivity
  Sensor for Bio-Photonics Applications}.
\newblock {\em Biosensors}, 12(10):793, September 2022.

\bibitem{Acerbi2019}
Fabio Acerbi and Stefan Gundacker.
\newblock {Understanding and simulating SiPMs}.
\newblock {\em Nuclear Instruments and Methods in Physics Research Section A:
  Accelerators, Spectrometers, Detectors and Associated Equipment},
  926:16–35, May 2019.

\bibitem{Pausch2020}
Guntram Pausch, Jonathan Berthold, Wolfgang Enghardt, Katja R\"{o}mer, Arno
  Straessner, Andreas Wagner, Theresa Werner, and Toni K\"{o}gler.
\newblock {Detection systems for range monitoring in proton therapy: Needs and
  challenges}.
\newblock {\em Nuclear Instruments and Methods in Physics Research Section A:
  Accelerators, Spectrometers, Detectors and Associated Equipment}, 954:161227,
  February 2020.

\bibitem{Krimmer2018}
J.~Krimmer, D.~Dauvergne, J.M. Létang, and E.~Testa.
\newblock {Prompt-gamma monitoring in hadrontherapy: A review}.
\newblock {\em Nuclear Instruments and Methods in Physics Research Section A:
  Accelerators, Spectrometers, Detectors and Associated Equipment},
  878:58–73, January 2018.

\bibitem{SiFiCC}
SiFi-CC.
\newblock {The SiFi-CC collaboration website}.
\newblock \url{https://bragg.if.uj.edu.pl/sificc} (Accessed: 2024-31-12).

\bibitem{Pinto2024}
Marco Pinto.
\newblock Prompt-gamma imaging in particle therapy.
\newblock {\em The European Physical Journal Plus}, 139(10), October 2024.

\bibitem{Schnfelder1973}
V.~Sch\"{o}nfelder, A.~Hirner, and K.~Schneider.
\newblock A telescope for soft gamma ray astronomy.
\newblock {\em Nuclear Instruments and Methods}, 107(2):385–394, March 1973.

\bibitem{Todd1974}
R.~W. Todd, J.~M. Nightingale, and D.~B. Everett.
\newblock A proposed $\gamma$ camera.
\newblock {\em Nature}, 251(5471):132–134, September 1974.

\bibitem{Phillips1997}
G.W. Phillips.
\newblock {Applications of Compton imaging in nuclear waste characterization
  and treaty verification}.
\newblock In {\em {1997 IEEE Nuclear Science Symposium Conference Record}},
  volume~31 of {\em NSSMIC-97}, page 362–364. IEEE, 1997.

\bibitem{Sweeney2014compton}
Anthony Sweeney.
\newblock {\em Compton imaging for homeland security}.
\newblock PhD thesis, University of Liverpool, 2014.

\bibitem{Roellinghoff2011}
F.~Roellinghoff, M.-H. Richard, M.~Chevallier, J.~Constanzo, D.~Dauvergne,
  N.~Freud, P.~Henriquet, F.~Le~Foulher, J.M. Létang, G.~Montarou, C.~Ray,
  E.~Testa, M.~Testa, and A.H. Walenta.
\newblock {Design of a Compton camera for 3D prompt-$\gamma$ imaging during ion
  beam therapy}.
\newblock {\em Nuclear Instruments and Methods in Physics Research Section A:
  Accelerators, Spectrometers, Detectors and Associated Equipment},
  648:S20–S23, August 2011.

\bibitem{Richard2009}
M.-H. Richard, M.~Chevallier, D.~Dauvergne, N.~Freud, P.~Henriquet,
  F.~Le~Foulher, J.~M. Letang, G.~Montarou, C.~Ray, F.~Roellinghoff, E.~Testa,
  M.~Testa, and A.~H. Walenta.
\newblock {Design study of a Compton camera for prompt gamma imaging during ion
  beam therapy}.
\newblock In {\em {2009 IEEE Nuclear Science Symposium Conference Record
  (NSS/MIC)}}, volume~99, page 4172–4175. IEEE, 2009.

\bibitem{Kurosawa2012}
Shunsuke Kurosawa, Hidetoshi Kubo, Kazuki Ueno, Shigeto Kabuki, Satoru Iwaki,
  Michiaki Takahashi, Kojiro Taniue, Naoki Higashi, Kentaro Miuchi, Toru
  Tanimori, Dogyun Kim, and Jongwon Kim.
\newblock Prompt gamma detection for range verification in proton therapy.
\newblock {\em Current Applied Physics}, 12(2):364–368, March 2012.

\bibitem{Llosa2013}
G.~Llosá, J.~Cabello, S.~Callier, J.E. Gillam, C.~Lacasta, M.~Rafecas,
  L.~Raux, C.~Solaz, V.~Stankova, C.~de~La~Taille, M.~Trovato, and J.~Barrio.
\newblock {First Compton telescope prototype based on continuous LaBr$_3$-SiPM
  detectors}.
\newblock {\em Nuclear Instruments and Methods in Physics Research Section A:
  Accelerators, Spectrometers, Detectors and Associated Equipment},
  718:130–133, 2013.

\bibitem{Barrientos2024}
L.~Barrientos, M.~Borja-Lloret, J.V. Casaña, P.~Dendooven, J.~García~López,
  F.~Hueso-González, M.C. Jiménez-Ramos, J.~Pérez-Curbelo, A.~Ros, J.~Roser,
  C.~Senra, R.~Viegas, and G.~Llosá.
\newblock {Gamma-ray sources imaging and test-beam results with MACACO III
  Compton camera}.
\newblock {\em Physica Medica}, 117:103199, January 2024.

\bibitem{Young1989}
Matt Young.
\newblock {The pinhole camera: Imaging without lenses or mirrors}.
\newblock {\em The Physics Teacher}, 27(9):648–655, December 1989.

\bibitem{MertzYoung1961}
L.~Mertz and N.O. Young.
\newblock {Fresnel transformations of images}.
\newblock Proc. Int. Conf. Optical Instruments and Techniques (Chapman and
  Hall, London), p. 305 (1961),
  \url{https://people.csail.mit.edu/bkph/courses/papers/Coded_Aperture/Fresnel_Transform_Mertz_Young.pdf}
  (Accessed: 2024-31-12).

\bibitem{Dicke1968}
R.~H. Dicke.
\newblock {Scatter-Hole Cameras for X-Rays and Gamma Rays}.
\newblock {\em The Astrophysical Journal}, 153:L101, August 1968.

\bibitem{Shepp1982}
L.~A. Shepp and Y.~Vardi.
\newblock {Maximum Likelihood Reconstruction for Emission Tomography}.
\newblock {\em IEEE Transactions on Medical Imaging}, 1(2):113–122, October
  1982.

\bibitem{Lange1984}
K.~Lange and R.~Carson.
\newblock {EM reconstruction algorithms for emission and transmission
  tomography}.
\newblock {\em IEEE Transactions on Medical Imaging}, 8(2):306--316, April
  1984.

\bibitem{Hudson1994}
H.M. Hudson and R.S. Larkin.
\newblock Accelerated image reconstruction using ordered subsets of projection
  data.
\newblock {\em IEEE Transactions on Medical Imaging}, 13(4):601– 609, 1994.

\bibitem{Mu2016}
Zhiping Mu, Lawrence~W. Dobrucki, and Yi-Hwa Liu.
\newblock {SPECT Imaging of 2-D and 3-D Distributed Sources with Near-Field
  Coded Aperture Collimation: Computer Simulation and Real Data Validation}.
\newblock {\em Journal of Medical and Biological Engineering}, 36(1):32–43,
  February 2016.

\bibitem{Beck2009}
Amir Beck and Marc Teboulle.
\newblock {A Fast Iterative Shrinkage-Thresholding Algorithm for Linear Inverse
  Problems}.
\newblock {\em SIAM Journal on Imaging Sciences}, 2(1):183–202, January 2009.

\bibitem{Busboom1998}
A.~Busboom, H.~Elders–Boll, and H.~D. Schotten.
\newblock {Uniformly Redundant Arrays}.
\newblock {\em Experimental Astronomy}, 8(2):97–123, 1998.

\bibitem{Gottesman1989}
Stephen~R. Gottesman and E.~E. Fenimore.
\newblock New family of binary arrays for coded aperture imaging.
\newblock {\em Applied Optics}, 28(20):4344, October 1989.

\bibitem{Rusiecka2021}
K.~Rusiecka, R.~Hetzel, J.~Kasper, M.~Kazemi~Kozani, N.~Kohlhase,
  M.~Kołodziej, R.~Lalik, A.~Magiera, W.~Migdał, M.~Rafecas, A.~Stahl,
  V.~Urbanevych, M.L. Wong, and A.~Wrońska.
\newblock {A systematic study of LYSO:Ce, LuAG:Ce and GAGG:Ce scintillating
  fibers properties}.
\newblock {\em Journal of Instrumentation}, 16(11):P11006, November 2021.

\bibitem{rusiecka2023sificc}
Katarzyna Rusiecka.
\newblock {\em {The SiFi-CC detector for beam range monitoring in proton
  therapy - characterization of components and a prototype detector module}}.
\newblock PhD thesis, Jagiellonian University, 2023.
\newblock \url{10.48550/arXiv.2306.10820}.

\bibitem{KasperPhD}
Jonas Kasper.
\newblock {\em {Optimisation of the SiFi-CC Compton camera for range
  verification in proton therapy through a genetic algorithm}}.
\newblock PhD thesis, RWTH Aachen University, 2022.
\newblock \url{10.18154/RWTH-2022-11279}.

\bibitem{KetekSiPMs}
{Broadcom}.
\newblock {AFBR-S4K11C0125B SiPM Data Sheet}.
\newblock \url{https://docs.broadcom.com/doc/AFBR-S4K11C0125B-SiPM-DS}
  (Accessed: 2025-01-01).

\bibitem{ElastosilPads}
Wacker.
\newblock {Elastosil RT 604 A/B}.
\newblock
  \url{https://www.wacker.com/h/en-us/c/elastosil-rt-604-ab/p/000009552}
  (Accessed: 2025-01-01).

\bibitem{DT5742}
CAEN.
\newblock {DT5742 desktop digitizer}.
\newblock \url{https://www.caen.it/products/dt5742/} (Accessed: 2025-01-01).

\bibitem{Frach2009}
T.~Frach, G.~Prescher, C.~Degenhardt, R.~de~Gruyter, A.~Schmitz, and
  R.~Ballizany.
\newblock {The digital silicon photomultiplier - Principle of operation and
  intrinsic detector performance}.
\newblock In {\em {2009 IEEE Nuclear Science Symposium Conference Record
  (NSS/MIC)}}. IEEE, 2009.

\bibitem{Weissler2015}
Bjoern Weissler, Pierre Gebhardt, Peter~M. Dueppenbecker, Jakob Wehner, David
  Schug, Christoph~W. Lerche, Benjamin Goldschmidt, Andre Salomon, Iris Verel,
  Edwin Heijman, Michael Perkuhn, Dirk Heberling, Rene~M. Botnar, Fabian
  Kiessling, and Volkmar Schulz.
\newblock {A Digital Preclinical PET/MRI Insert and Initial Results}.
\newblock {\em IEEE Transactions on Medical Imaging}, 34(11):2258–2270,
  November 2015.

\bibitem{AlFoil1060}
Ltd. Jiangsu Zhongzhilian Steel Industry~Co.
\newblock {1060 Aluminum Sheet}.
\newblock \url{https://www.zzlsteel.com/prodetails/49/417.html} (Accessed:
  2025-01-01).

\bibitem{TACwebsite}
{Taiwan Applied Crystal}.
\newblock {Taiwan Applied Crystal website}.
\newblock \url{https://www. tacrystal.com/} (Accessed: 2025-01-01).

\bibitem{BroadcomSiPMArrays}
Broadcom.
\newblock {AFBR-S4N44P164M SiPM Arrays}.
\newblock
  \url{https://docs.broadcom.com/doc/AFBR-S4N44P164M-DS-4x4-NUV-MT-Silicon-Photo-Multiplier-Array}
  (Accessed: 2025-01-01).

\bibitem{ResinAcquaClear}
{3D Jake}.
\newblock {3D Printer Resin Aqua-Clear Plus}.
\newblock
  \url{https://www.3djake.pl/phrozen/aqua-resin-clear-plus?sai=17691&gad_source=1&gclid=EAIaIQobChMIuLud9pbPiAMVdWlBAh1vPwbtEAQYASABEgJi8vD_BwE}
  (Accessed: 2025-01-01).

\bibitem{WongKolodziej2024}
M.L. Wong, M.~Kołodziej, K.~Briggl, R.~Hetzel, G.~Korcyl, R.~Lalik, A.~Malige,
  A.~Magiera, G.~Ostrzołek, K.~Rusiecka, A.~Stahl, V.~Urbanevych, M.~Wiebusch,
  and A.~Wrońska.
\newblock Comparison of readout systems for high-rate silicon photomultiplier
  applications.
\newblock {\em Journal of Instrumentation}, 19(01):P01019, January 2024.

\bibitem{fers5200}
{CAEN Technologies Inc.}
\newblock {FERS-5200 Front-End Readout System}.
\newblock \url{https://www.caentechnologies.com/product/fers-5200/} (Accessed:
  2025-01-01).

\bibitem{A5202_Manual_2023}
{CAEN}.
\newblock {A5202/DT5202: 64-Channel Citiroc-1A Unit for FERS-5200 (rev. 4, 05-
  2023)}.
\newblock \url{https://www.caen.it/products/a5202/} (Accessed: 2024-31-12).

\bibitem{DatasheetCitiroc1A}
{CAEN}.
\newblock {Citiroc-1A Datasheet (rev. 05.2019)}.
\newblock \url{https://www.caen.it/products/citiroc-1a/} (Accessed:
  2024-31-12).

\bibitem{CAEN_concentrator_board}
{CAEN}.
\newblock {DS7218 – FERS-5200 (rev.4, 03.2022)}.
\newblock \url{https://www.caen.it/?downloadfile=6627} (Accessed: 2025-01-01).

\bibitem{DT5742_chip}
Stefan Ritt.
\newblock {Design and performance of the 6 GHz waveform digitizing chip DRS4}.
\newblock In {\em {2008 IEEE Nuclear Science Symposium Conference Record
  (NSS/}{ MIC)}}, pages 1512--1515, 2008.

\bibitem{Wang2022}
K.~Wang, S.~Samaranayake, and A.~Estrade.
\newblock {Investigation of a digitizer for the plastic scintillation detectors
  of time-of-flight mass measurements}.
\newblock {\em Nuclear Instruments and Methods in Physics Research Section A:
  Accelerators, Spectrometers, Detectors and Associated Equipment},
  1027:166050, 2022.

\bibitem{Laudrain:2022fmh}
Antoine Laudrain.
\newblock {The CALICE AHCAL: a highly granular SiPM-on-tile hadron calorimeter
  prototype}.
\newblock {\em Journal of Physics: Conference Series}, 2374(1):012017, 2022.

\bibitem{CALICE:2013zlb}
K.~Francis et~al.
\newblock {Performance of the first prototype of the CALICE scintillator strip
  electromagnetic calorimeter}.
\newblock {\em Nuclear Instruments and Methods in Physics Research Section A:
  Accelerators, Spectrometers, Detectors and Associated Equipment},
  763:278--289, 2014.

\bibitem{CALICE}
CALICE.
\newblock {The CALICE Collaboration website}.
\newblock \url{https://twiki.cern.ch/twiki/bin/view/CALICE/} (Accessed:
  2025-01-01).

\bibitem{Briggl2014}
K.~Briggl, M.~Dorn, R.~Hagdorn, T.~Harion, H.~C. Schultz-Coulon, and W.~Shen.
\newblock {KLauS: an ASIC for silicon photomultiplier readout and its
  application in a setup for production testing of scintillating tiles}.
\newblock {\em Journal of Instrumentation}, 9(02):C02013–C02013, February
  2014.

\bibitem{Yuan2019}
Zhenxiong Yuan, Konrad Briggl, Huangshan Chen, Yonathan Munwes, Hans-Christian
  Schultz-Coulon, and Wei Shen.
\newblock {KLauS: A Low-power SiPM Readout ASIC for Highly Granular
  Calorimeters}.
\newblock In {\em {2019 IEEE Nuclear Science Symposium and Medical Imaging
  Conference (NSS/MIC)}}, volume~30, page 1–4. IEEE, 2019.

\bibitem{DIFRANCESCO2016194}
Agostino {Di Francesco}, Ricardo Bugalho, Luis Oliveira, Angelo Rivetti, Manuel
  Rolo, Jose~C. Silva, and Joao Varela.
\newblock {TOFPET 2: A high-performance circuit for PET time-of-flight}.
\newblock {\em Nuclear Instruments and Methods in Physics Research Section A:
  Accelerators, Spectrometers, Detectors and Associated Equipment},
  824:194--195, 2016.
\newblock Frontier Detectors for Frontier Physics: Proceedings of the 13th Pisa
  Meeting on Advanced Detectors.

\bibitem{FEBD}
PETsys.
\newblock {SiPM Readout System (v.15, 07.2023)}.
\newblock
  {\url{https://www.petsyselectronics.com/web/website/docs/products/product4/Long\%20Flyer\%20SiPM\%20readout\%20chain\_v15.pdf}
  (Accessed: 2025-01-01)}.

\bibitem{GSI_report_TwinPeaks}
{GSI Helmholtzzentrum f\"{u}r Schwerionenforschung, Darmstadt and FAIR GmbH,
  Darmstadt}.
\newblock {\em {GSI-FAIR Scientific Report 2021}}.
\newblock 2021.
\newblock \url{10.15120/GSI-2022-00454}.

\bibitem{TRBfamily}
{GSI Helmholtzzentrum für Schwerionenforschung}.
\newblock {The TRB family}.
\newblock \url{http://trb.gsi.de/} (Accessed: 2025-01-01).

\bibitem{Neiser:2013yma}
A.~Neiser et~al.
\newblock {TRB3: a 264 channel high precision TDC platform and its
  applications}.
\newblock {\em Journal of Instrumentation}, 8:C12043, 2013.

\bibitem{RUDIGIER2020163967}
M.~Rudigier, Zs. Podolyák, P.H. Regan, A.M. Bruce, S.~Lalkovski, R.L. Canavan,
  E.R. Gamba, O.~Roberts, I.~Burrows, D.M. Cullen, L.M. Fraile, L.~Gerhard,
  J.~Gerl, M.~Gorska, A.~Grant, J.~Jolie, V.~Karayonchev, N.~Kurz, W.~Korten,
  I.H. Lazarus, C.R. Nita, V.F.E. Pucknell, J.-M. Régis, H.~Schaffner,
  J.~Simpson, P.~Singh, C.M. Townsley, J.F. Smith, and J.~Vesic.
\newblock {FATIMA} — {FA}st {TIM}ing array for {DESPEC} at {FAIR}.
\newblock {\em Nuclear Instruments and Methods in Physics Research Section A:
  Accelerators, Spectrometers, Detectors and Associated Equipment}, 969:163967,
  2020.

\bibitem{Michel:2010ffa}
Jan Michel, Ingo Frohlich, Michael Bohmer, Grzegorz Korcyl, Ludwig Maier, Marek
  Palka, Joachim Stroth, Michael Traxler, and Sergey Yurevich.
\newblock {The HADES trigger and readout board network (TrbNet)}.
\newblock In {\em {2010 17th IEEE-NPSS Real Time Conference}}, page 1–5.
  IEEE, 2010.

\bibitem{Adamczewski-Musch:2015arx}
J.~Adamczewski-Musch, N.~Kurz, and S.~Linev.
\newblock {Developments and applications of DAQ framework DABC v2}.
\newblock {\em Journal of Physics: Conference Series}, 664(8):082027, 2015.

\bibitem{Anfre2007}
P.~Anfre, C.~Dujardin, J.-M. Fourmigue, B.~Hautefeuille, K.~Lebbou,
  C.~Ped-rini, D.~Perrodin, and O.~Tillement.
\newblock Evaluation of fiber-shaped {LYSO} for double readout gamma photon
  detection.
\newblock {\em {IEEE} Transactions on Nuclear Science}, 54(2):391--397, April
  2007.

\bibitem{PulseGeneratorWeb}
Keysight.
\newblock {81160A Pulse Function Arbitrary Noise Generator}.
\newblock
  \url{https://www.keysight.com/us/en/product/81160A/81160a-pulse-function-arbitrary-noise-generator.html}
  (Accessed: 2025-01-01).

\bibitem{blackoutFabric}
{Thorlabs, Inc.}
\newblock {BK5 Black Nylon, Polyurethane-Coated Fabric}.
\newblock \url{https://www.thorlabs.com/thorproduct.cfm?partnumber=BK5}
  (Accessed: 2025-01-01).

\bibitem{Kolodziej2025preprint}
Magdalena Kołodziej, Stephan Brons, Mikołaj Dubiel, George~N. Farah,
  Alexander Fenger, Ronja Hetzel, Jonas Kasper, Monika Kercz, Barbara
  Kołodziej, Linn Mielke, Gabriel Ostrzołek, Magdalena Rafecas, Jorge Roser,
  Katarzyna Rusiecka, Achim Stahl, Vitalii Urbanevych, Ming-Liang Wong, and
  Aleksandra Wrońska.
\newblock First experimental test of a coded-mask gamma camera for proton
  therapy monitoring, 2025.
\newblock arXiv preprint arXiv:2501.00666,
  \url{https://arxiv.org/abs/2501.00666}.

\bibitem{AlFoil8011}
Ltd Henan Tendeli Metallurgical Materials~Co.
\newblock {8011 Aluminum Sheet}.
\newblock
  \url{https://www.htmmgroup.com/aluminium-foil/alloy-8011-o-temper-household-aluminum-f.html}
  (Accessed: 2025-01-01).

\bibitem{Haberer2004}
Th. Haberer, J.~Debus, H.~Eickhoff, O.~J\"{a}kel, D.~Schulz-Ertner, and
  U.~Weber.
\newblock {The Heidelberg Ion Therapy Center}.
\newblock {\em Radiotherapy and Oncology}, 73:S186–S190, 2004.

\bibitem{HITCenterWeb}
{Heidelberg University Hospital}.
\newblock {Heidelberg Ion-beam Therapy Center}.
\newblock \url{https://www.klinikum.uni-heidelberg.de/} (Accessed: 2025-01-01).

\bibitem{PSTAR}
{National Institute of Standards and Technology}.
\newblock {The PSTAR program: Stopping Power and Range Tables for Protons}.
\newblock \url{https://physics.nist.gov/PhysRefData/Star/Text/PSTAR.html}
  (Accessed: 2025-01-01).

\bibitem{BroadcomOvervoltageApplicationNote}
Broadcom.
\newblock {AFBR-S4NxxPyy4M NUV-MT SiPM Performance Correlation} - application
  note.
\newblock
  \url{https://docs.broadcom.com/doc/AFBR-S4NxxPyy4M-NUV-MT-SiPM-Performance-Correlation}
  (Accessed: 2025-01-01).

\bibitem{AlvaSanchez2018}
H.~Alva-Sánchez, A.~Zepeda-Barrios, V.~D. Díaz-Martínez,
  T.~Murrieta-Rod-ríguez, A.~Martínez-Dávalos, and
  M.~Rodríguez-Villafuerte.
\newblock {Understanding the intrinsic radioactivity energy spectrum from
  $^{176}$Lu in LYSO/LSO scintillation crystals}.
\newblock {\em Scientific Reports}, 8(1), November 2018.

\bibitem{originalTOFPETSoftware}
{PETsys Electronics}.
\newblock {TOFPET2 Data Acquisition Software}.
\newblock \url{https://github.com/PETsys/sw_daq_tofpet2} (Accessed:
  2025-01-01).

\bibitem{githubDAQ}
{M. L. Wong}.
\newblock {Modified TOFPET2 Data Acquisition Software}.
\newblock \url{https://github.com/SiFi-CC/DAQ/tree/nonLinCorr} (Accessed:
  2025-01-01).

\bibitem{PetSYSSoftwareUserGuide}
{PETsys Electronics}.
\newblock {TOFPET2 ASIC SIPM Readout System software user guide, v.2022.04}.

\bibitem{githubSiFi}
{SiFi-CC collaboration}.
\newblock {SiFi framework}.
\newblock
  \url{https://github.com/SiFi-CC/sifi-framework/tree/4to1_classes_HIT_refactoring}
  (Accessed: 2025-01-01).

\bibitem{SRIM}
James~F. Ziegler.
\newblock {The Stopping and Range of Ions in Matter (SRIM) Software}.
\newblock \url{http://www.srim.org/} (Accessed: 2025-01-01).

\bibitem{Geant4}
S.~Agostinelli, J.~Allison, K.~Amako, J.~Apostolakis, H.~Araujo, P.~Arce,
  M.~Asai, D.~Axen, S.~Banerjee, G.~Barrand, F.~Behner, L.~Bellagamba,
  J.~Boudreau, L.~Broglia, A.~Brunengo, H.~Burkhardt, S.~Chauvie, J.~Chuma,
  R.~Chytracek, G.~Cooperman, G.~Cosmo, P.~Degtyarenko, A.~Dell'Acqua,
  G.~Depaola, D.~Dietrich, R.~Enami, A.~Feliciello, C.~Ferguson, H.~Fesefeldt,
  G.~Folger, F.~Foppiano, A.~Forti, S.~Garelli, S.~Giani, R.~Giannitrapani,
  D.~Gibin, J.J. {Gómez Cadenas}, I.~González, G.~{Gracia Abril},
  G.~Greeniaus, W.~Greiner, V.~Grichine, A.~Grossheim, S.~Guatelli,
  P.~Gumplinger, R.~Hamatsu, K.~Hashimoto, H.~Hasui, A.~Heikkinen, A.~Howard,
  V.~Ivanchenko, A.~Johnson, F.W. Jones, J.~Kallenbach, N.~Kanaya, M.~Kawabata,
  Y.~Kawabata, M.~Kawaguti, S.~Kelner, P.~Kent, A.~Kimura, T.~Kodama,
  R.~Kokoulin, M.~Kossov, H.~Kurashige, E.~Lamanna, T.~Lampén, V.~Lara,
  V.~Lefebure, F.~Lei, M.~Liendl, W.~Lockman, F.~Longo, S.~Magni, M.~Maire,
  E.~Medernach, K.~Minamimoto, P.~{Mora de Freitas}, Y.~Morita, K.~Murakami,
  M.~Nagamatu, R.~Nartallo, P.~Nieminen, T.~Nishimura, K.~Ohtsubo, M.~Okamura,
  S.~O'Neale, Y.~Oohata, K.~Paech, J.~Perl, A.~Pfeiffer, M.G. Pia, F.~Ranjard,
  A.~Rybin, S.~Sadilov, E.~{Di Salvo}, G.~Santin, T.~Sasaki, N.~Savvas,
  Y.~Sawada, S.~Scherer, S.~Sei, V.~Sirotenko, D.~Smith, N.~Starkov,
  H.~Stoecker, J.~Sulkimo, M.~Takahata, S.~Tanaka, E.~Tcherniaev, E.~{Safai
  Tehrani}, M.~Tropeano, P.~Truscott, H.~Uno, L.~Urban, P.~Urban, M.~Verderi,
  A.~Walkden, W.~Wander, H.~Weber, J.P. Wellisch, T.~Wenaus, D.C. Williams,
  D.~Wright, T.~Yamada, H.~Yoshida, and D.~Zschiesche.
\newblock {Geant4—a simulation toolkit}.
\newblock {\em Nuclear Instruments and Methods in Physics Research Section A:
  Accelerators, Spectrometers, Detectors and Associated Equipment},
  506(3):250--303, 2003.

\bibitem{PhysicsListGuide}
Geant4 Collaboration.
\newblock {Guide For Physics Lists}.
\newblock release 11.3,
  \url{https://geant4-userdoc.web.cern.ch/UsersGuides/PhysicsListGuide/fo/PhysicsListGuide.pdf}
  (Accessed: 2025-01-01).

\bibitem{GODDeSS}
E.~Dietz-Laursonn, T.~Hebbeker, A.~Künsken, M.~Merschmeyer, S.~Nieswand, and
  T.~Niggemann.
\newblock {GODDeSS: A Geant4 extension for easy modelling of optical detector
  components}.
\newblock {\em Journal of Instrumentation}, 12:P04026--P04026, 04 2017.

\bibitem{Rafecas2003}
M.~Rafecas, B.~Mosler, M.~Dietz, M.~Pogl, D.P. McElroy, and S.I. Ziegler.
\newblock {Use of a Monte-Carlo based probability matrix for 3D iterative
  reconstruction of MADPET-II data}.
\newblock In {\em {2003 IEEE Nuclear Science Symposium. Conference Record}},
  pages 1775--1779 Vol.3. IEEE, 2003.

\bibitem{Gueth2013}
P.~Gueth, D.~Dauvergne, N.~Freud, J.~M. Létang, C.~Ray, E.~Testa, and
  D.~Sarrut.
\newblock Machine learning-based patient specific prompt-gamma dose monitoring
  in proton therapy.
\newblock {\em Physics in Medicine and Biology}, 58(13):4563–4577, June 2013.

\bibitem{GaussianFilterPython}
{SciPy Documentation}.
\newblock {scipy.ndimage.gaussian\_filter function}.
\newblock
  \url{https://docs.scipy.org/doc/scipy/reference/generated/scipy.ndimage.gaussian_filter.html}
  (Accessed: 2025-01-01).

\bibitem{Efron1992}
Bradley Efron.
\newblock {\em Bootstrap Methods: Another Look at the Jackknife}, page
  569–593.
\newblock Springer New York, 1992.

\bibitem{TOFPET2overview}
A.~Di Francesco, R.~Bugalho, L.~Oliveira, L.~Pacher, A.~Rivetti, M.~Rolo, J.C.
  Silva, R.~Silva, and J.~Varela.
\newblock {TOFPET2: a high-performance ASIC for time and amplitude measurements
  of SiPM signals in time-of-flight applications}.
\newblock {\em Journal of Instrumentation}, 11(03):C03042–C03042, March 2016.

\bibitem{Bertschi2023}
Stefanie Bertschi, Kristin Stützer, Jonathan Berthold, Julian Pietsch, Julien
  Smeets, Guillaume Janssens, and Christian Richter.
\newblock {Potential margin reduction in prostate cancer proton therapy with
  prompt gamma imaging for online treatment verification}.
\newblock {\em {Physics and Imaging in Radiation Oncology}}, 26, 4 2023.

\end{thebibliography}
\end{document}